\documentclass[aps,pra,twocolumn,groupedaddress,asmsymb]{revtex4-1}

\usepackage{graphicx}
\usepackage{color}
\usepackage{subcaption}
\usepackage[caption=false]{subfig}
\usepackage{caption}
\usepackage[usenames,dvipsnames]{xcolor}

\newcommand{\be}{\begin{equation}}
\newcommand{\ee}{\end{equation}} 
\newcommand{\lb}{\label}

\newcommand{\br}{{\bf r}}
\newcommand{\bu}{{\bf u}}
\newcommand{\bx}{{\bf x}}

\newcommand{\bA}{{\bf A}}

\newcommand{\boeta}{\hbox{\boldmath $\eta$}}
\newcommand{\bomu}{\hbox{\boldmath $\mu$}}
\newcommand{\boxi}{\hbox{\boldmath $\xi$}}

\newcommand{\bdot}{\hbox{\boldmath {$\cdot$}}}
\newcommand{\grad}{\hbox{\boldmath {$\nabla$}}}
\newcommand{\btimes}{\hbox{\boldmath {$\times$}}}

\begin{document}


\title{Quantum Spontaneous Stochasticity}


\author{Gregory L. Eyink${\,\!}^{1,2}$ and Theodore D. Drivas${\,\!}^1$}
\affiliation{${\,\!}^1$Department of Applied Mathematics \& Statistics, The Johns Hopkins University, Baltimore, MD, USA}
\affiliation{${\,\!}^2$Department of Physics \& Astronomy, The Johns Hopkins University, Baltimore, MD, USA}


\date{\today}

\begin{abstract}
It is commonly believed that the quantum wave-function of a very massive particle with small initial 
uncertainties in both position and velocity (consistent with the uncertainty relation) will spread very 
slowly, so that the limiting dynamics is classical Newtonian and, in particular, deterministic. This 
argument assumes, however, that the classical motions for given initial data must be unique. It was 
discovered about two decades ago in the field of fluid turbulence that non-uniqueness due to 
``roughness'' of the (non-random) advecting velocity field may lead to stochastic motion of classical 
particles. Vanishingly small random perturbations are explosively magnified by turbulent Richardson 
diffusion in a ``nearly rough'' velocity field, so that motion remains stochastic as the noise disappears. 
This {\it classical spontaneous stochasticity} effect has great importance for magnetic reconnection and 
dynamo in astrophysics, for geophysical mixing problems, and for everyday manifestations of fluid 
turbulence. Simple analogies between stochastic particle motion in turbulence and quantum evolution 
(Wiener vs. Feynman path-integrals, Fokker-Planck vs. Schr\"odinger equations) suggest that there 
should also be {\it quantum spontaneous stochasticity} (QSS) driven by quantum fluctuations. We verify 
this idea for 1D models of a quantum particle in a repulsive potential which is ``nearly rough'', with 
$V(x)\sim -C|x|^{1+\alpha}$ at distances $|x|\gg\ell$ for some physical UV cut-off $\ell$, and for an initial Gaussian
minimum-uncertainty wave-packet centered at 0. We first consider the WKB limit with $\hbar/m\rightarrow 0$, 
then position-spread $\sigma\rightarrow 0.$ As in critical lattice systems, we show that the ``infrared'' limit of 
time $t\rightarrow \infty$ is equivalent to the ``continuum'' limit $\ell,\sigma\rightarrow 0,$ with $\mu=\ell/\sigma$ fixed. 
The corresponding scaling limit of the position density is non-deterministic, with equal probabilities of the 
particle following either of two, non-unique classical solutions and with super-ballistic, Richardson-like 
spreading of the wave-packet. The splitting of the wave-packet  into two peaks occurs in a very short time
$t\sim\tau_c(\mu) (\ell^{1+\alpha}/C)^{1/2}$ for small $\ell$. We derive the asymptotics of 
$\tau_c(\mu)$ for 
small and large $\mu$.  We also consider the scattering off the rough potential of minimum-uncertainty 
wave-packets with non-vanishing expectations $\langle x\rangle$, $\langle p\rangle$  and find that QSS 
occurs here too for careful fine-tuning of $\langle x\rangle$, $\langle p\rangle$, $\sigma$, $\ell$.  Beyond WKB, 
we also consider other semi-classical limits by a numerical solution of the Schr\"odinger equation, where we 
observe QSS both in position-space and in momentum-space.  Although the wave-function remains split into 
two widely separated branches in the classical limit, rapid phase oscillations within each branch prevent any 
coherent superposition. Thus, indeterminism persists in the classical limit for rough potentials, but not quantum 
entanglement.  These are unambiguous predictions of quantum mechanics which should be observable in 
laboratory experiments. We consider possible realizations by AMO, ultra-cold neutrons, and optical analogue 
experiments. We also briefly contrast QSS with Anderson localization.   
\end{abstract}


\maketitle

\section{Introduction}

The determinism of Newtonian particle dynamics is generally taken for granted. 
However, the uniqueness of solutions of classical initial-value problems for ordinary 
differential equations (ODE's) requires mathematical assumptions on smoothness of the 
velocity vector field, and there are simple counterexamples in every textbook on ODE's 
when these assumptions are not satisfied (e.g. see \cite{Hartman82}, section II.5.)
Such examples have engendered philosophical ruminations on acausality 
in classical dynamics \cite{Norton03}. This subject entered the realm of physics
in the fundamental work of Bernard, Gaw\c{e}dzki and Kupiainen on particle advection
in turbulent fluid flow \cite{Bernardetal98}, which predicted that Lagrangian particle 
trajectories must become non-unique and indeterministic for an exactly prescribed 
fluid velocity field and an exactly prescribed initial particle location, as a consequence of 
the ``roughness'' associated to a Kolmogorov energy spectrum. Those
authors pointed out that stochasticity in solutions of deterministic ODE's should occur 
in the situation physically realized by fluid turbulence with ``roughness" of the velocity field
only over a finite inertial-range of scales---and thus smooth velocity fields at sufficiently
small scales---when coupled with vanishingly small random perturbations to fluid trajectories.
This phenomenon was later dubbed {\it spontaneous stochasticity} \cite{Chavesetal03}.
It is, in retrospect, a direct consequence of the pioneering 1926 work of L. F. Richardson on 
turbulent particle dispersion \cite{Richardson26}, but long unappreciated. This phenomenon 
has profound consequences in astrophysics, for example, where spontaneous 
stochasticity of plasma fluid motions breaks the ideal constraint of magnetic flux-freezing even 
in the limit of infinite conductivity and allows magnetic reconnection to be fast \cite{LazarianVishniac99, 
VishniacLazarian99,Eyinketal11, Eyinketal13}. Analogous breaking of the  
Kelvin-Helmholtz theorems on vortex-line motions occurs in hydrodynamic turbulence 
\cite{Eyink06}, with consequences for turbulent mixing of vorticity in geophysical and engineering flows.  

The empirical verification of spontaneous stochasticity effects in fluid and plasma turbulence currently 
rests upon numerical simulations of the Navier-Stokes and magnetohydrodynamics equations 
\cite{Eyink11,Bitaneetal13,Eyinketal13}. The phenomenon has not yet been seen in controlled 
laboratory experiments of turbulent flows \cite{Bourgoinetal06}, presumably because of the 
technical difficulties in tracking advected particle pairs that start sufficiently close 
together. The best prospect for fluid turbulence experiments is to make a bigger flow apparatus 
with a larger Reynolds number. In this way, particles even at relatively large initial separations 
have sufficient time within the inertial-range to exhibit the effects. An alternative idea that we explore 
here is instead to go smaller and to search for analogous effects in the quantum world. There is in fact intrinsic interest in 
the question of how a quantum system will behave whose corresponding classical dynamics is (nearly) non-unique. 
As we shall see, this situation leads to quantum randomness surviving 
in the classical limit where one normally expects determinism to prevail. Before demonstrating 
this, we first review briefly the phenomenon of classical spontaneous stochasticity.

\section{Background}

The results of Richardson's theory of particle dispersion in turbulent flow can be most easily understood 
from a toy model for the growth of separation $r$ between a pair of particles advected by the flow:
\begin{equation}\label{eqn:traj}
dr/dt = C(\varepsilon r)^{1/3}, 
\end{equation}
where the righthand side represents the relative velocity $\delta u(r)$ of the particles at separation $r.$
The scaling with exponent 1/3 is consistent with the statistical theory of Kolmogorov \cite{Kolmogorov41}.
Solving by separation of variables gives the exact solution 
$$ r(t)=   \left(r_0^{2/3} + \frac{2}{3}C \varepsilon^{1/3} t\right)^{3/2} $$
which exhibits the Richardson scaling $r^2(t)\propto \varepsilon t^3$ \cite{Richardson26} 
for $t\gg r_0^{2/3}/\varepsilon^{1/3}.$ More surprisingly, however, 
$$ r(t) \rightarrow \left(\frac{2}{3}C \varepsilon^{1/3} t\right)^{3/2}>0 $$
as $r_0\rightarrow 0$. Thus, two particles at identically the same initial point, with $r_0=0,$ end up 
a finite distance apart at any positive time! One's natural expectation would be instead that 
$ r(t)\equiv 0  $ 
and that the two particles would move together for all time. In fact, these are just two examples 
of infinitely-many solutions of the initial value problem for equation (\ref{eqn:traj}) with $r_0=0.$ The most general 
solution has, for any waiting time $\tau\in [0,\infty],$ $r(t)=0$ for $t\leq \tau$ and then $r(t)=\left(\frac{2}{3}C 
\varepsilon^{1/3} (t-\tau)\right)^{3/2}$ for $t>\tau.$

The above toy problem is the standard example of general theorems on the non-uniqueness 
of solutions of the initial-value problem for an ODE:
\begin{equation}\label{eqn:particleTraj}
d\bx/dt=\bu(\bx,t), \,\,\,\,\, \bx(t_0)=\bx_0. 
\end{equation}
If the advecting velocity field satisfies a spatial {\it H\"{o}lder continuity condition}
\begin{equation}\label{eqn:holder}
 |\bu(\bx+\br)-\bu(\bx)| \leq C |\br|^\alpha
 \end{equation}
for $\alpha=1$ (a so-called Lipschitz condition) then solutions are unique, but examples like the previous one with $\alpha=1/3$ show that 
solutions need not be unique for $0<\alpha<1$ \cite{Hartman82}. The latter situation is the expected one for 
turbulence.  At the level of individual flow realizations, it was shown by Onsager 
\cite{Onsager49,EyinkSreenivasan06}
that the observed energy dissipation anomaly for turbulent flow requires that in the limit of infinite Reynolds 
number the fluid velocity field must have a H\"older exponent $\alpha\leq 1/3$ and, in particular, $\bu$ cannot 
remain space-differentiable as $Re\rightarrow\infty.$

Real fluid turbulence has, of course, a finite Reynolds number albeit possibly very large. Thus, the velocity 
field is ultimately smooth and satisfies a bound of the form (\ref{eqn:holder}) with $\alpha=1$ for $|\br|<\eta,$ 
a small length-scale set by viscosity $\nu$. For example, in the Kolmogorov theory, $\eta=\nu^{3/4}\varepsilon^{-1/4}.$  
Thus, particle trajectories are unique. However, it was realized in \cite{Bernardetal98} that the non-uniqueness
resurfaces for the fluid trajectories with small random perturbations to the dynamics or to the initial data. 
That paper considered the specific problem of stochastic particle trajectories satisfying a Langevin equation
\begin{equation}\label{eqn:stochTraj}
d\tilde{\bx}/dt= \bu_\nu(\tilde{\bx},t) + \sqrt{2\kappa}\ \tilde{\boeta}(t), 
\,\,\,\, \tilde{\bx}(t_0)=\bx_0+ \rho\ \tilde{\boxi}
\end{equation}
with $\bu_\nu(\bx,t)$ the turbulent fluid velocity field, with $\tilde{\boeta}(t)$ a vector white-noise, and with $\tilde{\boxi}$ 
any random vector with zero mean and finite variance. The subscript $\nu$ stands for the (kinematic) viscosity 
of the fluid, which is a finite, positive value. This stochastic evolution is relevant to the problem 
of an advected passive scalar $\theta$ (e.g. temperature or dye concentration) satisfying   
\begin{equation}\label{eqn:passiveScalar}
 \partial_t\theta(\bx,t) +\bu_\nu(\bx,t)\bdot\grad\theta(\bx,t)=\kappa\triangle\theta(\bx,t). 
 \end{equation}
In fact, if $p_{\nu}(\bx',t'|\bx,t)$ is the transition probability density for the stochastic evolution, then 
$$ \theta(\bx,t) = \int d^dx_0 \ \theta(\bx_0,t_0) \ p_{\nu}(\bx_0,t_0|\bx,t), \,\,\,\, t_0<t. $$
It is convenient to express the transition probability by a Wiener path-integral 
\cite{ShraimanSiggia94,Bernardetal98,Eyink11}
\begin{eqnarray} \label{eqn:pathInt}
&& p_\nu(\bx,t|\bx_0,t_0) =\int_{\boxi(t_0)=\bx_0} {\mathcal D}\boxi \ \delta^d(\bx-\boxi(t)) \cr
&& \hspace{20pt} \times \exp\left(-\frac{1}{4\kappa}\int_{t_0}^{t} ds |\dot{\boxi}(s)-\bu_\nu(\boxi(s),s)|^2\right) 
\end{eqnarray} 
If the velocity field $\bu$ remains smooth (i.e. if the viscosity $\nu$ is held fixed), then a steepest descent 
argument on the path-integral shows that as $\kappa\rightarrow 0$
$$ p_\nu(\bx,t|\bx_0,t_0) \rightarrow \delta^d(\bx-\bx_t(\bx_0)) $$
where $\bx_t(\bx_0)$ is the unique solution of the initial-value problem (\ref{eqn:particleTraj}). However, if the turbulent velocity 
field becomes only H\"{o}lder continuous, as when taking the limit $\nu,\kappa\rightarrow 0$ together, then 
it was shown in \cite{Bernardetal98} that a non-trivial limit $p_*(\bx,t|\bx_0,t_0)$ may result.  Thus, the particle 
trajectories remain random as the external noise is removed and the deterministic problem (\ref{eqn:particleTraj}) is recovered. 

There is a close analogy of this ``spontaneous stochasticity'' with the zero-temperature phase transition in 
the 1D Ising model \cite{EyinkDrivas15}. We may write the 1D Ising model in finite-volume $[-N,...,N]$
as a Gibbs distribution $ P_N[\sigma]=\frac{1}{Z} \exp\left(-\frac{1}{k_BT} H[\sigma]\right)$ for Hamiltonian 
$$ H[\sigma]= \frac{J}{2}\sum_{{\tiny \begin{array}{l}
                                                                                                                                                     i=-N\cr
                                                                                                                                                     \sigma_{-N}=+1
                                                                                                                                                     \end{array}}}^N (\sigma_i-\sigma_{i+1})^2$$
with boundary condition $\sigma_{-N}=+1$ and free boundary condition on $\sigma_{N}$. To emphasize the analogy 
with the path-integral (\ref{eqn:pathInt}),  we have rewritten the Ising model Hamiltonian as a sum of squares and imposed a 
boundary condition only at the left endpoint. In the zero-temperature limit at fixed $N$, 
$$ P_N[\sigma] \rightarrow \prod_{i=-N}^N \delta_{\sigma_i,+1} \,\,\,\, {\rm as} \,\,\,\, T\rightarrow 0,$$ 
the unique ground-state with $\sigma_{-N}=+1.$  On the other hand, if the infinite-volume limit $N\rightarrow\infty$
is taken first before taking the zero-temperature limit, then 
$$ P_\infty[\sigma] \rightarrow \frac{1}{2}\prod_i \delta_{\sigma_i,+1} +\frac{1}{2}\prod_i \delta_{\sigma_i,-1} \,\,\,\, {\rm as} \,\,\,\, T\rightarrow 0,$$ 
the symmetric mixture of infinite-volume ground-states. This is the well-known zero-temperature phase transition
in the 1D Ising model. The phenomenon discovered in \cite{Bernardetal98} for non-smooth dynamical systems 
is exactly analogous, with the viscosity $\nu$ corresponding to $N,$ the noise $\kappa$ to temperature $T,$
and the non-unique solutions of the initial-value problem (\ref{eqn:particleTraj}) corresponding to the non-unique 
ground states of the infinite-volume Ising model.            

The results of \cite{Bernardetal98} were obtained for a soluble turbulence model, the Kraichnan model
\cite{Kraichnan68,Falkovichetal01}, in which the fluid velocity field is replaced by a ``synthetic turbulence" 
consisting of a Gaussian random field which is delta-correlated in time. The results on spontaneous
stochasticity hold with probability one for every realization of the Gaussian velocity field. These 
results in the Kraichnan model have been confirmed and further elaborated in many subsequent works
\cite{GawedzkiVergassola00,Chavesetal03,EvandenEijnden00,EvandenEijnden01,LeJanRaimond02,LeJanRaimond04}.  
The current evidence of these effects in fluid turbulence described by the Navier-Stokes equations 
consists of numerical studies of mean-square pair-dispersion $\langle r^2(t)\rangle,$ where the average is 
over the initial space position $\bx$ of the pair in a single flow realization. These studies confirm the 
Richardson prediction $\langle r^2(t)\rangle\propto \varepsilon t^3$ and, even more importantly, 
provide evidence that the pair-dispersion for particles evolving under the stochastic evolution (\ref{eqn:stochTraj}) 
becomes independent of $\kappa$ after a short time of order $(\kappa/\varepsilon)^{1/2}$ \cite{Eyink11},
or, similarly, that the pair-dispersion for particles evolving under the deterministic equation (\ref{eqn:traj}) 
becomes independent of the initial separation $r_0$ after a time of order $r_0^{2/3}/\varepsilon^{1/3}$ 
\cite{Bitaneetal13}. Indirect evidence for spontaneous stochasticity is also afforded by 
experimental observation of anomalous dissipation of passive scalars in fluid turbulent flows \cite{DrivasEyink15}. 

There is a close similarity of the path-integral (\ref{eqn:pathInt}) for the transition probability of the 
stochastic advection problem and the Feynman path-integral for the transition amplitude in quantum 
theory \cite{Feynman48}.  Consider a non-relativistic quantum-mechanical particle of mass $m$ and 
electric charge $q$ moving in an electric field ${\bf E}=-\grad\Phi-(1/c)\partial_t\bA$ 
and magnetic field ${\bf B} =\grad\btimes\bA $ governed by the Schr\"odinger equation 
$$ i\hbar \partial_t\psi = \frac{1}{2m}\left(-i\hbar\grad-\frac{q}{c}\bA\right)^2\psi+q\Phi\psi. $$
Then Feynman represented the transition amplitudes by the formula
$$ \langle \bx,t|\bx_0,0\rangle =\int_{\bx(0)=\bx_0}^{\bx(t)=\bx} {\mathcal D}\bx 
\exp\left(\frac{i}{\hbar}\int_0^t ds\ L(\bx(s),\dot{\bx}(s),s)\right), $$
where the classical Lagrangian is 
$$ L(\bx,\dot{\bx},t)=\frac{1}{2}m|\dot{\bx}|^2+\frac{q}{c}\bA(\bx,t)\cdot\dot{\bx}-q\Phi(\bx,t). $$
Not only is Feynman's formula analogous to (\ref{eqn:pathInt}), but it can be used to derive (\ref{eqn:pathInt})  
(\cite{Eyink11}, Appendix). 
This close relation naturally raises the question whether {\it quantum-spontaneous stochasticity} effects
can arise in the classical limit, given formally by  $\hbar\rightarrow 0.$ 
When ${\bf E},{\bf B}$ are Lipschitz, then the classical equations of motion 
$$ m\ddot{\bx} =q\left[{\bf E}(\bx,t) + \frac{1}{c}\dot{\bx}\ \btimes\ {\bf B}(\bx,t)\right] $$
have unique solutions and the stationary phase argument of Feynman \cite{Feynman48} 
yields classical dynamics for $\hbar\rightarrow 0.$ However, if the electromagnetic fields are non-Lipschitz, 
then quantum randomness could possibly persist in the classical limit. 

Similar questions have been raised previously in the literature. An interesting recent 
mathematical study \cite{AthanassoulisPaul12,Paul13} has considered the classical limit for the simplest 
1D Hamiltonian system whose classical solutions are non-unique. See eqs. (\ref{eqn:pot})-(\ref{eqn:classicalSol}) below. In 
this example a power-law potential leads to a force field which is H\"older continuous with an exponent 
$\alpha<1.$ It was proved in \cite{AthanassoulisPaul12} that the Wigner function for the quantum particle 
converges in the formal classical limit $\hbar\rightarrow 0$ to a non-trivial probability measure, so that 
indeterminacy persists in the limit. This mathematical result assumes, however, an exact power-law potential down 
to zero-length scale, i.e. to the Planck length and below! The physical validity of the nonrelativistic 
Schr\"odinger equation with a classical potential will break down well before this scale, e.g. quantum electrodynamic
effects such as particle pair-production will become relevant at the Compton wavelength $h/mc.$ In 
any case, an exact power-law down to zero length-scales should not be necessary to obtain such effects, 
just as \cite{Bernardetal98} showed for the problem of turbulent advection. The main aim of our work will
be to introduce physical short-distance cut-offs into the model of \cite{AthanassoulisPaul12}  and to show 
that quantum spontaneous stochasticity effects exist in experimentally accessible regimes. The need for 
such a short-distance cut-off was realized in  previous work of Bl\"umel \cite{Blumel98}, who studied a related 
phenomenon of ``quantum ray-splitting'' in discontinuous step-potentials (see also \cite{Couchmanetal92,Blumeletal96}).  
In this case also, quantum randomness persists in the classical limit of vanishing de Broglie wavelength $\lambda$ 
as the short-distance cut-off $\ell$ is taken simultaneously to zero. Such step-potentials lead, however, to very 
singular delta-function forces. The innovation of our work will be to show that such effects can occur with 
much smoother force fields which, even as the cut-off $\ell$ is removed, remain bounded and continuous.

\section{Models}\lb{models} 

Just as in \cite{AthanassoulisPaul12}, we consider the model of a single particle moving in a repulsive, scale-invariant 
cusp (or conic) potential energy field
\begin{equation} \label{eqn:pot}
V(x)=- \frac{C}{1+\alpha} |x|^{1+\alpha}, \,\,\,\, 0<\alpha<1. \lb{cusp} 
\end{equation}
In fact, we take $V$ to be the potential energy per mass $m$ of the particle, so the constant $C>0$ absorbs 
a factor of $1/m.$ The classical dynamics is thus given by the ODE
\begin{equation}\label{eqn:classicalODE}
 \dot{x}=v, \,\,\,\, \dot{v}=C |x|^{\alpha} {\rm sign}(x), 
 \end{equation}
which are Hamiltonian with $H=\frac{1}{2}v^2 + V(x).$ The flow vector field ${\bf U}(x,v)=(v,C |x|^{\alpha} {\rm sign}(x))$ 
is infinitely differentiable except at points of the form $(x,v)=(0,v_0),$ where it is only H\"older continuous with 
exponent $\alpha.$ In particular, 
$$|{\bf U}(r,0)-{\bf U}(0,0)|=C|r|^{\alpha}, $$
at point $(x,v)=(0,0).$ It is not hard to show that the solutions of the initial-value problem are unique at all
starting points except those that lead to solutions passing through $(x,v)=(0,0),$ where there are infinitely 
many solutions. The two extremal solutions starting at $(x_0,v_0)=(0,0)$ for initial time $t_0=0$ (which 
reach the farthest distance at each later time $t$) are 
\begin{equation}\label{eqn:classicalSol}
x_\pm(t) =\pm \left[\frac{1}{2}(1-\alpha)\left(\frac{2C}{1+\alpha}\right)^{1/2} t \right]^{2/(1-\alpha)}. \lb{xtreme}
\end{equation}
There are infinitely many other solutions for this same initial condition in which the particle waits at 
$x=0$ for a time $\tau\in [0,\infty],$ before moving off in the same manner. 

The general solution of the initial-value problem for (\ref{eqn:classicalODE}) can be obtained by 
quadrature using energy conservation, which reduces the system of two equations 
to a single first-order ODE for $|x(t)|$: 
\begin{equation}\label{eqn:cuspODE}
 \frac{d}{dt}|x|= \pm \sqrt{ \frac{2C}{1+\alpha}|x|^{1+\alpha}+H_0}, \,\,\,\,x(t_0)=x_0 
\end{equation}
with $\pm$ representing ${\rm sign}(x\cdot v)$ and $H_0=\frac{1}{2}v_0^2+V(x_0).$
For any $H_0<0,$ one can write $H_0=- \frac{C}{1+\alpha} |\bar{x}_0|^{1+\alpha}$ where $\bar{x}_0$ 
satisfying $|\bar{x}_0|<|x_0|$ is the position of a particle instantaneously at rest 
with the same energy.  Separating variables yields \cite{SupplMat} an implicit 
solution in terms of hypergeometric functions ( \cite{AbramowitzStegun12}, Ch.15; \cite{Batemanetal07}, Ch.II):
\begin{eqnarray}\label{eqn:cuspSol}
&& |x|^{(1-\alpha)/2}{\!\,}_2F_1\left(\frac{1}{2},-\frac{(1-\alpha)}{2(1+\alpha)}; \frac{1+3\alpha}{2(1+\alpha)}; +\left|\frac{\bar{x}_0}{x}\right|^{1+\alpha}\right) \cr
&& -|x_0|^{(1-\alpha)/2}{\!\,}_2F_1\left(\frac{1}{2},-\frac{(1-\alpha)}{2(1+\alpha)}; \frac{1+3\alpha}{2(1+\alpha)}; +\left|\frac{\bar{x}_0}{x_0}\right|^{1+\alpha}\right)\cr
&& \hspace{40pt}
=\pm \frac{1}{2}(1-\alpha)\left(\frac{2C}{1+\alpha}\right)^{1/2} (t-t_0). 
\end{eqnarray}
If ${\rm sign}(x_0\cdot v_0)<0,$ then the $-$ sign holds initially and $|x|$ decreases to $|\bar{x}_0|$,
whereupon the velocity reverses direction, $|x|$ increases and the $+$ sign holds.   
For $H_0>0$ there is a similar formula obtained by writing $H_0=\frac{C}{1+\alpha} |\bar{x}_0|^{1+\alpha}$, but with the $+$ sign 
in the main argument of the hypergeometric function changed to a $-$ sign. In this case, $v$ never changes sign, but 
$|x|$ may pass through 0 with  a corresponding of change of overall sign on the righthand side from $-$ to $+$.  
It is easy to see from the exact solution (\ref{eqn:cuspSol}) that, for all choices of initial conditions $(x_0,v_0)$, 
the ratio $x(t)/x_\pm(t)\rightarrow 1$ as $t\rightarrow\infty$ for one choice of $\pm$. This ``forgetting'' 
of the initial data is the essence of spontaneous stochasticity. 

A physically realizable potential will however exhibit a power-law scaling as in (\ref{eqn:pot}) only over a limited range of $x$-values,
with an ``inner" or short-distance cutoff $\ell$ and an ``outer" or large-distance cut-off $L.$ We shall not 
introduce an explicit cutoff $L$ at large-distances, but we can use our results with $L=\infty$ to estimate 
the practical effects of such a cut-off. We introduce two concrete models with a short-distance
cut-off $\ell.$ The first is a globally analytic potential given by a confluent hypergeometric 
or Kummer function
(\cite{AbramowitzStegun12}, Ch.13; \cite{Batemanetal07}, Ch.VI; \cite{Batemanetal54}, Ch.IV):
\begin{eqnarray} \label{eqn:KummerPot}
&& V_\ell(x)= - C\ell^{1+\alpha} \frac{2^{(1+\alpha)/2}\Gamma((2+\alpha)/2)}{(1+\alpha)\sqrt{\pi}} \cr 
&& \hspace{50pt} \times\  {\,\!}_1F_1\left(-\frac{1}{2}(1+\alpha), \frac{1}{2}; -\frac{x^2}{2\ell^2}\right)
\end{eqnarray} 
We shall refer to this as the ``Kummer potential". It can be obtained by convolving the power-law 
potential in (\ref{eqn:pot}) with a Gaussian smoothing kernel $\exp(-x^2/2\ell^2)/\sqrt{2\pi\ell^2},$ and thus becomes 
indistinguishable from the power-law for $|x|\gg \ell.$ The second model potential is chosen to coincide exactly 
with the power-law for $|x|\geq\ell$ but is only piecewise smooth with continuous first-derivatives at $|x|=\ell,$
given by the elementary formulas:
\begin{equation} \label{eqn:splicedPot}
V_\ell(x) = \left\{\begin{array}{ll}
                               -\frac{1}{2}C\ell^{1+\alpha}\left(\frac{1-\alpha}{1+\alpha}+\frac{x^2}{\ell^2}\right), & |x|<\ell \cr
                               -\frac{C}{1+\alpha}|x|^{1+\alpha}, & |x|\geq \ell
                              \end{array}  .
\right.
\end{equation}
We shall refer to this as the ``first-order spliced potential''. It has less generic behavior because of its lower
degree of spatial smoothness but it is more tractable for analytic calculations.   

The main problem that we shall consider is the quantum evolution in these potentials 
of an initial minimum-uncertainty Gaussian wave-packet 
\begin{equation}\label{eqn:GaussianWP}
 \psi_0(x)=\frac{1}{(2\pi\sigma^2)^{1/4}}\exp\left(-\frac{(x-\langle x\rangle)^2}{4\sigma^2}+\frac{1}{\hbar}i\langle p\rangle x\right)
 \end{equation}
with mean $\langle x\rangle$ and uncertainty $\sigma$ in position-space and mean $\langle p\rangle$ and uncertainty $\hbar/2\sigma$ in momentum-space. There is a clear analogy with the problem of turbulent 
transport, which we make explicit in the table below:  

\vspace{20pt}

\hspace{-15pt}\scalebox{.9}{
\vspace{20pt}
\begin{tabular}{llll}
\multicolumn{4}{c}{Turbulence-Quantum Analogy} \\
\hline
\hline
Turbulent Advection &  Quantum Mechanics \\
\hline
\hline
Fokker-Planck equation & Schr\"odinger equation \\
$ \partial_t p = \kappa\triangle p-\grad\bdot(\bu p)$ & $ i\partial_t\psi = \frac{\hbar}{2m}\triangle \psi+\frac{V}{\hbar} \psi, $ \\
\vspace{-5 pt} & \vspace{-5pt} \\ 
\hline
Velocity integral length $L$ & Outer length $L$ of potential \\
\hline
Dissipation length $\eta$  & Inner length $\ell$ of potential \\
\hline
Scalar diffusivity $\kappa$ & Quantum of circulation $h/m$ \\
\hline
Position uncertainty $\rho$  & Wavepacket spread $\sigma$ \\
\hline
Reynolds number $Re=(L/\eta)^{4/3}$  & Scale ratio $L/\ell$ of potential \\
\hline 
& & & \\
\end{tabular}
}

\vspace{10pt}

\noindent Note that the Fokker-Planck equation for the probability distribution $p(\bx,t)$ of an 
advected particle is the adjoint to the passive scalar equation (\ref{eqn:passiveScalar}).  Similar analogies between 
high-Reynolds-number turbulence and the classical limit of Schr\"odinger evolution have been 
pointed out already some time ago (e.g. \cite{Kraichnan75}, section 6), but little pursued 
up until now.

\section{WKB Analysis}

We shall first analyze our model problem in the limit where the de Broglie wavelength 
$\lambda=h/mv$ of the particle is much smaller than the inner length $\ell$ of the potential.
This is the usual semi-classical or WKB limit. In the turbulence-quantum analogy sketched 
above, this corresponds to the limit of very large Prandtl number $Pr=\nu/\kappa$. 
As discussed elsewhere (\cite{Eyink11}, section IV) classical spontaneous stochasticity 
effects are expected to be observable for Lagrangian particle dispersion in high-Prandtl number 
turbulent flows at sufficiently long times. This is an importance astrophysical regime appearing 
in the interstellar \cite{Lazarianetal04} and intracluster \cite{Ryuetal08} media and relevant to 
problems of star formation \cite{Lazarian14}.  It should be stressed that the WKB limit is 
not necessary for quantum spontaneous stochasticity but it is probably one of the easiest 
regimes to study experimentally and, also, it has the conceptual advantage that the relation
between the quantum and classical dynamics is transparent.  

We apply the standard time-dependent WKB approximation \cite{GottfriedYan03,ORaifeartaighWipf88,Goldfarbetal08},
which corresponds to making the ansatz 
$$ \psi(x,t) = \exp(i \Theta(x,t)/\hbar) $$
for the solution of the Schr\"odinger equation, together with the asymptotic expansion
$$ \Theta(x,t) = S(x,t) +  (\hbar/i) S_1(x,t) + (\hbar/i)^2 S_2(x,t) + \cdots .$$
Dropping quadratic terms and higher, this gives  
$$ \psi(x,t) = \sqrt{\rho(x,t)} \exp(i S(x,t)/\hbar) $$
with $\rho(x,t)=\exp(2 S_1(x,t)),$ and plugging into the Schr\"odinger equation yields 
the Hamilton-Jacobi equation (or eikonal equation) for the classical action $S(x,t):$
$$ \partial_t S(x,t)+ \frac{1}{2m}|\nabla S(x,t)|^2 +V(x)=0, $$
together with the conservation (transport) equation for the position probability density $\rho(x,t):$
$$ \partial_t\rho(x,t) + \nabla\cdot( v(x,t) \rho(x,t)) =0 $$
with $v(x,t)=\nabla S(x,t)/m$ the classical velocity. These equations are most easily solved 
by method of characteristics, using Hamilton's equations of motion 
\begin{equation} \label{eqn:HamiltonsEqn}
\dot{x}=p/m, \,\,\,\, \dot{p}= -\nabla V(x) 
\end{equation}
with the initial conditions $x(0)=x_0$ and $p(0)=\nabla S_0(x_0).$ Calling the resulting solutions
$x_t(x_0)$ and $p_t(x_0),$ it follows that $S(x_t(x_0),t)=S_t(x_0)$ with
\be S_t(x_0) = S_0(x_0) + \int_0^t ds\ [ \frac{1}{2m}p_s^2(x_0) - V(x_s(x_0))]. \lb{action} \ee
The probability conservation equation can likewise be integrated along characteristics as 
\begin{equation}\label{eqn:WKBprob}
 \rho(x_t(x_0),t) = \frac{\rho_0(x_0)}{J_t(x_0)} 
 \end{equation}
where we have introduced the Jacobian 
$$J_t(x_0) = \partial x_t(x_0)/\partial x_0=x_t'(x_0). $$
To calculate the latter it is most convenient to differentiate Hamilton's equations to obtain 
a set of ODE's 
\begin{equation} \label{eqn:JacobianODE}
\dot{J}=K, \,\,\,\, \dot{K}= -V''(x)J 
\end{equation}
for $J_t(x_0)$ and $K_t(x_0)=p_t'(x_0).$

We apply these standard results to our model problems. Our scaling of $V$ corresponds to setting 
$m=1.$  The initial Gaussian wave-packet (\ref{eqn:GaussianWP}) corresponds to taking
$$ S_0(x) = \langle p\rangle\cdot x, \,\,\,\, \rho_0(x)=\frac{1}{\sqrt{2\pi\sigma^2}}\exp\left(-\frac{(x-\langle x\rangle)^2}{2\sigma^2}\right). $$ 
In that case, we have initial conditions in the ODE's
$$ x(0)=x_0, \,\, p(0)=\langle p\rangle, \,\, J(0)=1, \,\, K(0)=0. $$
There are no caustics for this problem, so that the WKB approximation is globally valid 
in time. As usual, the WKB approximation is locally valid in space when
$$ \left| \lambda \frac{\partial}{\partial x}\left( \frac{p^2}{2m}\right)\right| \ll 
\max\left\{ \frac{p^2}{2m}, \left|\frac{\partial S}{\partial t}\right|,|V|  \right\}
$$
where $\lambda(x,t)=h/p(x,t)$  is a local de Broglie wavelength with $p(x,t)=\partial S(x,t)/\partial x.$ 
E.g. see \cite{GottfriedYan03}. This corresponds to our starting assumption $\lambda\ll \ell$ when 
the classical action $S(x,t)$ develops no smaller length-scales than in the potential $V(x).$ 
Note finally that the Schr\"odinger equation divided by $m$ is
\be i\frac{\hbar}{m}\partial_t\psi=-\frac{1}{2}\left(\frac{\hbar}{m}\right)^2\triangle\psi + V\psi \lb{Schrod-by-m} \ee 
where $V/m\longrightarrow V$ now represents in reality the potential energy per mass. 
Thus, holding $V$ fixed, the only dependence of $\hbar$ is through the ratio $\hbar/m$ and one may, 
by varying $m$, study the problem formally with $m=1$ and varying $\hbar.$  (Of course, the dependence of $V$ 
on $m$ must be remembered in comparing results with experiment.) 
\vspace{4pt}

\subsection{Scaling Limit}

We shall begin by studying in this and the following section the symmetrical problem with 
$\langle x\rangle=\langle p\rangle =0.$ We are interested here primarily 
in the probability density $\rho(x,t)$ for the location of the particle. Since in the WKB limit there is 
no dependence of $\rho(x,t)$ upon Planck's constant $\hbar,$ it contains only dimensional parameters 
$\sigma,\ell$ and $C$, in addition to $x,t$. It is natural to introduce a dimensionless {\it scaled position} 
$$ \hat{x} = x/x_+(t). $$
Then simple dimensional analysis gives
$$ \rho(x,t; \sigma,\ell,C)=\frac{1}{x_+(t)} \hat{\rho}(x/x_+(t),\tau; \mu) $$
where $\hat{\rho}$ is the probability density of $\hat{x}$ and
$$ \tau= t/(\ell^{1-\alpha}/C)^{1/2}, \,\,\,\, \mu=\ell/\sigma. $$  
This result tells us that while keeping the ratio $\mu=\ell/\sigma$ fixed, the limits 
$t\rightarrow\infty$ and $\ell,\sigma\rightarrow 0$ are identical. Quantum spontaneous 
stochasticity will be present if
\begin{equation}\label{eqn:rhoLim}
\hat{\rho}(\hat{x},\tau;\mu) \stackrel{\tau\rightarrow \infty}{\longrightarrow} \frac{1}{2}\delta(\hat{x}+1)
+ \frac{1}{2}\delta(\hat{x}-1),
\end{equation} 
or, equivalently, if at fixed time $t$ 
$$ \rho(x,t; \sigma,\ell,C) \longrightarrow\frac{1}{2}\delta(x-x_-(t))
+ \frac{1}{2}\delta(x-x_+(t)) $$ 
in the limit $\sigma,\ell\ll (C t^2)^{1/(1-\alpha)}$. Such a limit was obtained in \cite{AthanassoulisPaul12} 
for the exact power-law potential. We now establish it here for the model with UV cut-off $\ell.$

To prove this result, we smear $\hat{\rho}$ with a smooth test function $\varphi(\hat{x})$, giving
$$ \int d\hat{x} \ \varphi(\hat{x}) \hat{\rho}(\hat{x},\tau) = \int dx \ \varphi(x/x_+(t)) \rho(x,t). $$
Using the WKB formula (\ref{eqn:WKBprob}), this can furthermore be written as 
$$ \int d\hat{x} \ \varphi(\hat{x}) \hat{\rho}(\hat{x},\tau) = \int dx_0 \ \varphi(x_t(x_0)/x_+(t)) \rho_0(x_0),  $$ 
where we used the definition $J_t=\partial x_t/\partial x_0$ to change the integration variable to $x_0.$
The essence of the result is now to show that 
\begin{equation}\label{eqn:trajLim}
\lim_{t\rightarrow \infty} x_t(x_0)/x_+(t) = \pm 1, 
\end{equation}
with $\pm={\rm sign}(x_0).$ In fact, this yields a more general result than (\ref{eqn:rhoLim}), for any choice of $\rho_0(x)$
which contains no delta function part $\propto \delta(x).$ In that case, we can write 
\begin{eqnarray*}
\int d\hat{x} \ \varphi(\hat{x}) \hat{\rho}(\hat{x},\tau) &=& \int_{-\infty}^0 dx_0 \ \varphi(x_t(x_0)/x_+(t)) \rho_0(x_0). \cr
&+& \int_0^\infty dx_0 \ \varphi(x_t(x_0)/x_+(t)) \rho_0(x_0),
\end{eqnarray*}
and apply dominated convergence to infer that 
$$ \lim_{\tau\rightarrow\infty} \int d\hat{x} \ \varphi(\hat{x}) \hat{\rho}(\hat{x},\tau) 
     =\varphi(-1) p_- + \varphi(1) p_-, $$
with      
$$ p_-= \int_{-\infty}^0 dx_0 \ \rho_0(x_0), \,\,\,\, p_+= \int_0^\infty dx_0 \ \rho_0(x_0). $$  
This gives as a special case the result (\ref{eqn:rhoLim}) for any symmetrical choice of $\rho_0$ with $p_-=p_+=1/2.$

\begin{figure*}[!ht] 
  \begin{subfigure}[b]{0.5\linewidth}
    \centering
    \includegraphics[width=.9\columnwidth,height=0.5\columnwidth]{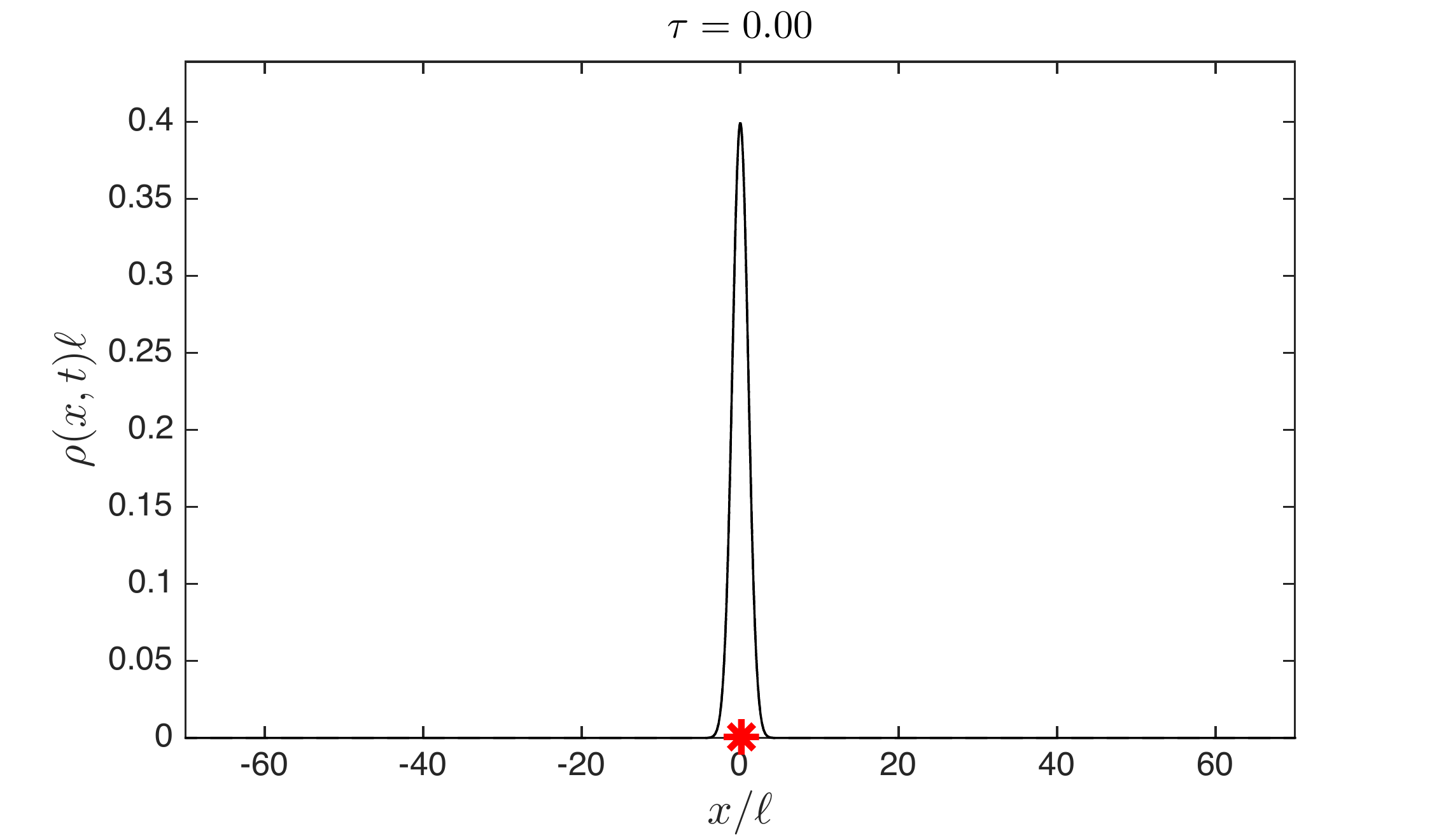} 
   \caption{} 
    \label{fig7:a} 
  \end{subfigure}
  \begin{subfigure}[b]{0.5\linewidth}
    \centering
    \includegraphics[width=.9\columnwidth,height=0.5\columnwidth]{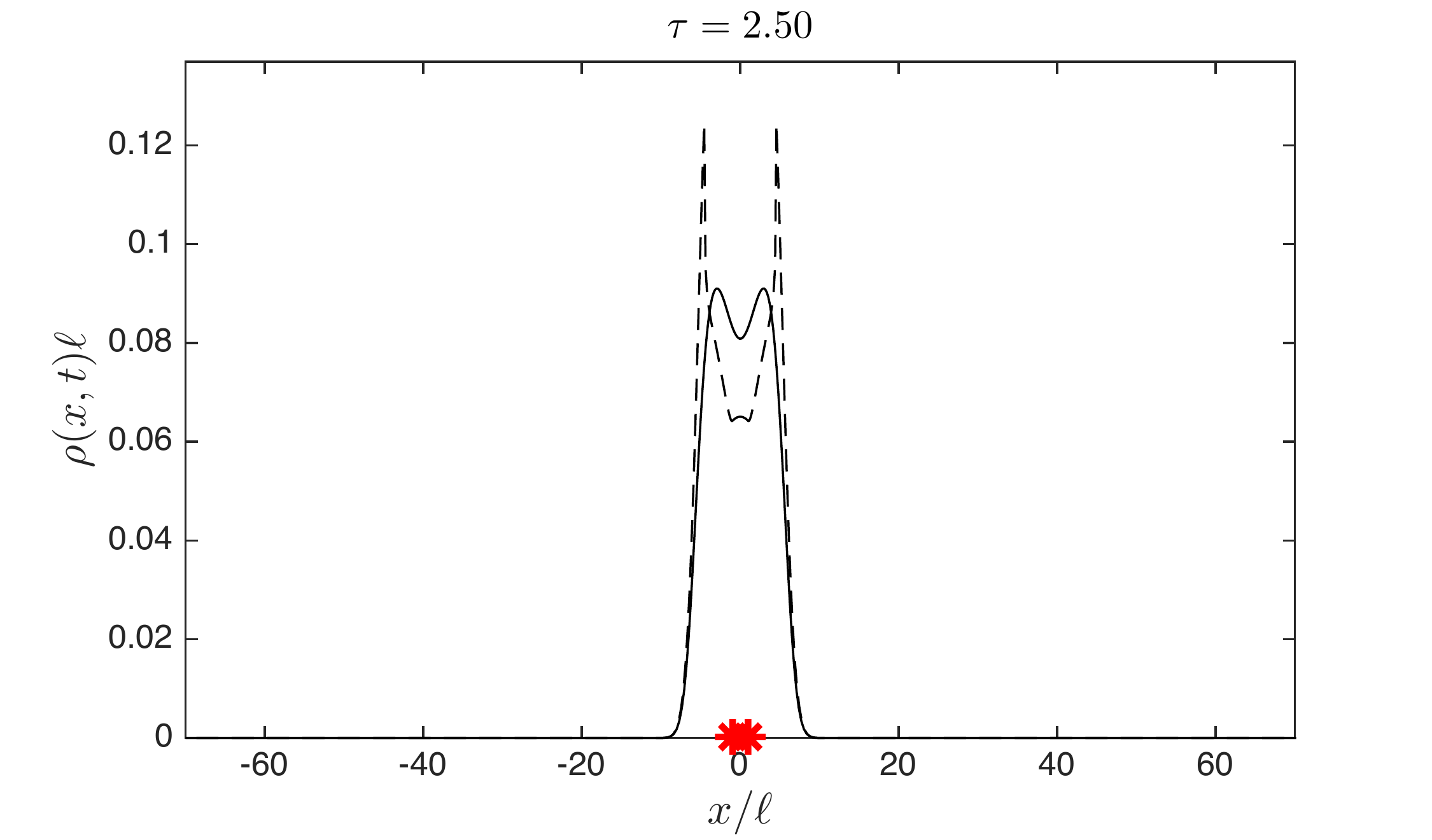} 
   \caption{} 
    \label{fig7:b} 
  \end{subfigure} 
  \begin{subfigure}[b]{0.5\linewidth}
    \centering
    \includegraphics[width=.9\columnwidth,height=0.5\columnwidth]{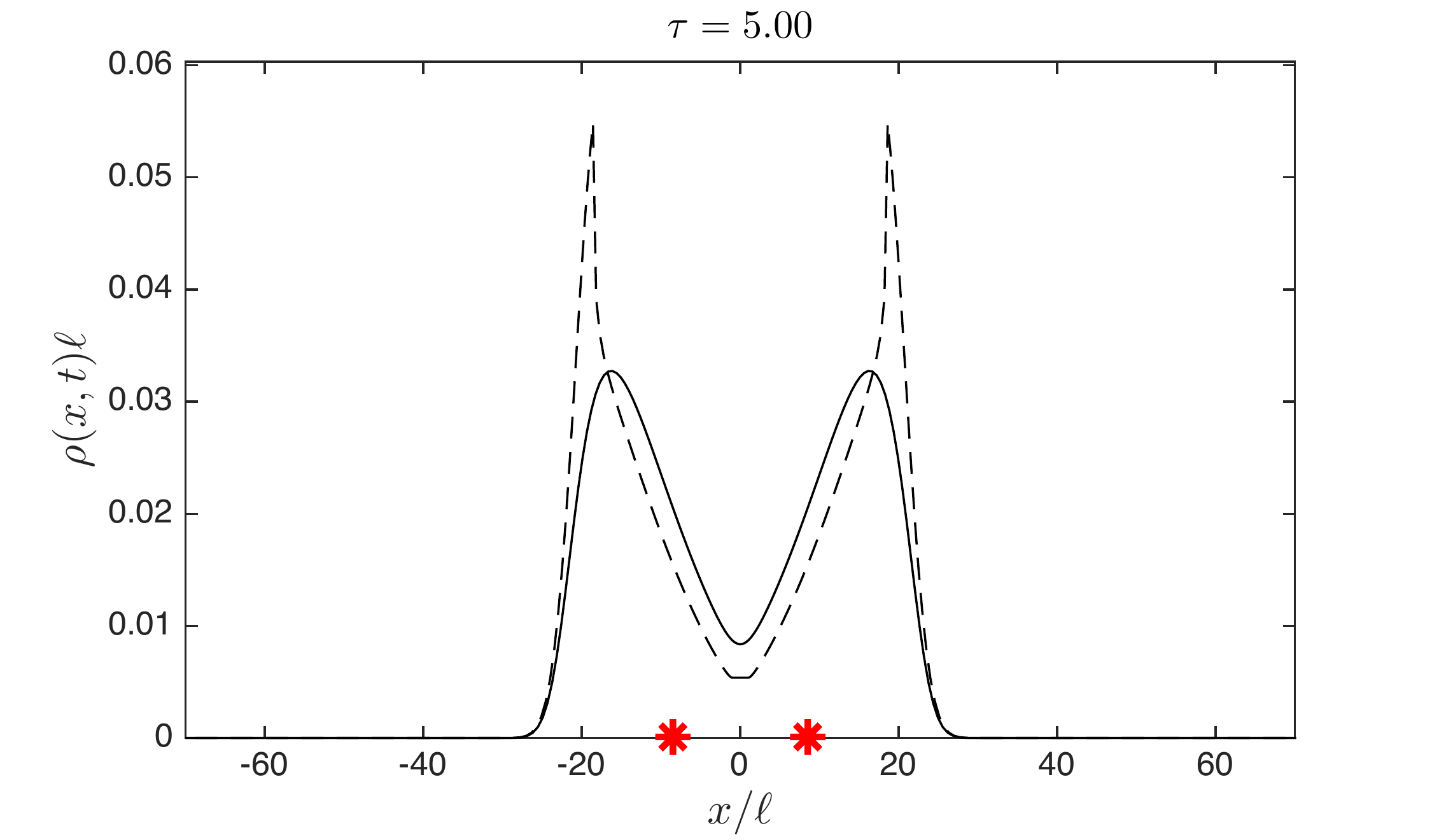} 
   \caption{} 
    \label{fig7:c} 
  \end{subfigure}
  \begin{subfigure}[b]{0.5\linewidth}
    \centering
    \includegraphics[width=.9\columnwidth,height=0.5\columnwidth]{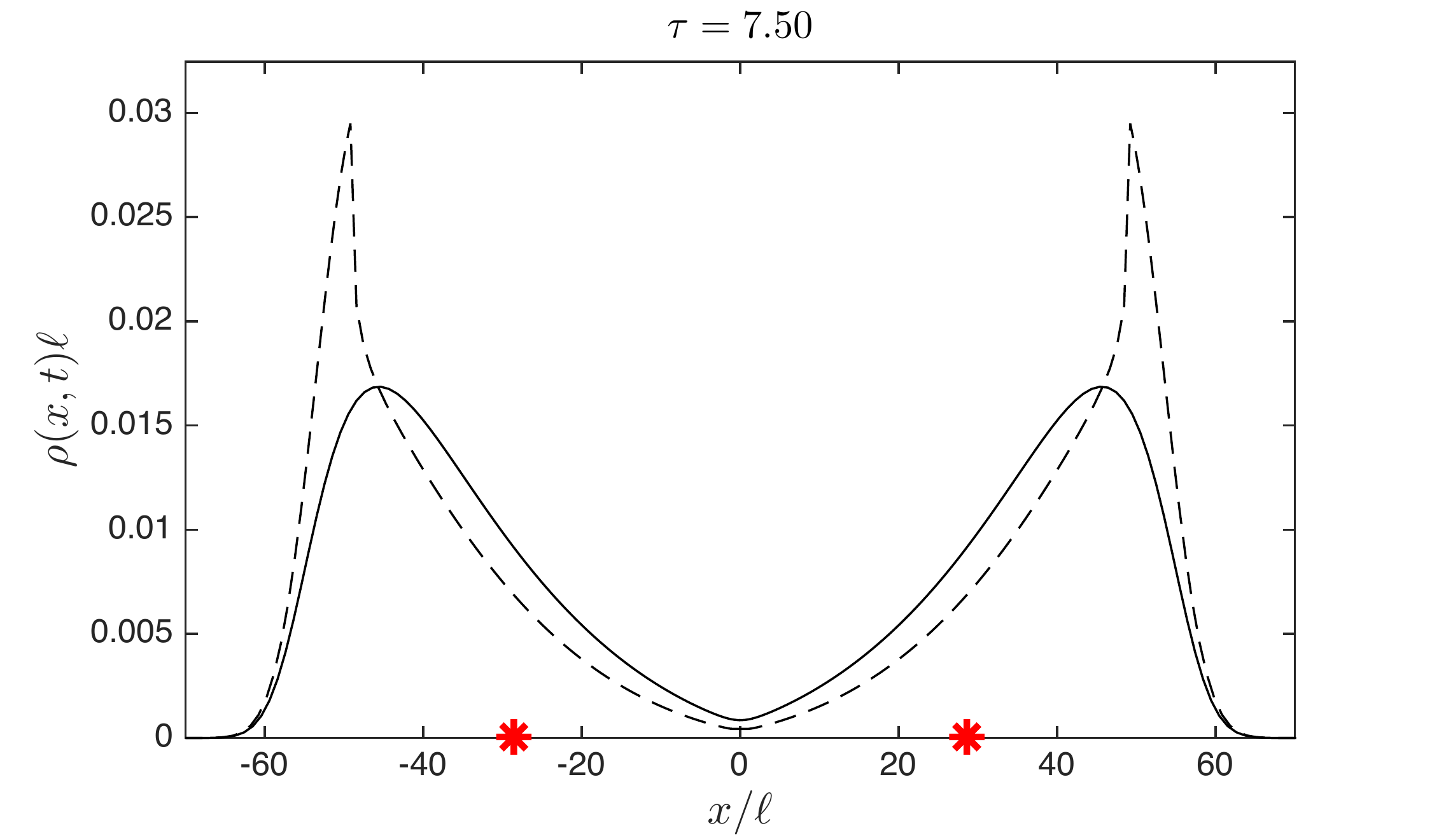} 
   \caption{} 
    \label{fig7:d} 
  \end{subfigure} 
  \caption{ \raggedright WKB position-space probability densities for $\mu=1$ and  dimensionless
   times $\tau= 0.0, 2.5, 5.0, 7.5$.  Solid is for the Kummer potential, dashed for the spliced potential, 
  both with $\alpha =1/3.$ The two asterisks on the abscissa indicate the locations of the extremal classical 
  solutions at the particular time of each snapshot. For movie, see \cite{SupplMat}.}
  \label{WKB_fixed_ell} 
\end{figure*}


\begin{figure*}[!ht] 
  \begin{subfigure}[b]{0.5\linewidth}
    \centering
    \includegraphics[width=.9\columnwidth,height=0.5\columnwidth]{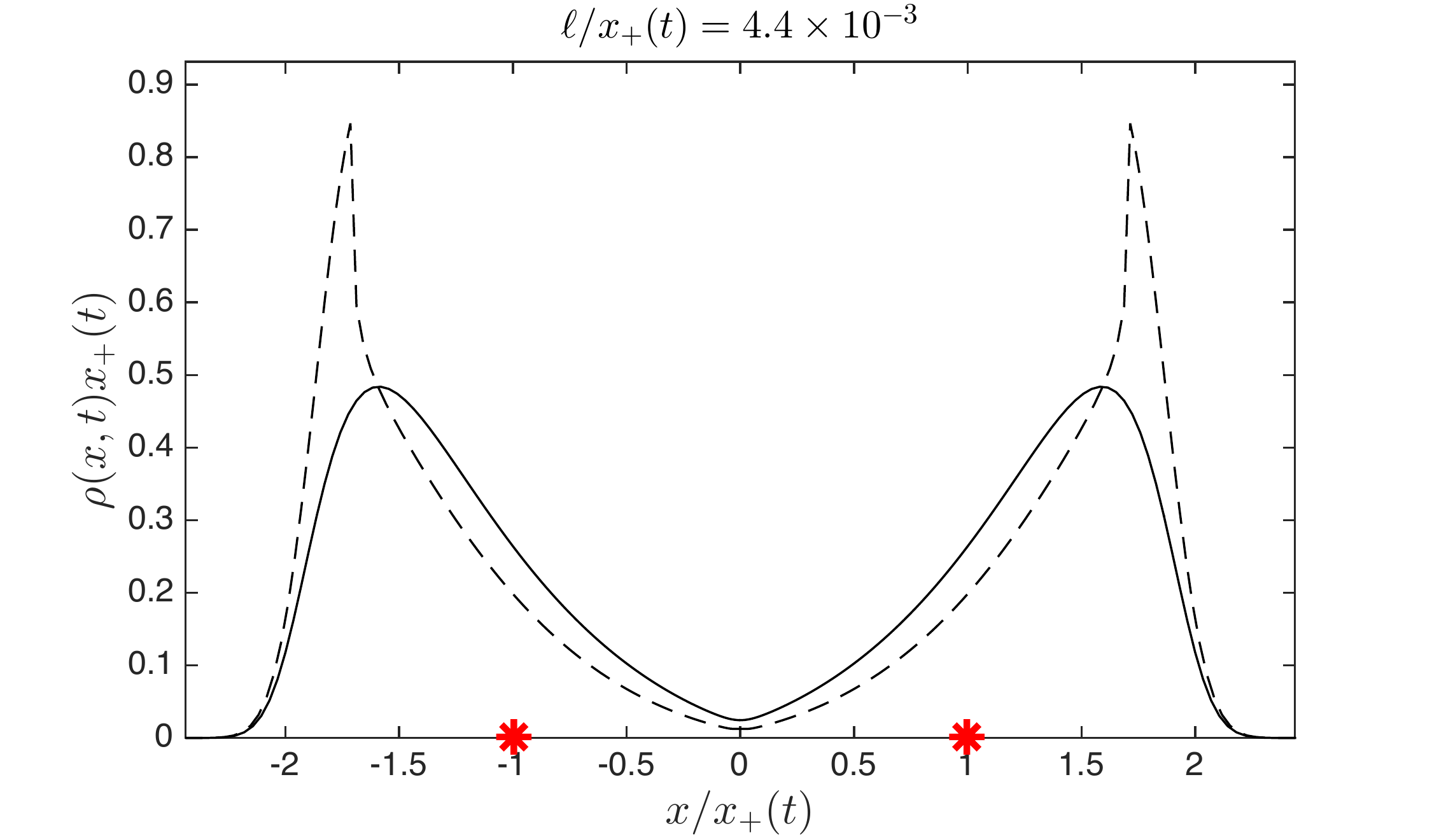} 
   \caption{} 
    \label{fig7:a} 
  \end{subfigure}
  \begin{subfigure}[b]{0.5\linewidth}
    \centering
    \includegraphics[width=.9\columnwidth,height=0.5\columnwidth]{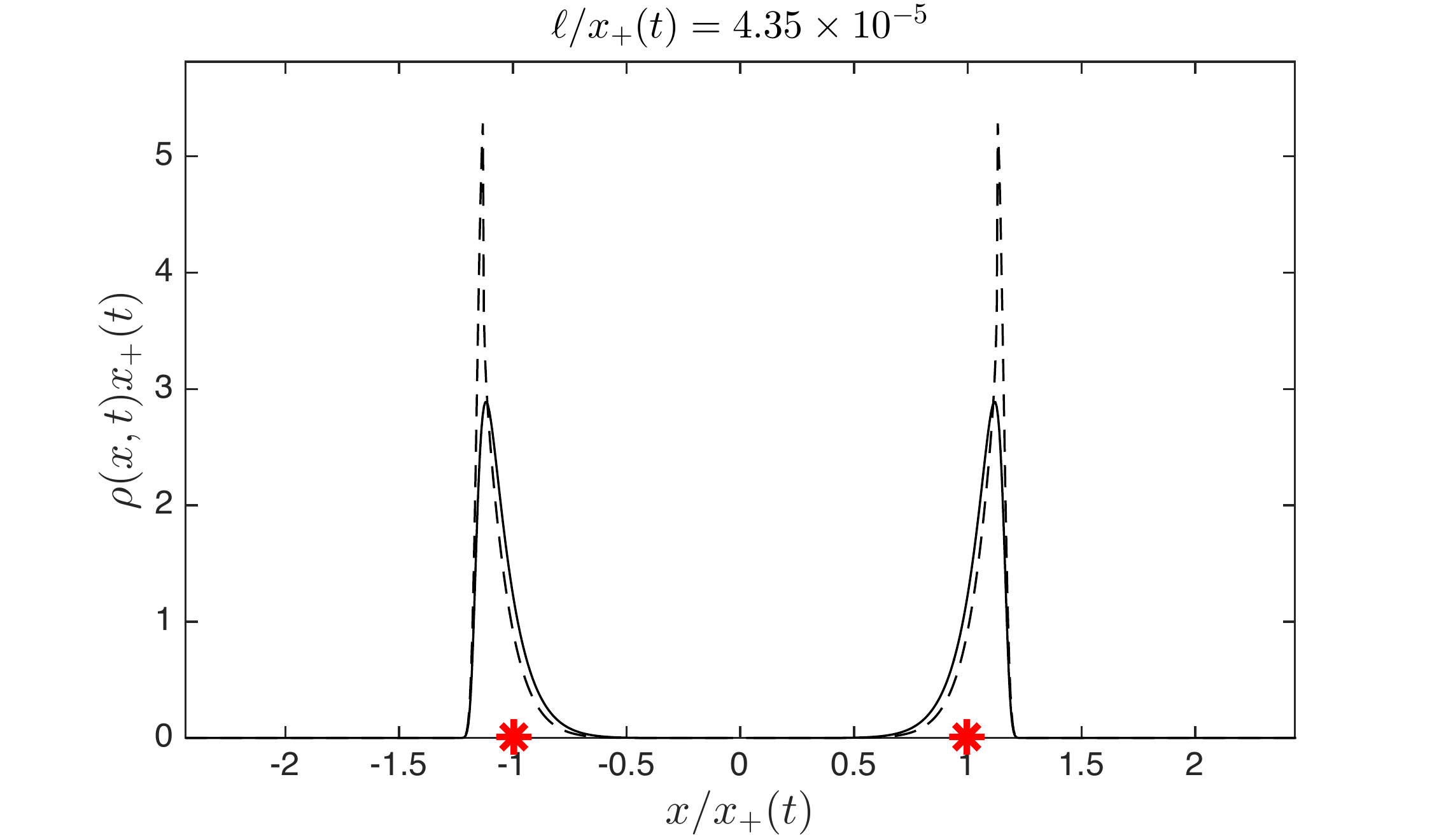} 
   \caption{} 
    \label{fig7:b} 
  \end{subfigure} 
  \begin{subfigure}[b]{0.5\linewidth}
    \centering
    \includegraphics[width=.9\columnwidth,height=0.5\columnwidth]{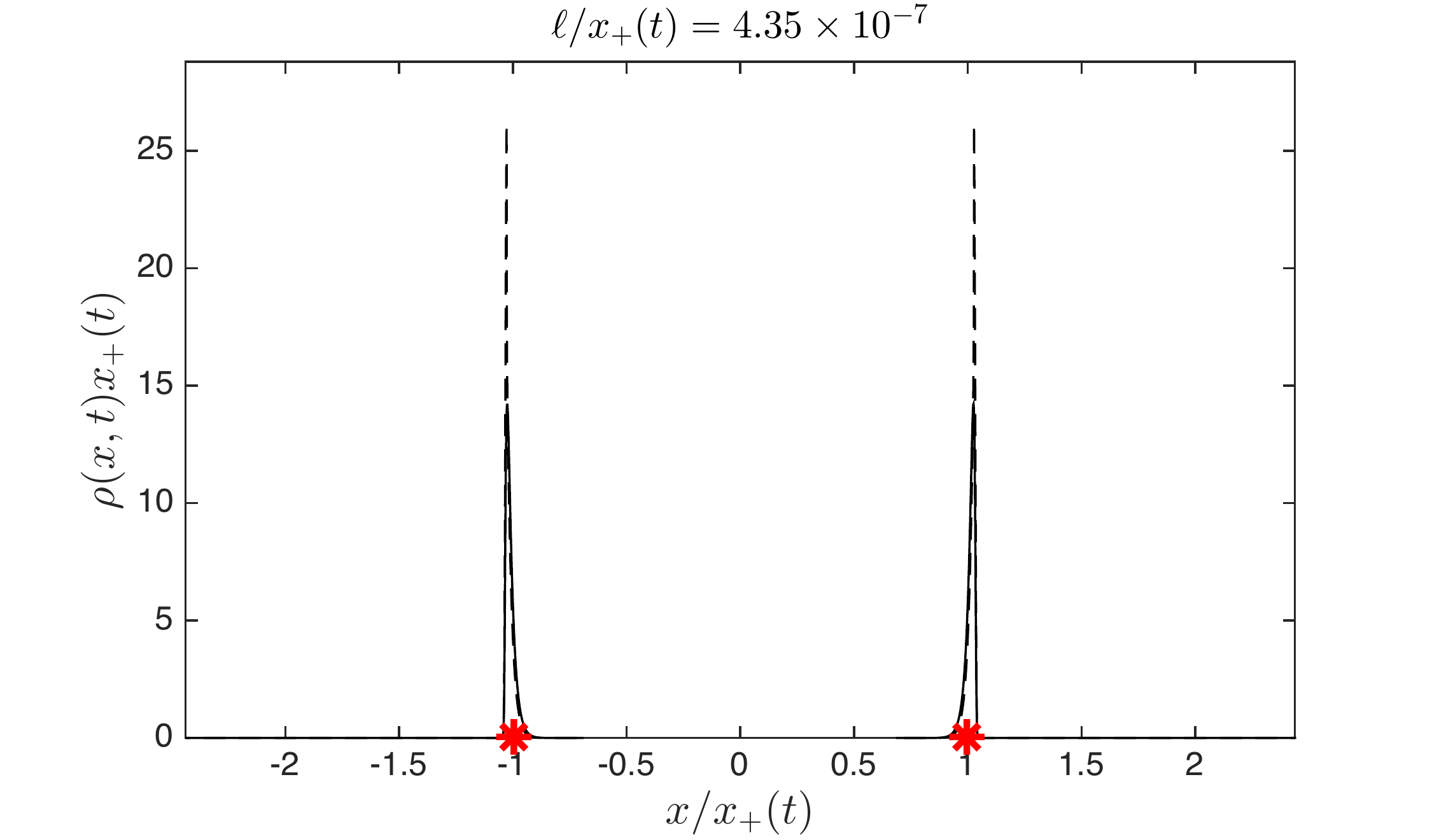} 
   \caption{} 
    \label{fig7:c} 
  \end{subfigure}
  \begin{subfigure}[b]{0.5\linewidth}
    \centering
    \includegraphics[width=.9\columnwidth,height=0.5\columnwidth]{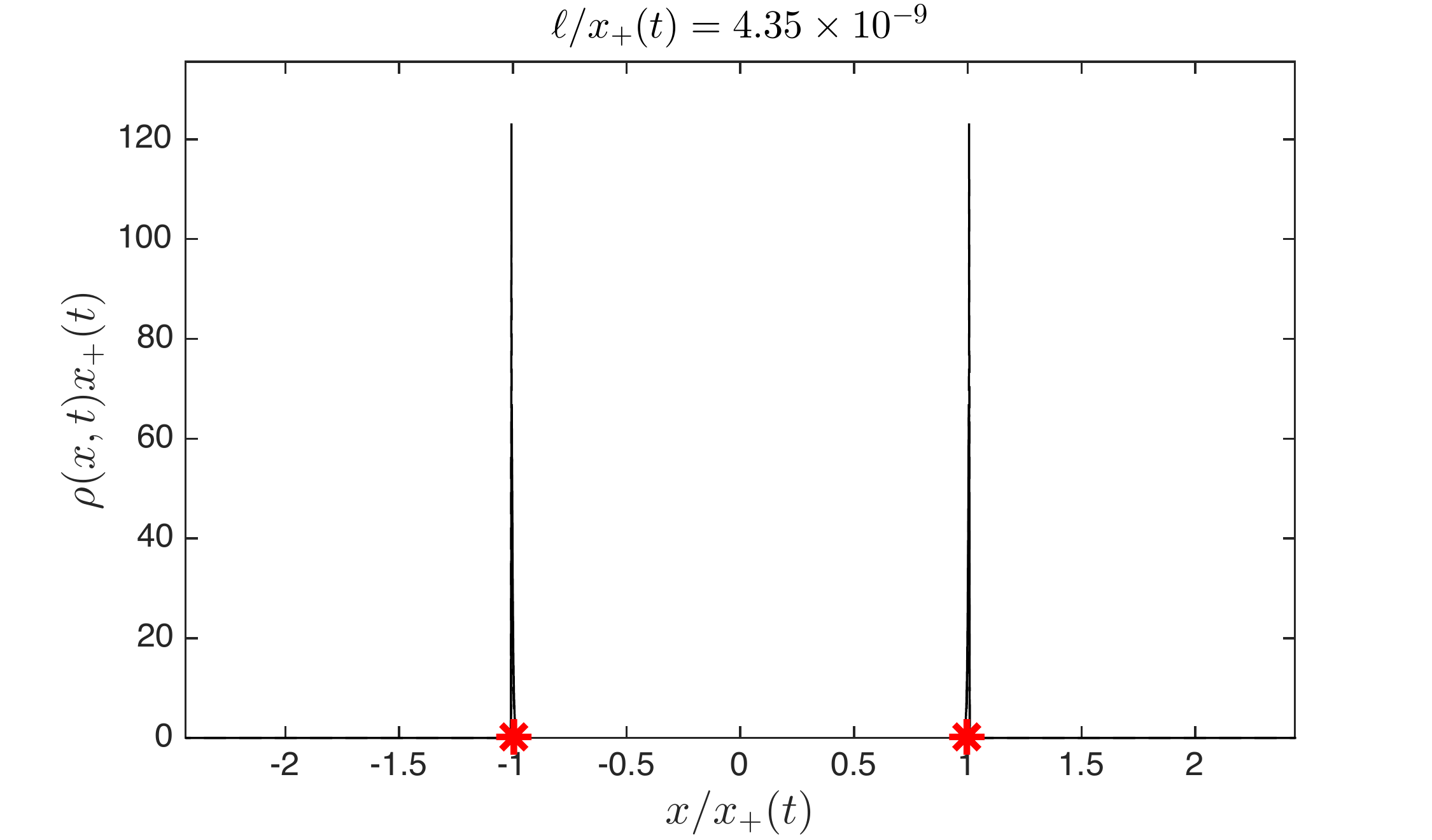} 
   \caption{} 
    \label{fig7:d} 
  \end{subfigure} 
  \caption{ \raggedright  WKB position-space probability densities at a fixed time $t$ for $\mu=1$ and $\ell/x_+(t)=4.35\times 10^{-3}$,
   $10^{-5}$, $10^{-7}$, $10^{-9}$. Notations as in Fig. \ref{WKB_fixed_ell}. For movie, see \cite{SupplMat}.}
  \label{WKB_fixed_t} 
\end{figure*}

We now demonstrate the crucial result (\ref{eqn:trajLim}) for the first-order spliced potential. Here we 
can develop explicit solutions for $x_t$ and $v_t$. For $x_0<\ell$ it is straightforward to show that
\begin{equation}\label{eqn:innerSol}
 x_t(x_0) = x_0 \cosh\left(\frac{C^{1/2}t }{\ell^{(1-\alpha)/2}}\right)
 \end{equation}
when $t<t_\ell(x_0)$ with the definition 
$$ t_\ell(x_0)= \frac{\ell^{(1-\alpha)/2}}{C^{1/2}}{\rm arccosh}(\ell/x_0), $$
at which time $x_t=\ell.$ Thereafter one can use (\ref{eqn:cuspSol}) to write 
\begin{eqnarray}\label{eqn:innerSol}
&& |x_t|^{(1-\alpha)/2}{\!\,}_2F_1\left(\frac{1}{2},-\frac{(1-\alpha)}{2(1+\alpha)}; \frac{1+3\alpha}{2(1+\alpha)}; +\left|\frac{\bar{x}_0}{x_t}\right|^{1+\alpha}\right) \cr
&& -\ell^{(1-\alpha)/2}{\!\,}_2F_1\left(\frac{1}{2},-\frac{(1-\alpha)}{2(1+\alpha)}; \frac{1+3\alpha}{2(1+\alpha)}; +\left|\frac{\bar{x}_0}{\ell}\right|^{1+\alpha}\right)\cr
&& \hspace{40pt}
= \frac{1}{2}(1-\alpha)\left(\frac{2C}{1+\alpha}\right)^{1/2} (t-t_\ell(x_0)). 
\end{eqnarray}
for $t>t_\ell(x_0)$ where it follows directly from the definition of the spliced potential (\ref{eqn:splicedPot}) that 
$$ |\bar{x}_0|^{1+\alpha}=\ell^{1+\alpha} \left(\frac{1}{2}(1-\alpha)+\frac{1}{2}(1+\alpha)\left(\frac{x_0}{\ell}\right)^2\right), $$
with $\bar{x}_0$ as an ``effective initial position.''
On the other hand, if $x_0>\ell,$ then the particle starts in the region where the cusp potential is exact and 
for all times $t>0$ the result (\ref{eqn:cuspSol}) gives 
\begin{eqnarray}\label{eqn:outerSol}
&& |x_t|^{(1-\alpha)/2}{\!\,}_2F_1\left(\frac{1}{2},-\frac{(1-\alpha)}{2(1+\alpha)}; \frac{1+3\alpha}{2(1+\alpha)}; +\left|\frac{x_0}{x_t}\right|^{1+\alpha}\right) \cr
&& -|x_0|^{(1-\alpha)/2}{\!\,}_2F_1\left(\frac{1}{2},-\frac{(1-\alpha)}{2(1+\alpha)}; \frac{1+3\alpha}{2(1+\alpha)}; +1\right)\cr
&& \hspace{40pt}
= \frac{1}{2}(1-\alpha)\left(\frac{2C}{1+\alpha}\right)^{1/2} t. 
\end{eqnarray}
Here we have used the fact that $\bar{x}_0=x_0$ in this region. To prove (\ref{eqn:trajLim}) we use these exact formulas
and the fact that, for $\langle p\rangle =0,$ solutions $x_t(x_0)$ preserve the sign of $x_0$. Since the righthand sides 
of both (\ref{eqn:innerSol}) and (\ref{eqn:outerSol}) are asymptotic to $(x_+(t))^{(1-\alpha)/2}$ as $t\rightarrow\infty,$ one gets easily that 
$x_t(x_0)/x_+(t)\rightarrow {\rm sign}(x_0).$ This is also true for the fixed point $x_0=0,$ with the convention
that ${\rm sign}(0)=0.$

The above results should carry over qualitatively to every potential which has the power-law form
(\ref{eqn:pot}) for some long range of $x$, $\ell\ll |x|\ll L,$ but smoother behavior at $x\ll \ell$ preserving the repulsive 
 character of the potential. For $x_0$ near 0 there will be an initial period where the classical particle position 
 $x_t(x_0)$ moves away from the origin exponentially quickly, until the particle enters the region of 
 power-law behavior of the potential. The particle will then move outward ``explosively'', approaching 
one of the extremal solutions $x_\pm(t)$ at long times for every initial condition $x_0.$ We illustrate 
this with WKB results on the position probability density (PDF) for both Kummer and spliced potentials. 
The default values will be $\alpha=1/3$ and $\mu=1$ in these and all later figures, unless 
stated otherwise. The results were obtained by numerically integrating the ODE's (\ref{eqn:HamiltonsEqn}),(\ref{eqn:JacobianODE}) 
for logarithmically evenly spaced initial $x_0=10^{i/N_F}$ and integers $i= N_{S}N_F$,...,$N_BN_F$ with 
$N_F,N_S<N_B$ chosen to provide suitable resolution and then using formula (\ref{eqn:WKBprob}) to 
evaluate the position PDF. Because of the symmetry of the problem, we could consider only $x_0>0.$
The convergence of the results was tested by varying $N_F$,$N_S$,$N_B$ and also by verifying that the total probability 
was unity to within at least a percent (and usually much closer). For details of the numerical methods, see \cite{SupplMat}.  
Fig.~\ref{WKB_fixed_ell} shows the time evolution of the position PDF's and their splitting from an initial unimodal 
function into bimodal densities at later times. The peaks follow rather closely the two extremal classical solutions,
indicated by asterisks on the abscissa. It is notable that there are jump-discontinuities in the PDF's for the spliced 
potential, which arise from similar jumps in the Jacobians $J_t(x_0)$ for that potential, due to its lower smoothness.
See \cite{SupplMat} for details. However, qualitatively the PDF evolutions for the first-order spliced potential and the 
smoother Kummer potential are quite similar. In Fig.~\ref{WKB_fixed_t} we illustrate instead the convergence of 
the PDF's at fixed time $t$ to a pair of delta functions for $\ell=\sigma\ll x_+(t),$ which is the key 
signature of QSS. Again the behaviors for the spliced potential and the Kummer potential are quite similar. 

These results have interesting implications for the position-space spreading of the initial wavepacket, 
defined as usual by 
\begin{equation}\label{eqn:disp}
 (\Delta x(t))^2= \int dx \ (x-\langle x\rangle)^2 |\psi(x,t)|^2 . 
 \end{equation}
Using our WKB results, we find that asymptotically at long times $t\gg \left(\ell^{1-\alpha}/C\right)^{1/2}$
the spreading is given by 
\begin{equation}\label{eqn:LimDisp}
 (\Delta x(t))^2 \sim \left[\frac{1}{2}(1-\alpha)\left(\frac{2C}{1+\alpha}\right)^{1/2} t \right]^{4/(1-\alpha)}. 
 \lb{QSS-spread} 
 \end{equation}
This is ``Richardson-like" not only in that it is a super-ballistic power-law but, more importantly, 
in that it is independent of both $\hbar$ and $\sigma,$ and thus non-vanishing in the mathematical limit of 
$\hbar,\sigma$ going to zero. This should be contrasted with the standard result for ballistic 
position-space spreading of the same initial wavepacket under free-particle dynamics:
$$ (\Delta x(t))^2 = \sigma^2 + (h t/2m\sigma)^2. $$
In this case, the spreading vanishes in the limit $h/m\rightarrow 0$ \\ first (WKB limit) and then $\sigma\rightarrow 0$.
In fact, this is the usual textbook explanation for classical behavior of very massive particles. It should be 
remembered that the constant $C$ in (\ref{eqn:LimDisp}) is inversely proportional to $m,$ so that the spreading in the cusp potential
is also slower for large $m,$ but one can always observe a large spread by waiting a time proportional to $m^{1/2}$
but independent of $\hbar$. 

It is also instructive to compare with the spreading of the Gaussian wave-packet 
in the power-law potential for $\alpha=1$, that is, the inverted oscillator potential, 
$$ V(x)= -\frac{1}{2}Cx^2, $$
which has been very well-studied \cite{Barton86,Yuceetal06}. 
The spreading here is: 
$$ (\Delta x(t))^2 = \sigma^2\cosh^2(C^{1/2}t) + \left(\frac{\hbar }{2mC^{1/2}\sigma}\right)^2\sinh^2(C^{1/2}t). $$
This result is easily obtained by analytic continuation of the well-known result for the harmonic 
oscillator to imaginary frequency. If one takes first the WKB limit $h/m\rightarrow 0$ (or $h/\sqrt{m}\rightarrow 0,$
since $C\propto 1/m$) and then $\sigma\rightarrow 0,$ the spreading vanishes just as for the free particle. 
Furthermore, the wavepacket remains unimodal and even Gaussian in the WKB limit since the Jacobian $J_t(x_0) = \cosh(C^{1/2} t)$
is independent of $x_0.$ In fact, this is exactly true for the full Schr\"odinger evolution as well \cite{Barton86},
and no maxima of $|\psi(x,t)|^2$ at non-zero $x$ ever develop in that case. This example shows that the 
novel features associated to the cusp potential are due to its ``roughness" or non-smoothness and not 
merely due to strong repulsion from the origin.  

We have ignored so far the effect of a large-scale cut-off $L$ to the power-law range of the potential. 
However, at a time of the order of $(L^{1-\alpha}/C)^{1/2}$ the maxima of $\rho=|\psi|^2$ at 
$x_\pm(t)$ will reach $|x|\simeq L$ and the effects of that cut-off will begin to be felt. At that time 
the power-law spreading (\ref{eqn:LimDisp}) will terminate. The behavior subsequently will depend upon the 
specific details of the potential at length scales $|x|>L.$

\subsection{Splitting Time}

The results of the previous section demonstrate that indeterminacy survives as $\sigma,\ell\rightarrow 0$ or
as $t\rightarrow \infty$ in the WKB limit. 
However, in order to design possible experiments, it is necessary to know exactly how long one must wait to observe 
such effects.  The simplest measure of this is the time of first appearance of new local maxima of $\rho$ at non-zero 
values of $x,$ i.e. when the wave-function ``splits" into multiple peaks. By our previous dimensional analysis, this will
occur at a dimensionless time $\tau_c(\mu)$ which will depend upon the parameter $\mu$, which we refer to as the 
{\it splitting time}. We can think of the position $x$ of the new maximum as the ``order parameter'' for transition to QSS 
at the critical time $\tau_c(\mu)$. 
All local extrema of the  WKB formula $\rho=\rho_0/J$ will occur at $x=x_t(x_0)$ for $x_0$ satisfying $(\rho_0/J)'=0,$  at points 
where $\rho$ is differentiable. This gives the condition for a local extremum as 
$$ J_t'(x_0)/J_t(x_0)= \rho_0'(x_0)/\rho_0(x_0), $$
or as the vanishing of the Wronskian $W(J_t,\rho_0).$  

Because of the symmetry of the classical dynamics around the origin $J_t'(0)=0$ and because of the symmetry of the initial
wavepacket $\rho_0'(0)=0$ also, so that there is an extremum of $\rho(x,t)$ at $x=0$  for all times. For the generic 
case of a symmetric, smooth potential the appearance of the new maxima should appear by a (supercritical) pitchfork bifurcation  
at $x=0,$ with the origin changing from maximum to minimum and two new maxima appearing to either side. 
We shall verify this expectation for the Kummer potential.  Of course, the pitchfork bifurcation is unstable 
to any slight asymmetry,  in either the potential or initial wavepacket, and will usually be modified then to a saddle-node bifurcation 
in which a new maximum-minimum pair appear near (but not exactly at) $x=0$. The observable consequences are hardly distinguishable 
from the pitchfork bifurcation for slight asymmetry, so that we analyze only the exactly symmetrical problem here. 
The first-order spliced potential, although symmetric, exhibits also a saddle-node bifurcation \cite{SupplMat}. 
In that case the Jacobian $J_t(x_0)$ is constant in $x_0$ very near the origin
for all times $t$ (see eq.(\ref{eqn:innerSol})) and thus the local maximum in $\rho_0$ at $x=0$ is preserved in $\rho=\rho_0/J$ . 
This non-generic behaviour occurs because of the lower smoothness of the spliced potential. The peak of $\rho$ at $x=0$ cannot 
be easily seen in the numerical solutions presented in the previous section because the height of the peak decreases exponentially in time.

To identify the critical time $\tau_c(\mu)$ one can apply the condition for a degenerate extremum, which is here    
$$ J_t''(x_0)/J_t(x_0)= \rho_0''(x_0)/\rho_0(x_0). $$
However, in the present case there is a more convenient approach, due to the fact that $-J_t'/J$ increases
monotonically in time for $x_0>0.$ Because of the symmetry of the problem, we can restrict our attention 
only to positive $x_0,x$. To demonstrate the monotonicity for $x_0>0$, first note using the evolution equations 
(\ref{eqn:JacobianODE}) for $J,K$ and similar evolution equations for $J',K'$ \cite{SupplMat} that 
$$ \frac{d}{dt}(K'J-J'K)=-V'''(x)J^3.$$
Since particles in 1D preserve their order, $J>0$ and thus $d/dt(K'J-J'K)\leq 0$ for potentials of the 
sort considered here with $V'''(x)>0$ for $x>0$. Also, $K'J-J'K=0$ at $t=0$ so that 
$$ \frac{d}{dt}( -J'/J)= -(K'J-J'K)/J^2\geq 0. $$
Note that generally $-\rho_0'/\rho_0\geq 0$ at $x>0,$ for initial wave-packets with probability decaying monotonically
away from the origin. Also, at time zero $J'/J\equiv 0$, since $J_0(x_0)\equiv 1.$ Thus, a simple procedure is to evolve $-J_t'/J_t$ in time and, as it rises upward from zero, identify the first time that its graph touches the graph of $-\rho_0'/\rho_0.$  

\begin{figure}[htp]
    	\includegraphics[width=1.\columnwidth,height=.7\columnwidth]{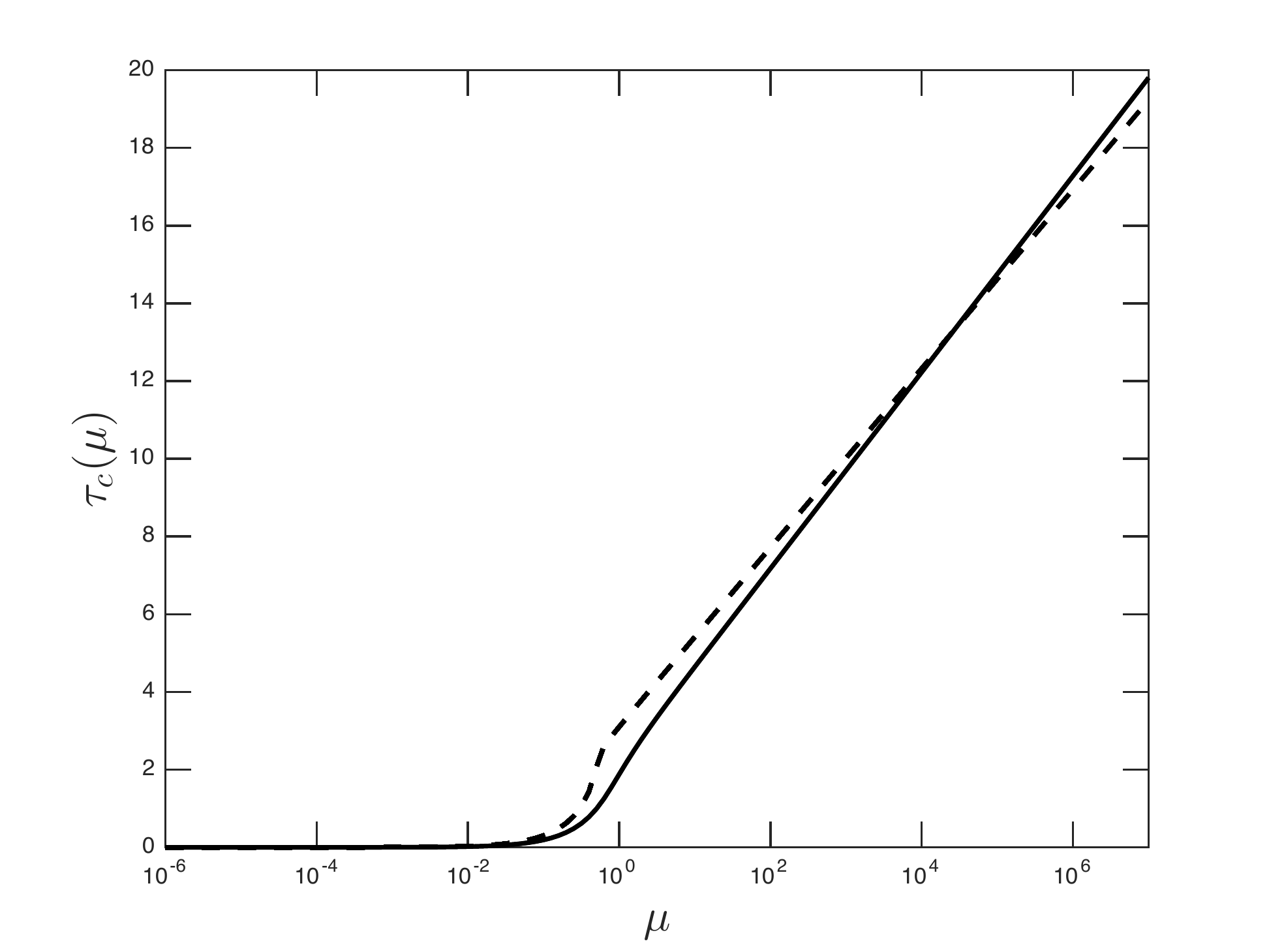}
\textbf{(a)}
\includegraphics[width=1.\columnwidth,height=.7\columnwidth]{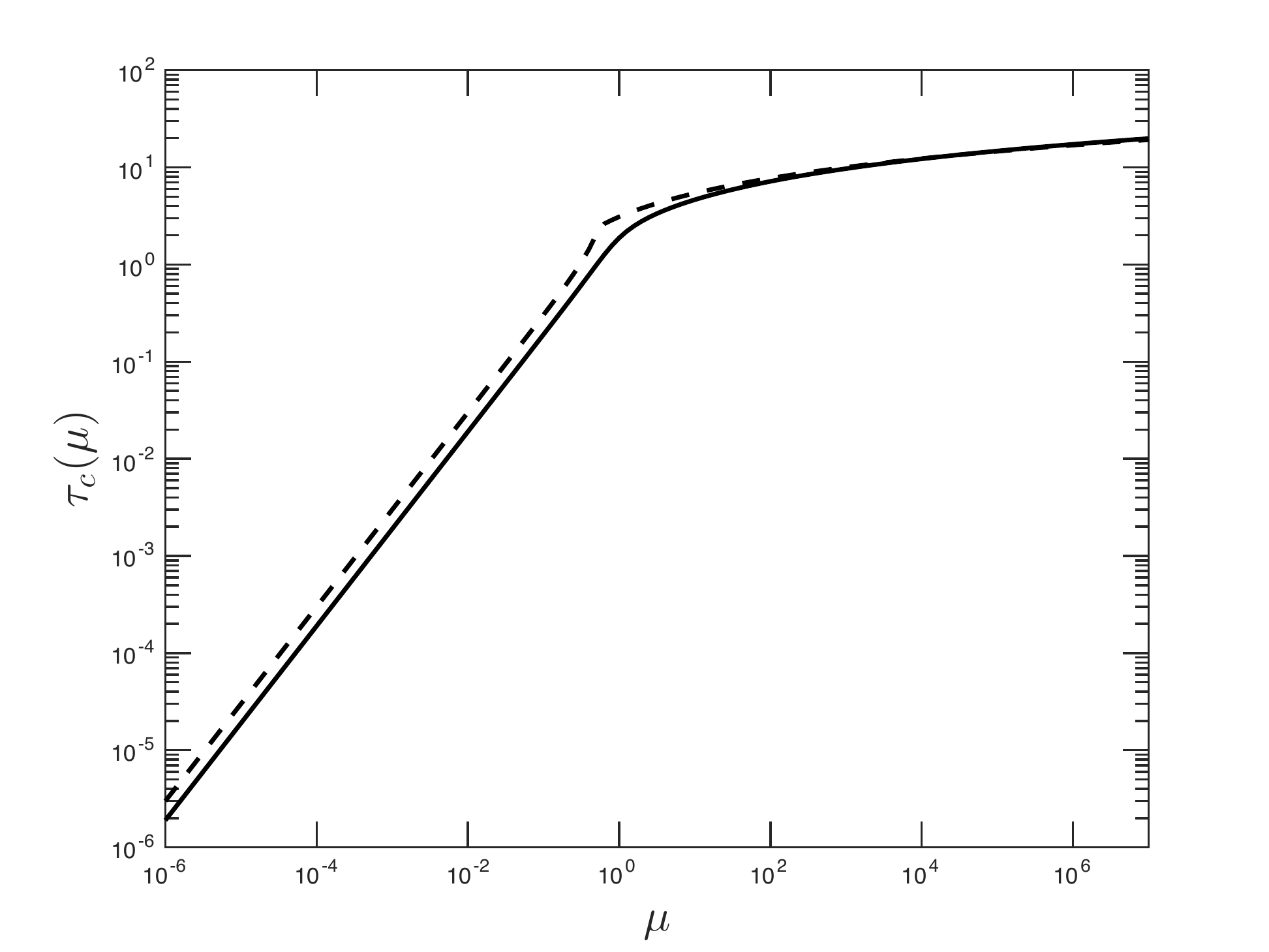}
\textbf{(b)}
  \caption{\raggedright Splitting time $\tau_c(\mu)$ plotted versus $\mu,$ (a) in log-linear and (b) in log-log,  
for both Kummer potential \\ and 1st-order spliced potential. Notations as in Fig. \ref{WKB_fixed_ell}.} 
\label{fig:Splitting}
\end{figure}

We can now apply these results to evaluate the splitting time $\tau_c(\mu)$ for the case of a 
Gaussian wave-packet. Since $-\rho_0'(x_0)/\rho_0(x_0)=x_0/\sigma^2,$  the condition for extrema 
of $\rho$ becomes
$$ -J_t'(x_0)/J_t(x_0)=x_0/\sigma^2. $$
It is most convenient to use this condition in a dimensionless form for $u_0=x_0/\ell$ and 
$\tau=t/(\ell^{1-\alpha}/C)^{1/2}$, or  
$$ -J_\tau'(u_0)/J_\tau(u_0)=\mu^2 u_0, $$
with $\mu=\ell/\sigma.$ 
The full analysis of this equation for all $\mu$ to find the mimimum $\tau$ with 
solutions at positive $u_0$ requires a numerical study using the method discussed above
\cite{SupplMat}, the results of which are presented in Fig.\ref{fig:Splitting}.
However, analytical results can be obtained asymptotically for $\mu\ll 1$ and $\mu\gg 1.$  

\begin{figure}[h!]
\includegraphics[width=1.\columnwidth,height=.7\columnwidth]{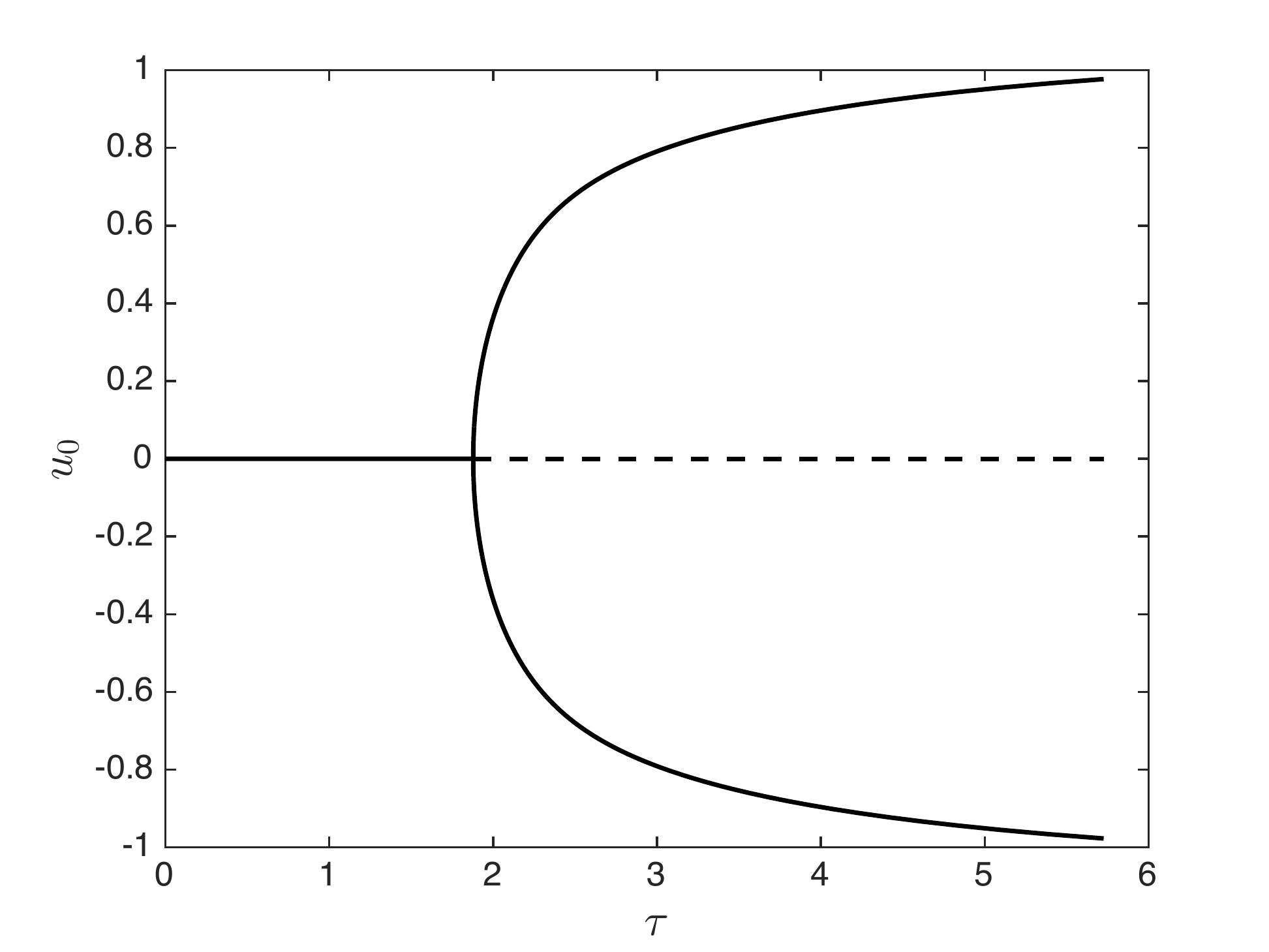}
\caption{\raggedright Bifucation diagram for the Kummer potential with $\alpha = 1/3$ and for $\mu = 1$, plotting 
$u_0$ for all extrema of the WKB position PDF versus dimensionless time $\tau$. Solid lines denote local maxima 
of the PDF,  dashed lines local minima.} 
\label{fig:BifurcDiag}
\end{figure}

We first consider $\mu\ll 1$ for the Kummer potential.  One can then just Taylor-expand to get
$$ x_t(x_0) =
x_0 + \dot{x}_0 t + \frac{1}{2}\ddot{x}_0 t^2 + \cdots = x_0 - \frac{1}{2} V'(x_0) t^2+ \cdots . $$
We note here the convenient formulas \cite{SupplMat}
$$ V_\ell'(x)= -A_\alpha C\ell^{\alpha-1}  x \cdot  {\,\!}_1F_1\left(
\text{{\footnotesize $\frac{1-\alpha}{2}$}}, \text{{\footnotesize $\frac{3}{2}$}}; -\frac{x^2}{2\ell^2}\right) $$
$$ V_\ell''(x)= -A_\alpha C\ell^{\alpha-1}  {\,\!}_1F_1\left(
\text{{\footnotesize $\frac{1-\alpha}{2}$}}, \text{{\footnotesize $\frac{1}{2}$}}; -\frac{x^2}{2\ell^2}\right) $$
\hspace{-3pt} $$ V_\ell'''(x)=  (1-\alpha) A_\alpha C\ell^{\alpha-3} x\ \cdot \  {\,\!}_1F_1\left(
\text{{\footnotesize $\frac{3-\alpha}{2}$}}, \text{{\footnotesize $\frac{3}{2}$}}; -\frac{x^2}{2\ell^2}\right) $$
with the dimensionless constant 
$$ A_\alpha = 2^{(1+\alpha)/2}\Gamma((2+\alpha)/2)/\sqrt{\pi}. $$
Differentiation using the formulas above and non-dimensionalization yields for $\tau\ll 1$  
$$ -\frac{J_\tau'(u_0)}{J_\tau(u_0)}= \frac{1}{2}V'''(u_0)\tau^2 +\cdots. $$
Using further the Taylor expansion of the Kummer function ${\,\!}_1F_1(a,b;z)=1+\frac{a}{b}z+\cdots,$
gives the result to linear order in $\tau^2$ and cubic order in $u_0$:
$$ \frac{1}{4}(1-\alpha)(1-\frac{\alpha}{3})A_\alpha \tau^2 u_0^3
= \left(\frac{1}{2}(1-\alpha) A_\alpha \tau^2-\mu^2 \right)u_0. $$
Up to a rescaling of the variables, this is the standard normal form of the pitchfork bifurcation, which 
is verified by the numerically obtained bifurcation diagram in Fig.~\ref{fig:BifurcDiag}. There 
is always the solution $u_0=0,$ but a symmetrical pair of nonzero solutions appears when 
the righthand side is positive, i.e. for $\tau>\tau_c(\mu)$ with  
$$ \tau_c(\mu)\sim \left(\frac{2}{(1-\alpha)A_\alpha}\right)^{1/2}\mu, \,\,\,\,\mu\ll 1. $$
This asymptotic behavior is verified in the numerical results plotted in Fig.\ref{fig:Splitting}(b). 
Note that the linear scaling of $\tau_c(\mu)$ for small $\mu$ (but with a different prefactor) will hold for a generic, 
smooth symmetric potential with $V'''(x)>0$ for $x>0,$ by the same argument. We also observe in 
Fig.\ref{fig:Splitting}(b) the same linear scaling of $\tau_c(\mu)$ for the spliced potential, even 
though that potential has lower-order smoothness and the bifurcation in that case is an off-origin saddle-node type, as discussed in \cite{SupplMat}.   

We now turn to the limit $\mu\gg 1.$ We have found no convenient analytic approach here for the 
Kummer potential, so that we consider instead the first-order spliced potential, where we can obtain 
closed expressions for $J_t$ and $J_t'/J_t.$ For $x_0<\ell$ and $t<t_\ell(x_0),$ one sees easily that 
$$ J_t(x_0)=  \cosh\left(\frac{C^{1/2}t }{\ell^{(1-\alpha)/2}}\right), \,\,\,\, J_t'(x_0)\equiv 0. $$ 
For $x_0<\ell$ and $t>t_\ell(x_0),$ we find that 
\begin{eqnarray*}
&& J_t(x_0) = \frac{x_0x_t}{\bar{x}_0^{1+\alpha}\ell^{1-\alpha}}+ \sqrt{x_t^{1+\alpha}-\bar{x}_0^{1+\alpha} }\times \cr 
&&  \hspace{30pt} \left[-\frac{1}{2}(1-\alpha)\left(\frac{2C}{1+\alpha}\right)^{1/2}(t-t_\ell(x_0))\frac{x_0}{\bar{x}_0^{1+\alpha}\ell^{1-\alpha}} \right. \cr
&&  \hspace{30pt} \left. +\frac{1}{2}(1-\alpha)\left(\frac{2}{1+\alpha}\right)^{1/2}\frac{\ell^{(1+\alpha)/2}\sqrt{\ell^2-x_0^2}}{x_0\bar{x}_0^{1+\alpha}} \right].
\end{eqnarray*} 
Finally, for $x_0>\ell$ and all times $t>0$
$$ J_t(x_0) = \frac{1}{x_0}\left[x- (x_+(t))^{(1-\alpha)/2}\sqrt{x_t^{1+\alpha}-x_0^{1+\alpha}}\right]. $$
Before proceeding, we note that these latter two expressions exhibit a jump-discontinuity in $J_t(x_0)$
at $x_0=\ell,$ which leads to the discontinuity in the WKB probability density $\rho$ which was observed 
in the previous section. We also note here that $J_t$ at long times appears to grow $\propto x_+(t),$ but 
there is in fact a near-cancellation between two terms with such growth. The actual growth is   
$$ J_t(x_0) \propto [x_+(t)/\ell]^{(1+\alpha)/2} G(x_0/\ell), \,\,\,\, t\gg (\ell^{1-\alpha}/C)^{1/2} $$
for some explicit function $G(u_0)$ \cite{SupplMat}. This result has the interesting implication that the peaks in $\rho(x,t)$
at $x_\pm(t)$ have widths $\propto[x_+(t)]^{(1+\alpha)/2}$ as $t\rightarrow \infty.$ However, the 
scaled density $\hat{\rho}(\hat{x},\tau)$ has peaks at $\pm 1$ with widths $\propto [x_+(t)]^{-(1-\alpha)/2}\rightarrow 0$ 
as $\tau\rightarrow\infty.$ 

\begin{figure}[h!]
\includegraphics[width=1.\columnwidth,height=.7\columnwidth]{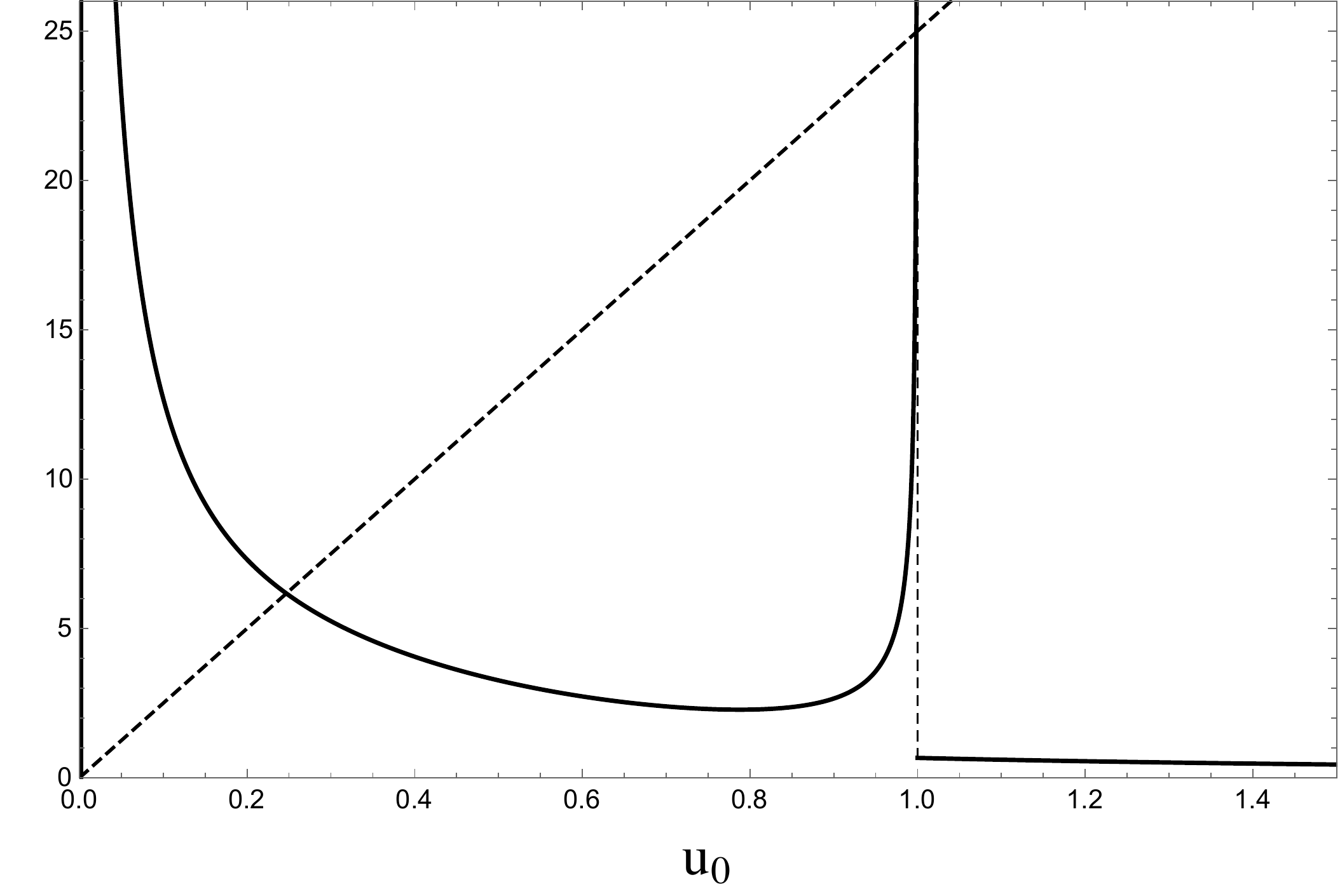}
\caption{\raggedright Plotted versus $u_0$ are the function $-J_\infty'(u_0)/J_\infty(u_0)$ for the spliced potential with $\alpha= 1/3,$
as solid line,  and the linear function $\mu^2 u_0$ for $\mu = 5$, as dashed line.} 
\label{fig:longTime}
\end{figure}

It is quite straightforward by differentiating the previous results with respect to $x_0$ to obtain
expressions for $-J_t'(x_0)/J_t(x_0).$ Complete formulas are given in \cite{SupplMat}, but here 
we write only the finite limiting result for $\tau\rightarrow\infty:$
\begin{eqnarray}
&& -J_\infty'(u_0)/J_\infty(u_0) = -\frac{1}{u_0}+\frac{1+3\alpha}{(1-\alpha)+(1+\alpha)u_0^2}\cr
&& \vspace{-10pt} \cr
&& +\frac{2}{u_0\left[ 1-u_0^2 + \frac{2}{1-\alpha}\left(\frac{1+\alpha}{2}\right)^{1/2}u_0^2\sqrt{1-u_0^2}F(u_0)\right]}
\lb{JpJ-1}
\end{eqnarray}
with 
$$ \hspace{-3pt} F(u_0) = {\,\!}_2F_1\left(\text{{\footnotesize $\frac{1}{2}$}},
\text{{\footnotesize{ $-\frac{1-\alpha}{2(1+\alpha)}$}}}; 
\text{{\footnotesize $\frac{1+3\alpha}{2(1+\alpha)}$}}; 
\text{{\footnotesize $\frac{(1-\alpha)+(1+\alpha)u_0^2}{2}$}} \right), $$
for $u_0<1,$ and 
\be -J_\infty'(u_0)/J_\infty(u_0) = \frac{1+\alpha}{2}\frac{1}{u_0} \lb{JpJ-2} \ee
for $u_0>1.$ This function is plotted in Fig.\ref{fig:longTime}.  

We note that the straight line $\mu^2 u_0$ for $\mu\gg 1$ intersects the curve $-J_\infty'(u_0)/J_\infty(u_0)$
at two non-zero values of $u_0$ in the range $0<u_0<1,$
one near $u_0=0$ and another near $u_0=1$. Because of the monotonic
increase of $-J_\tau'(u_0)/J_\tau(u_0)$ with $\tau,$ there are no solutions at any time between these values of $u_0$.
We can determine the solutions near $u_0=0$ and $u_0=1$ by considering the behavior of $-J_\infty'(u_0)/J_\infty(u_0)$
near these points. Near $u_0=0,$  
$$ -J_\infty'(u_0)/J_\infty(u_0) \sim \frac{1}{u_0}$$
so the solution is at $u_0\simeq 1/\mu$ and a height $\mu.$ On the other hand, near $u_0=1,$
$$ -J_\infty'(u_0)/J_\infty(u_0) \sim \frac{c}{\sqrt{1-u_0^2}}$$
for a constant $c,$ so the solution is at $u_0\simeq 1-c^2/2\mu^4$ and a height $\mu^2.$
Since the height of the first intersection 
is much smaller than the height of the second for $\mu\gg 1,$ the graph of $-J_\tau'(u_0)/J_\tau(u_0)$ 
touches the line $\mu^2 u_0$ for a smaller value of $\tau$ near the first point $u_0\simeq 1/\mu.$ 
This cannot happen at any time $\tau<{\rm arccosh}(1/u_0),$ because $-J_\tau'(u_0)/J_\tau(u_0)\equiv 0$ before then. 
This argument yields an exact low bound $\tau_c(\mu)\geq {\rm arccosh}(\mu).$
In fact, the graph rises sufficiently rapidly after that time so that 
$$\tau_c(\mu)\simeq {\rm arccosh}(\mu)\sim \ln\mu, \,\,\,\,\mu\gg 1; \,\,\,\, $$
This asymptotic formula agrees with the numerical results on splitting time $\tau_c(\mu)$ for the 
spliced potential at $\mu\gg 1$ observed in Fig.\ref{fig:Splitting}(a). 

Furthermore, the same logarithmic asymptotics $\tau_c(\mu)\sim \ln\mu $ at large $\mu$ appear 
in the plot for the Kummer potential in Fig.\ref{fig:Splitting}(a). 
It is interesting that the same large-$\mu$ asymptotics of $\tau_c(\mu)$ hold for the pitchfork bifurcation in the Kummer   
potential and for the saddle-node bifurcation in the spliced potential. Although we have no proof of this asymptotics 
for the Kummer potential, there is a simple heuristic argument which leads to this conclusion in both cases. The 
regime $\mu\gg 1$ corresponds to an initial wavepacket with an extremely narrow  spread in position space,
$\sigma\ll \ell,$  very well localized in the region where the potential appears as an inverted oscillator. However, 
it is intuitively clear that the splitting of the wavefunction cannot occur until it has spread out of the parabolic region
of the potential. In the WKB limit the spreading is due to the classical trajectories, which escape that region exponentially 
quickly. The time $\tau_c(\mu)\sim \ln\mu $ thus corresponds to the time it takes for classical particles at distance
$\sim\sigma$ from the origin to reach a distance $\sim\ell$.   

These findings for $\tau_c(\mu)$ are key results of our paper, providing the time of first onset of 
quantum spontaneous stochasticity phenomena. The implication is that in the WKB regime this is 
a short time of order $(\ell^{1-\alpha}/C)^{1/2}.$ The longest onset times are for $\mu\gg 1$, but even here the growth 
of $\tau_c(\mu)$ is only logarithmic, so that QSS phenomena are observable quickly unless the initial wave-packet spread 
is  exponentially small, with $\sigma=e^{-M}\ell$ for $M=\ln\mu\gg 1$. 

\vspace{4pt}

\subsection{Scattering}



We now consider the problem of the scattering of a wavepacket with $\langle x\rangle, \langle p\rangle \neq 0$ 
when directed toward a UV-cutoff version of the cusp potential (\ref{cusp}), which may be easier
to study experimentally.  To observe QSS in this setting 
it is clear that the classical dynamics for the initial condition $(\langle x\rangle,\langle p\rangle)$ 
must reach the phase-point $(x,p)=(0,0),$ since only here is the flow vector field ${\bf U}(x,p)$ for the 
cusp potential non-Lipschitz. Unlike the problem of turbulent advection where the fluid velocity field is everywhere 
(nearly) non-Lipschitz,  a very careful, fine-tuned choice of $\langle x\rangle, \langle p\rangle$ is here required 
to observe QSS. The only initial conditions for the unregulated cusp-potential which reach $(0,0)$ 
are of the form $(x_+(t),v_-(t))$ and $(x_-(t),v_+(t))$ for some $t>0$, where $x_{\pm}(t)$ are the extremal 
solutions given by (\ref{xtreme}) and $v_\pm(t)=dx_\pm/dt.$ This condition imposes a relation 
between $x$ and $v$:
\be  v=-{\rm sign}(x) \left(\frac{2C}{1+\alpha}\right)^{1/2}|x|^{\frac{1+\alpha}{2}} \lb{scat-cond} \ee 
It is natural to expect that the pair 
$(\langle x\rangle,\langle v\rangle)$ must be chosen close to a point satisfying (\ref{scat-cond}) in order 
to observe QSS also for a UV-cutoff version of the cusp-potential.

To verify that this is so, we apply a simple criterion for presence of QSS. In order to have wave-packet 
splitting in a regulated cusp-potential it is enough to verify that the probabilities for the particle to be found 
to the left and right of the origin 
$$p_-(t)=\int_{-\infty}^0 dx\ |\psi(x,t)|^2, \quad p_+(t)=\int_0^{+\infty} dx\ |\psi(x,t)|^2, $$
both remain strictly positive for all times $t\geq t_*,$ the splitting time, in the limit $\ell,\sigma\rightarrow 0.$    
One expects corrections to $t_*$ of order $O((\ell^{1-\alpha}/C)^{1/2}.$ These probabilities are easy to evaluate
in the WKB limit as 
$$
p_-(t)  =  \frac{1}{ \sqrt{2\pi}} \int_{-\infty}^{y_*(t)} d y \ e^{-\frac{y^2}{2}}, \ \ \  p_+(t) =  \frac{1}{ \sqrt{2\pi}} \int_{y_*(t)}^\infty  d y \ e^{-\frac{y^2}{2}},
$$
assuming for simplicity a minimum-uncertainty Gaussian initial wavepacket, where 
\begin{equation}
y_*(t) = \lim_{\ell,\sigma \to 0 } \left(\frac{x_t^{-1}(0)- \langle x\rangle}{\sigma}\right).
\end{equation}
and $x_t^{-1}(0)=x_0$ for the initial condition $(x_0,\langle v\rangle)$ which arrives to the origin 0 under 
the classical dynamics exactly at time $t.$ It follows that, to see wavepacket splitting within WKB, one must 
have $|y_*(t)|<\infty$ for all $t\geq t_*$.

We can calculate the above quantities analytically for the spliced potential. Note that $x_t^{-1}(0)$ must lie between
the origin 0 and $x_\infty\equiv x_\infty^{-1}(0),$ the point at which a particle with initial velocity $\langle v\rangle$ arrives 
to the origin only after an infinite time. Energy conservation gives $x_\infty$ as  
$$ \frac{C}{1+\alpha}|x_\infty|^{1+\alpha} = \frac{1}{2}|\langle v\rangle|^2+ \frac{1}{2}C\ell^{1+\alpha}
\left(\frac{1-\alpha}{1+\alpha}\right),$$
since the particle must arrive to 0 with vanishing velocity. If we introduce the position $x_*$ which satisfies
the constraint (\ref{scat-cond}) with the velocity $\langle v\rangle,$ then this becomes
\be |x_\infty|^{1+\alpha}=|x_*|^{1+\alpha}+\frac{1}{2}(1-\alpha)\ell^{1+\alpha}. \lb{xinf} \ee
As expected, $x_\infty\rightarrow x_*$ as $\ell\rightarrow 0.$ The above derivation has assumed 
that $|\langle v\rangle|>C\ell^{1+\alpha},$ so that $|x_\infty|>\ell.$ If instead, $|\langle v\rangle|<C\ell^{1+\alpha},$
then the point $x_\infty$ lies in the inverted-oscillator region of the spliced potential. In that case it is easy 
to show again by energy conservation that 
$$ x_\infty= - \langle v\rangle/\gamma, $$
with $\gamma=(C/\ell^{1-\alpha})^{1/2}$.

The classical solution which starts at $x_\infty$ must spend an infinite amount of time
in the inverted-oscillator region $|x|<\ell$ before reaching $0.$ This can be verified by exactly solving 
the dynamics in the inverted-oscillator region as 
$$ x= x_0 \cosh(\gamma t)  + \frac{1}{\gamma} v_0 \sinh(\gamma t), $$
so that the particle reaches the origin when $|v_0|>\gamma |x_0|$ and $x_0,v_0$ have opposite signs, 
with this process taking a time $t=(1/\gamma){\rm arctanh} |\gamma x_0/v_0|$. To apply this result to solutions 
starting at $x_0$ with $|x_0|>\ell,$ we note that such solutions reach distance $\ell$ from the origin with a velocity 
$v_\ell(x_0)$ given by energy conservation as
$$ v_\ell^2=\frac{2C}{1+\alpha}\left[|x_*|^{1+\alpha}-|x_0|^{1+\alpha}+\ell^{1+\alpha}\right], $$
or, using (\ref{xinf}), as   
$$ v_\ell^2 = (\gamma \ell)^2 + \frac{2C}{1+\alpha}\left[|x_\infty|^{1+\alpha}-|x_0|^{1+\alpha}\right]. $$
Hence the time that classical solutions for such initial positions $x_0$ spend in the inverted-oscillator region is 
$$ \tau_\ell(x_0) = \frac{1}{\gamma}{\rm arctanh}\left(\frac{\gamma\ell}{v_\ell(x_0)}\right). $$
For the special case $x_0=x_\infty,$ one has $v_\ell=\gamma\ell$ and thus the time is infinite,
as expected.

\begin{figure*}[!ht] 
  \begin{subfigure}[b]{0.5\linewidth}
    \centering
    \includegraphics[width=.9\columnwidth,height=0.5\columnwidth]{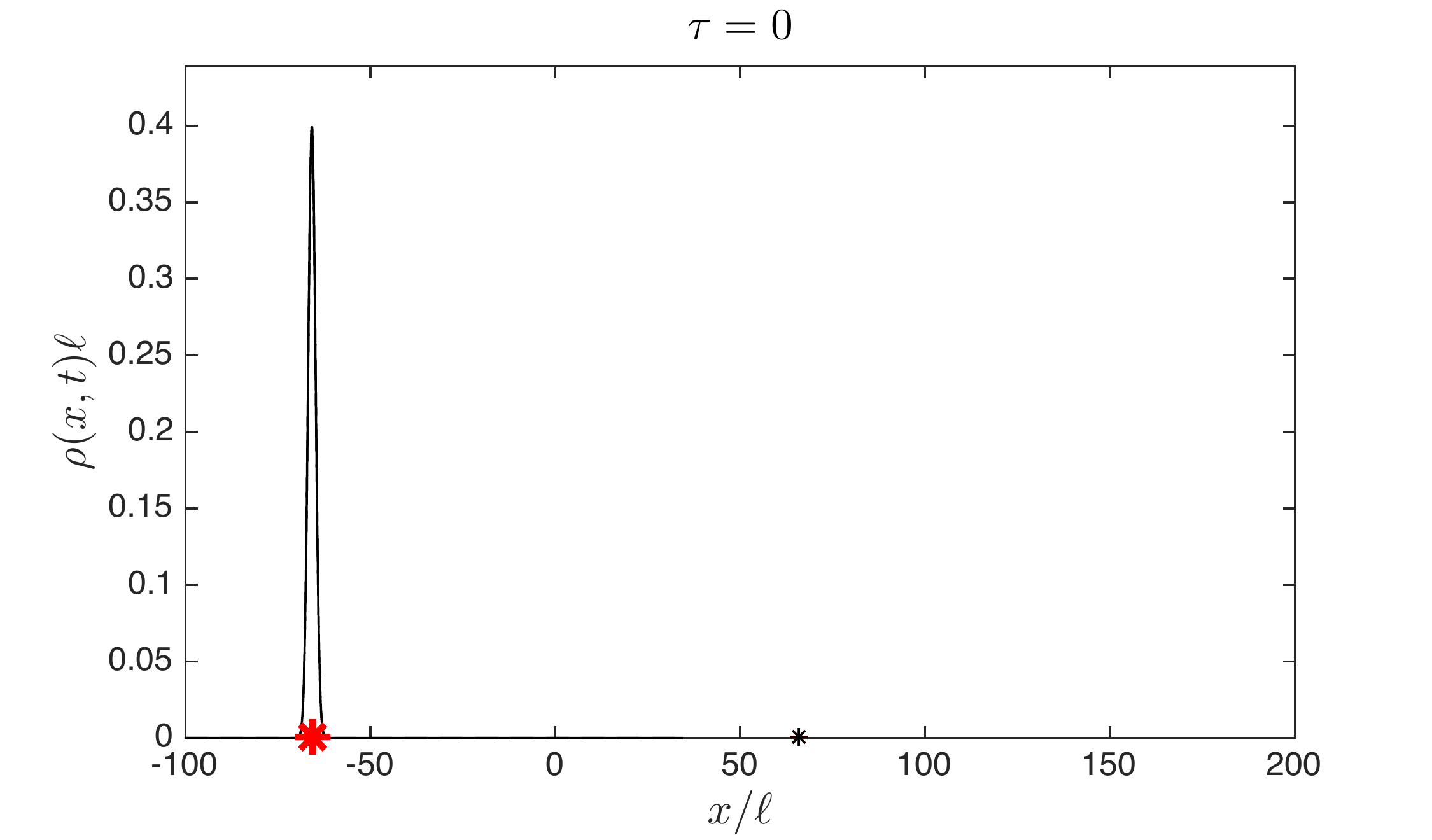}
   \caption{} 
    \label{fig6:a} 
  \end{subfigure}
  \begin{subfigure}[b]{0.5\linewidth}
    \centering
   \includegraphics[width=.9\columnwidth,height=0.5\columnwidth]{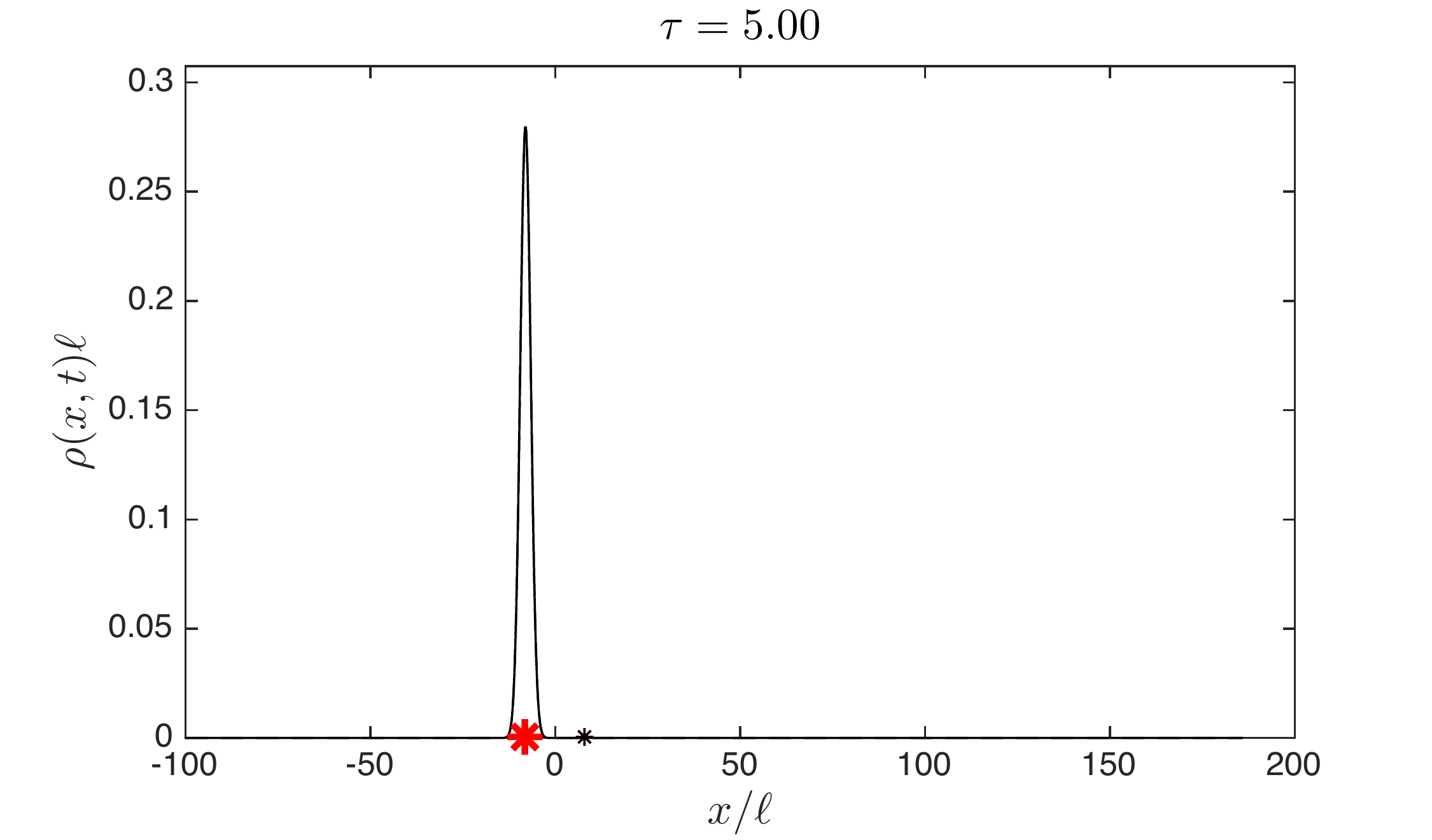} 
   \caption{} 
    \label{fig6:b} 
  \end{subfigure} 
  \begin{subfigure}[b]{0.5\linewidth}
    \centering
    \includegraphics[width=.9\columnwidth,height=0.5\columnwidth]{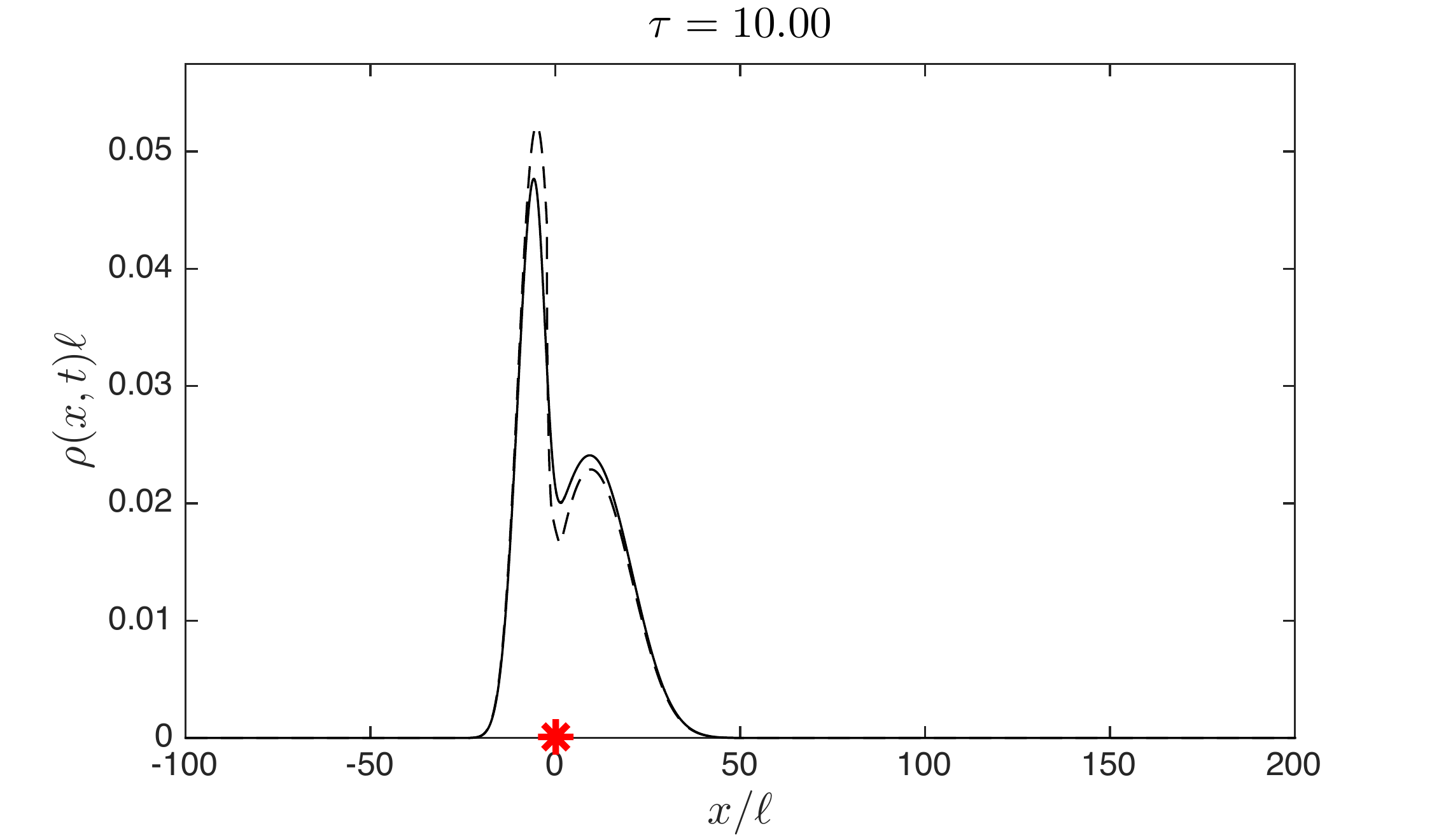} 
   \caption{} 
    \label{fig6:c} 
  \end{subfigure}
  \begin{subfigure}[b]{0.5\linewidth}
    \centering
    \includegraphics[width=.9\columnwidth,height=0.5\columnwidth]{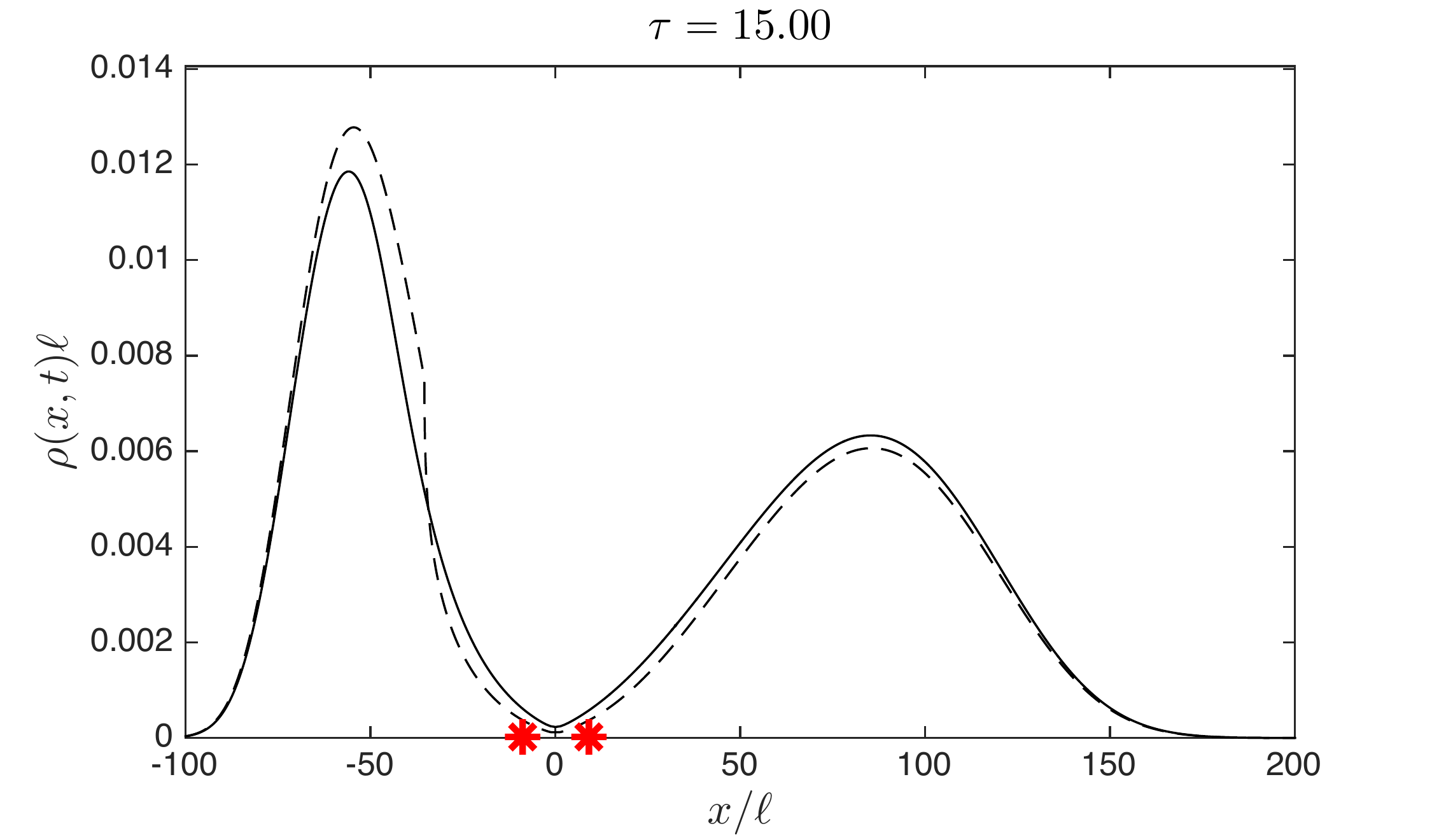} 
   \caption{} 
    \label{fig6:d} 
  \end{subfigure} 
  \caption{ \raggedright WKB position-space probability densities for an incoming wave-packet with 
    $\langle v\rangle/(C\ell^{1+\alpha})^{1/2}=20,$ $\langle x\rangle=x_\infty,$ and $\sigma=\ell$ at 
    dimensionless times $\tau=$ 0, 5, 10, 15. Here $\alpha=1/3$ and notations are as in Fig.~\ref{WKB_fixed_ell}.
   For movie, see \cite{SupplMat}.} 
  \label{fig:scatt-timeEvol} 
\end{figure*}

\begin{figure*}[!ht] 
  \begin{subfigure}[b]{0.5\linewidth}
    \centering
    \includegraphics[width=.9\columnwidth,height=0.5\columnwidth]{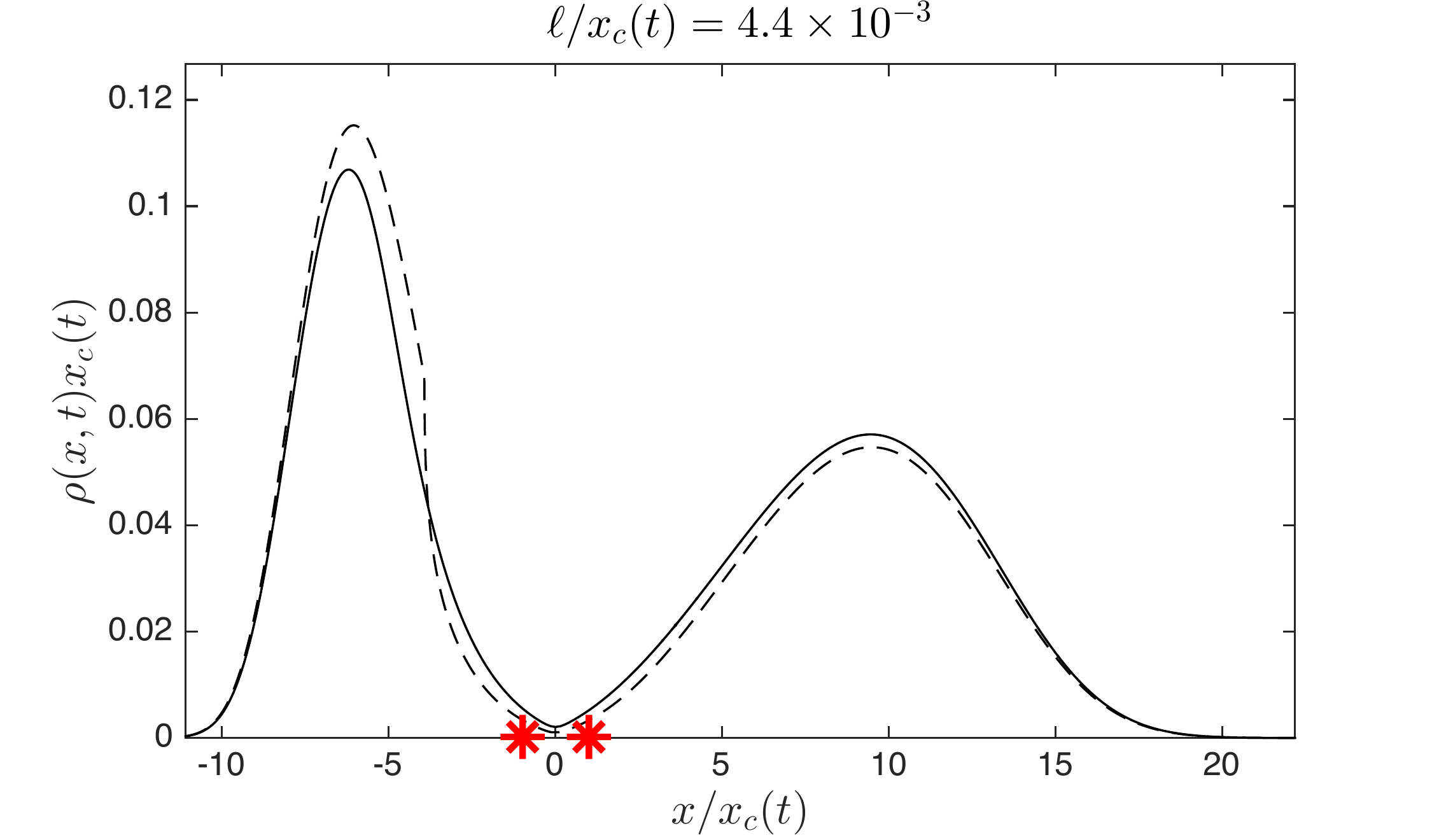} 
   \caption{} 
    \label{fig7:a} 
  \end{subfigure}
  \begin{subfigure}[b]{0.5\linewidth}
    \centering
    \includegraphics[width=.9\columnwidth,height=0.5\columnwidth]{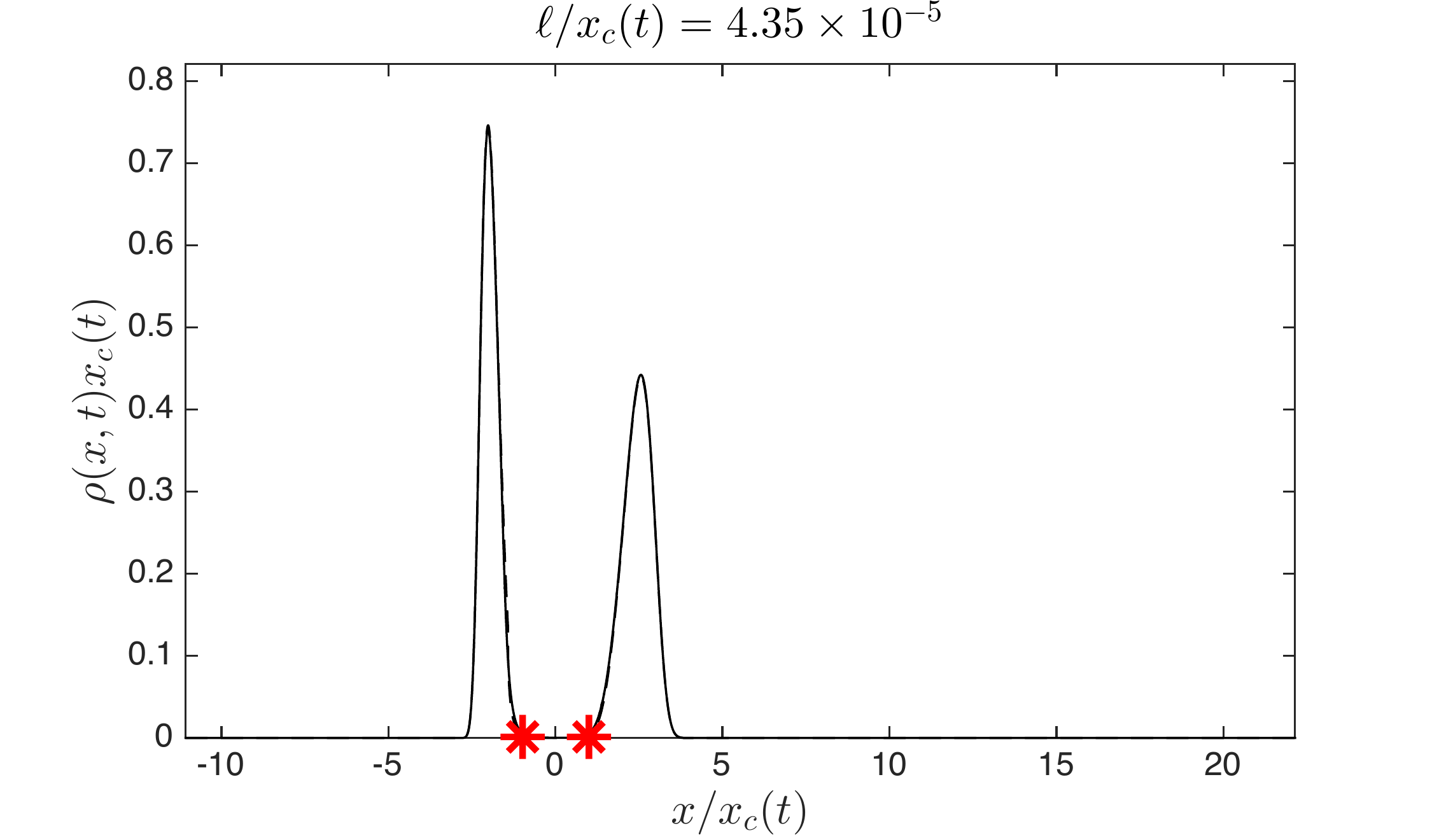} 
   \caption{} 
    \label{fig7:b} 
  \end{subfigure} 
  \begin{subfigure}[b]{0.5\linewidth}
    \centering
    \includegraphics[width=.9\columnwidth,height=0.5\columnwidth]{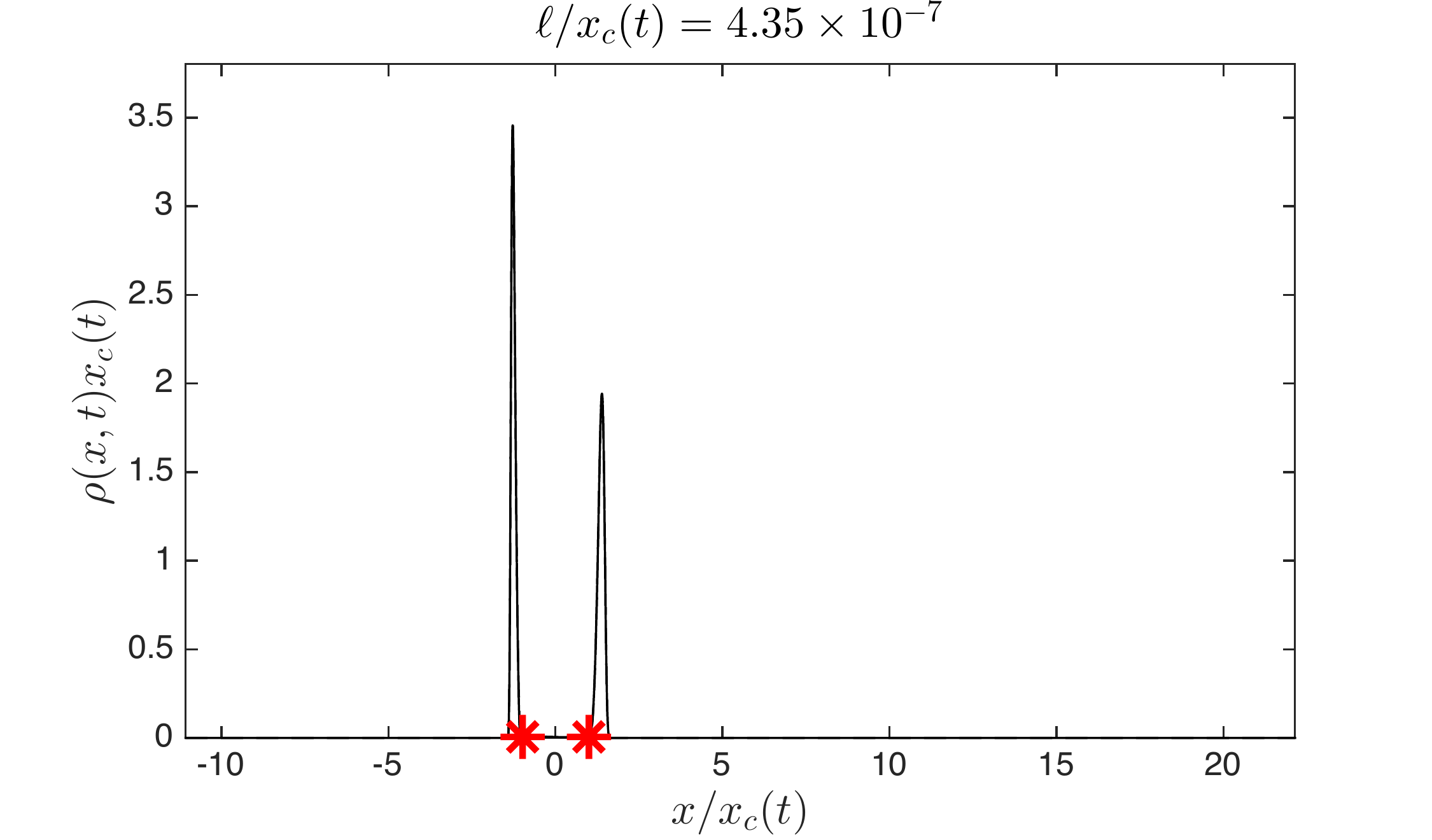} 
   \caption{} 
    \label{fig7:c} 
  \end{subfigure}
  \begin{subfigure}[b]{0.5\linewidth}
    \centering
    \includegraphics[width=.9\columnwidth,height=0.5\columnwidth]{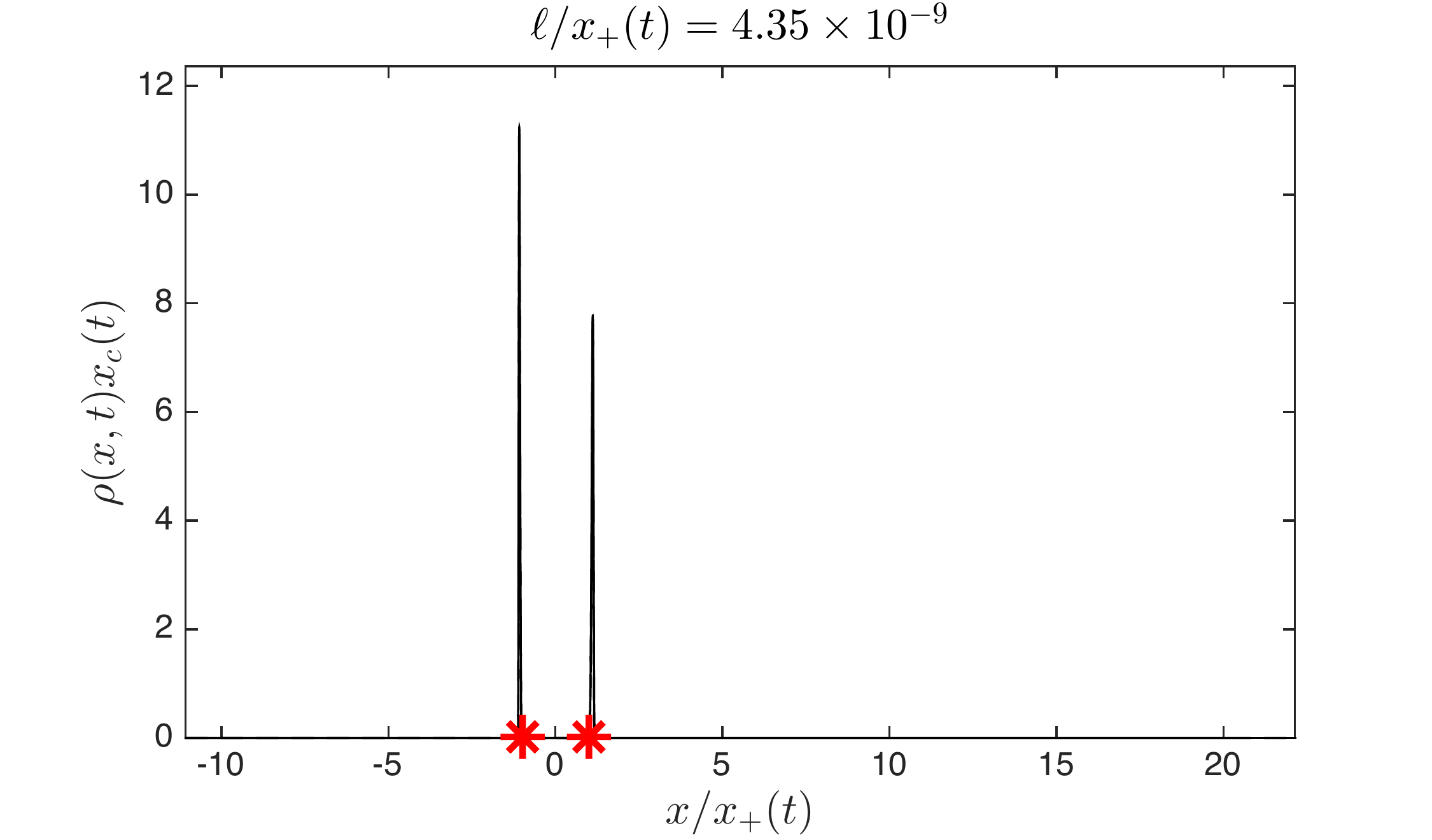} 
   \caption{} 
    \label{fig7:d} 
  \end{subfigure} 
   \caption{  \raggedright 
   WKB position-space probability densities for incoming wave-packet with 
    $\langle v\rangle$, $\langle x\rangle$ and time $t$ as in Fig.~\ref{fig:scatt-timeEvol}(d), but  
    now with $\ell$ reduced to give $\ell/x_c(t)=4.35\times 10^{-3}$, $10^{-5},$ $10^{-7}$, $10^{-9},$ 
    with notations as in Fig.~\ref{WKB_fixed_ell}. For movie, see \cite{SupplMat}.}
  \label{fig:scatt-ellEvol} 
\end{figure*}


These considerations fully characterize the value $x_0\equiv x_t^{-1}(0).$ The important constraint 
comes from the solution (\ref{eqn:cuspSol}) for the dynamics in the cusp-region $|x|>\ell$ of the potential.
Using this equation for final position $|x|=\ell$ and time $t-\tau_\ell(x_0)$ to reach that point yields
\begin{eqnarray}\label{xtinv-cond}
&& \ell^{(1-\alpha)/2}{\!\,}_2F_1\left(\frac{1}{2},-\frac{(1-\alpha)}{2(1+\alpha)}; \frac{1+3\alpha}{2(1+\alpha)}; +\left|\frac{\bar{x}_0}{\ell}\right|^{1+\alpha}\right) \cr
&& -|x_0|^{(1-\alpha)/2}{\!\,}_2F_1\left(\frac{1}{2},-\frac{(1-\alpha)}{2(1+\alpha)}; \frac{1+3\alpha}{2(1+\alpha)}; +\left|\frac{\bar{x}_0}{x_0}\right|^{1+\alpha}\right)\cr
&& \hspace{40pt}
= - \frac{1}{2}(1-\alpha)\left(\frac{2C}{1+\alpha}\right)^{1/2} (t-\tau_\ell(x_0)). 
\end{eqnarray}
Here we have assumed a negative-energy solution appropriate to a point near $x_\infty$ and taken a minus sign 
on the righthand side for the case ${\rm sign}(x_0\cdot\langle v\rangle)=-1.$ 
To evaluate $x_0=x_t^{-1}(0)$ asymptotically for small $\ell$ it is convenient to parameterize this solution by 
the dimensionless quantity $\xi>0$ introduced through the formula
\be  |x_0|^{1+\alpha}=|x_\infty|^{1+\alpha}-\xi \ell^{1+\alpha}. \lb{xi-eq} \ee 
In terms of this parameter $v_\ell^2=C\frac{1+\alpha+2\xi}{1+\alpha}$ and 
\be \tau_\ell(\xi)=\frac{1}{\gamma}{\rm arctanh}\sqrt{\frac{1+\alpha}{1+\alpha+2\xi}}. \lb{tau-eq} \ee
Also the quantity $\bar{x}_0$ appearing in (\ref{xtinv-cond}) must be chosen according to the condition 
$$ H_0=-\frac{C}{1+\alpha} |\bar{x}_0|^{1+\alpha}=\frac{1}{2}v_\ell^2-\frac{C}{1+\alpha}\ell^{1+\alpha}$$
which gives 
\be  |\bar{x}_0|^{1+\alpha} = \left[\frac{1}{2}(1-\alpha)-\xi\right]\ell^{1+\alpha}. \lb{xbar-eq} \ee
With these choices $x_0=x_t^{-1}(0)$ is the solution of the fixed-point condition (\ref{xtinv-cond}).  

We can now evaluate the solution asymptotically as $\ell\rightarrow 0.$ 
Because $\gamma\rightarrow\infty$ in that limit, the only way to satisfy (\ref{tau-eq}) 
for a fixed finite value $\tau_\ell(\xi)=\tau>0$ is to choose the parameter $\xi$ exponentially small:
$$ \xi \doteq 2(1+\alpha) e^{-2\gamma \tau}. $$
This implies that $x_0$ in (\ref{xi-eq}) must be exponentially close to $x_\infty$ and, in particular, 
$ x_0\rightarrow x_*, $
verifying our original expectation. Furthermore from (\ref{xbar-eq})
$$ \left|\frac{\bar{x}_0}{\ell}\right|^{1+\alpha} \rightarrow  \frac{1}{2}(1-\alpha), $$
and from (\ref{xbar-eq}),(\ref{xi-eq})
$$ \left|\frac{\bar{x}_0}{x_0}\right|^{1+\alpha} =O\left(\left|\frac{\ell}{x_*}\right|^{1+\alpha}\right)\rightarrow 0. $$
These results in the fixed-point condition (\ref{xtinv-cond}) give
$$ O(\ell^{(1-\alpha)/2}) + |x_*|^{(1-\alpha)/2}= \frac{1}{2}(1-\alpha)\left(\frac{2C}{1+\alpha}\right)^{1/2} (t-\tau_\ell),$$
which can only hold if $t-\tau_\ell = t_* +O((\ell^{1-\alpha}/C)^{1/2}),$ where $t_*$ is the time given by the relation 
$$|x_*|= \left[\frac{1}{2}(1-\alpha)\left(\frac{2C}{1+\alpha}\right)^{1/2} t_*\right]^{2/(1-\alpha)}, $$
or the time $t_*$ required for the extremal solutions (\ref{xtreme}) of the cusp-potential to transit 
from $x_*$ to 0. We conclude that $x_0=x_t^{-1}(0)$ for the spliced-potential is given by 
\be  |x_0|^{1+\alpha}\doteq |x_\infty|^{1+\alpha}-2(1+\alpha) e^{-2\gamma (t-t_*)}\ell^{1+\alpha} \lb{xtinv}  \ee
in the limit $\ell\rightarrow 0,$ for $x_\infty$ in (\ref{xinf}) and $\gamma=(C/\ell^{1-\alpha})^{1/2}, $
for all $t\geq t_*.$ The simple picture that emerges here is that solutions starting at $x_0=x_t^{-1}(0)$ 
given by (\ref{xtinv}) reach a distance $\ell$ from the origin in a time $\simeq t_*$,  nearly independent 
of $\ell,$ and then take an additional time $\simeq \tau=t-t_*$ to reach the origin, for a total time $t.$ This is 
made possible by the fact that extremal solutions (\ref{xtreme}) of the cusp-potential
transit between $x_*$ and 0 in a finite time $t_*,$ as a consequence of the non-smoothness of that potential.

With these results in hand, we can now discuss the wavepacket scattering in the WKB limit.  
With the mean velocity $\langle v\rangle$ fixed, there are various possible choices of $\langle x\rangle$ to 
observe splitting of the wavepacket and QSS. For example, one may choose $\langle x\rangle=x_\infty.$ 
Since (\ref{xtinv}) implies
$$ x_t^{-1}(0) \doteq x_\infty \left[1-2 e^{-2\gamma(t-t_*)}\left|\frac{\ell}{x_*}\right|^{1+\alpha}\right], $$ 
almost any reasonable choice of $\sigma,$ e.g. $\sigma\propto \ell^p$ for any power $p,$ will give
$$ y_*(t)=0, \quad t\geq t_*. $$
This corresponds to a symmetrical splitting, with both $p_-(t)=p_+(t)=\frac{1}{2}$ for $t\geq t_*.$ 
Another natural choice is $\langle x\rangle=x_*.$ Using (\ref{xtinv}),(\ref{xinf}), it follows that
$$ x_t^{-1}(0) \doteq x_* \left[1+\frac{1-\alpha}{2(1+\alpha)}\left|\frac{\ell}{x_*}\right|^{1+\alpha}\right], $$ 
so that taking $\sigma=\beta \ell^{1+\alpha}/|x_*|^\alpha$ gives
$$ y_*(t)=\frac{1}{2\beta} \frac{1-\alpha}{1+\alpha}\ {\rm sign} (x_*), \quad t\geq t_*. $$
Hence $p_+(t)=\frac{1}{2}{\rm erfc}(\frac{\pm1}{2^{3/2}\beta} \frac{1-\alpha}{1+\alpha})$ for $t\geq t_*$
in terms of the complementary error function, corresponding to an asymmetrical splitting. Still another 
possible choice is $\langle x\rangle=x_*-\kappa \sigma$ for any $\sigma$ vanishing more slowly than 
$\ell^{1+\alpha},$ so that $\ell^{1+\alpha}/\sigma\rightarrow 0$ as $\ell\rightarrow 0.$ In this case
$$ y_*(t)=\kappa, \quad t\geq t_*. $$
and $p_+(t)=\frac{1}{2}{\rm erfc}(\kappa)$ for $t\geq t_*.$ This would hold with the choice 
$\sigma=\ell/\mu$ of the previous section, for $\mu$ fixed. Although there seem to be a plethora 
of possibilities, they are actually quite restricted since one must always make a careful fine-tuning 
so that $\langle x\rangle\rightarrow x_*$ as $\ell\rightarrow 0$ in order to observe QSS in the scattering process
for fixed $\langle v\rangle.$ 

We now present numerical WKB results for scattering of a wave-packet off both the Kummer potential and the 
spliced potential. Here we shall consider only the choice $\langle x\rangle=x_\infty$ for fixed $\langle v\rangle$
and $\sigma=\ell,$ based upon the analytical results for the sliced potential. Using the same choices for the 
Kummer potential is a good test on the robustness of the derived selection criteria for small changes 
in the regularization of the potential. (Note that one could also make an alternative choice of $x_\infty$ 
for the Kummer potential by using its own energy conservation law to choose the position of the particle 
with velocity $\langle v\rangle$ which arrives to the origin with zero velocity.) We consider first the choice 
$\alpha=1/3$ and $\langle v\rangle/(C\ell^{1+\alpha})^{1/2}=20$ so that $x_\infty/\ell\doteq -66.05$ and 
$\tau_*=t_*/(\ell^{1-\alpha}/C)^{1/2}\doteq 9.902.$ In Fig.~\ref{fig:scatt-timeEvol} 
we show the time-evolution of the WKB position PDF by plotting it at four successive choices of the dimensionless 
time $\tau$. The classical solution with which we now compare is $x_c(t)=x_\pm(|t-t_*|),$ $\pm={\rm sign}(t-t_*),$
which is incoming as $x_-$ and outgoing as $x_+.$ We see that the PDF's for both potentials evolve very 
similarly and split at a time near $t_*.$ It is notable that in the final frame for $\tau=15,$ the 
probabilities $p_-,p_+$ are nearly equal to the symmetrical value 1/2 but the shapes of the peaks 
on the left and right are different. Next  we study the effect of decreasing values of $\sigma=\ell.$ In 
Fig.~\ref{fig:scatt-ellEvol} we fix $\langle v\rangle$, $\langle x\rangle=x_\infty$,  final time $t$, and position $x_c(t)$ 
from the last panel in Fig.~\ref{fig:scatt-timeEvol} and successively reduce $\ell/x_c(t)$.
There is clear evidence of QSS, with the PDF converging to a pair of delta-functions. Even though
$p_+,p_-\rightarrow1/2,$ the shape asymmetry of the PDF peaks remains. The most important observation
is that the results for the Kummer and spliced potentials are quite similar, giving hope that our selection
criterion, although derived rigorously only for the spliced potential, will apply more generally. 
 Other choices of the fine-tuning of the initial wave-packet 
are implemented by the codes in \cite{SupplMat}.    

\section{Numerical Schr\"odinger Evolution}\lb{SchrEq}

\begin{figure*}[!ht] 
  \begin{subfigure}[b]{0.5\linewidth}
    \centering
    \includegraphics[width=1\columnwidth,height=0.5\columnwidth]{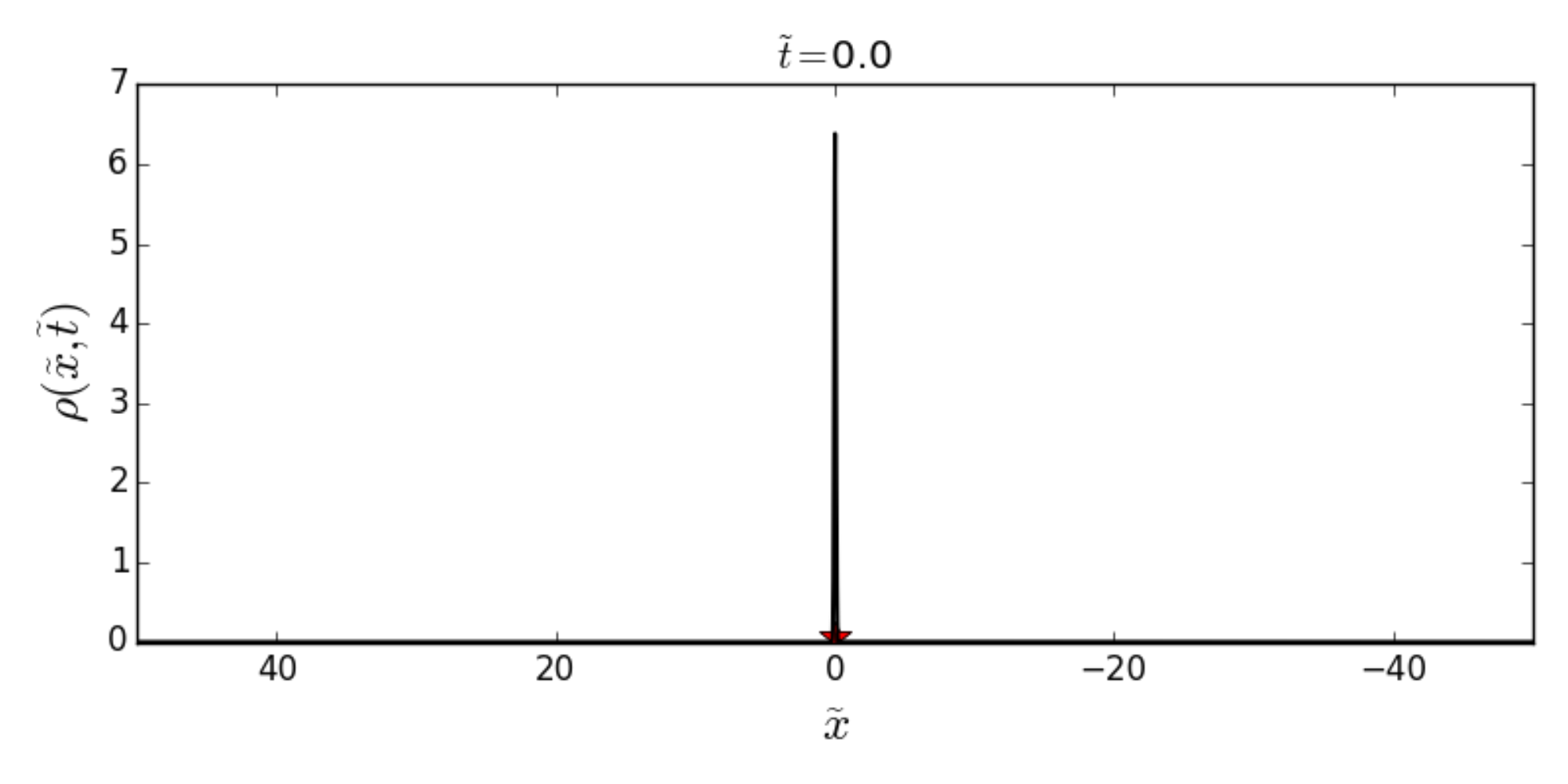} 
        \vspace{-8mm}
   \caption{} 
    \label{fig8:a} 
        \vspace{-1mm}
  \end{subfigure}
  \begin{subfigure}[b]{0.5\linewidth}
    \centering
    \includegraphics[width=1\columnwidth,height=0.5\columnwidth]{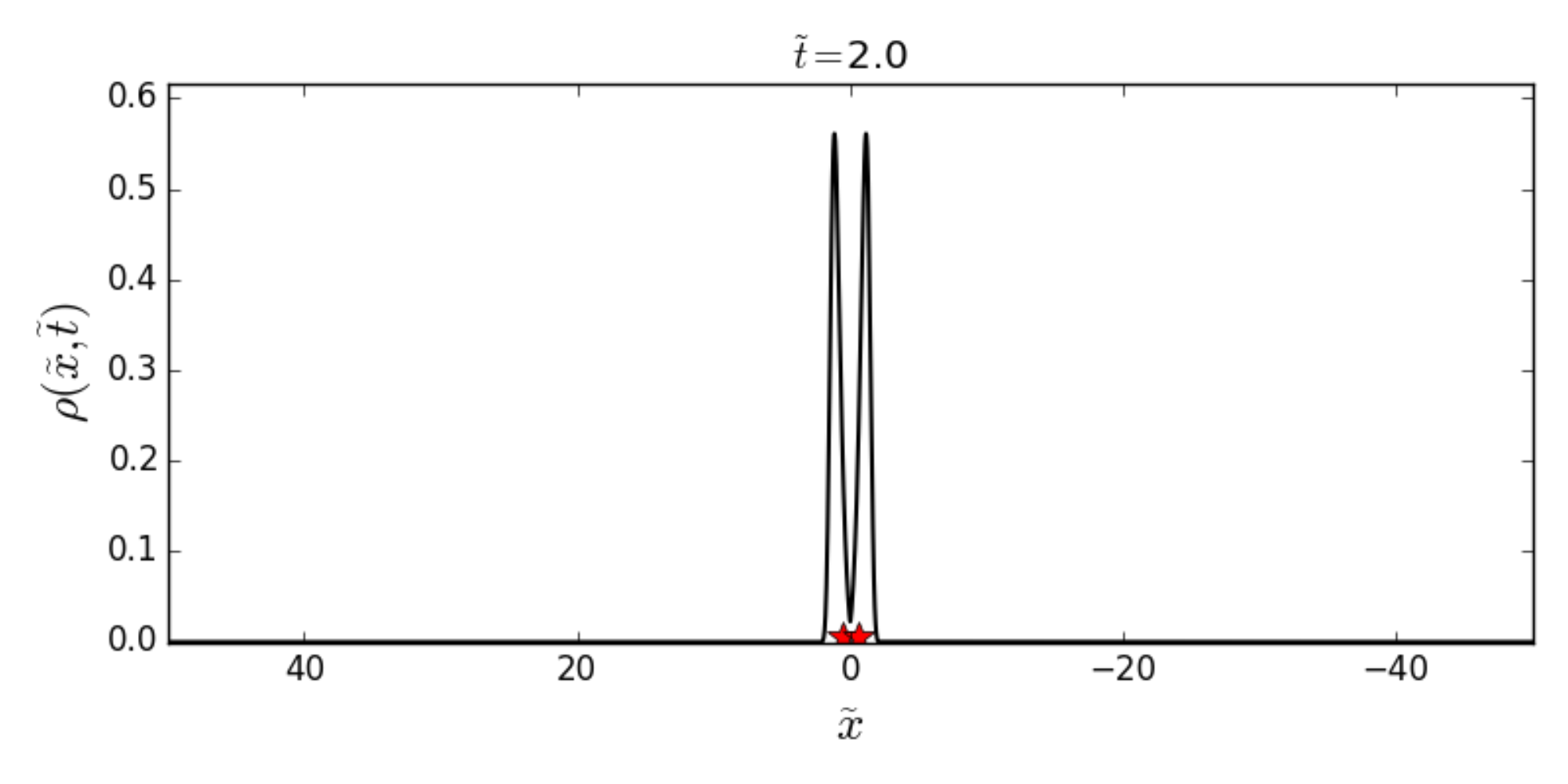} 
   \vspace{-8mm}
   \caption{} 
    \label{fig8:b} 
        \vspace{-1mm}
  \end{subfigure} 
  \begin{subfigure}[b]{0.5\linewidth}
    \centering
    \includegraphics[width=1\columnwidth,height=0.5\columnwidth]{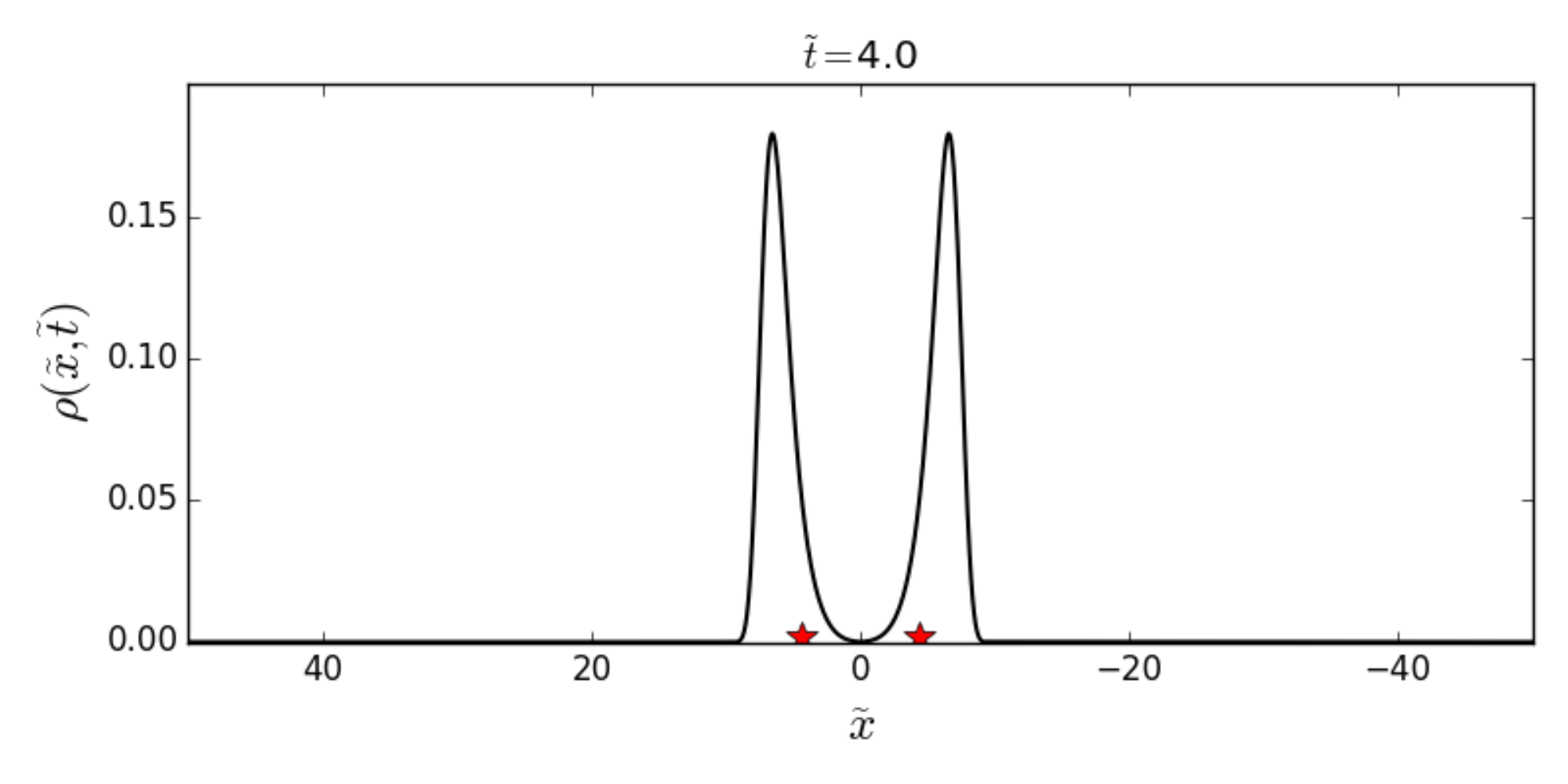} 
        \vspace{-8mm}
   \caption{} 
    \label{fig8:c} 
  \end{subfigure}
  \begin{subfigure}[b]{0.5\linewidth}
    \centering
    \includegraphics[width=1\columnwidth,height=0.5\columnwidth]{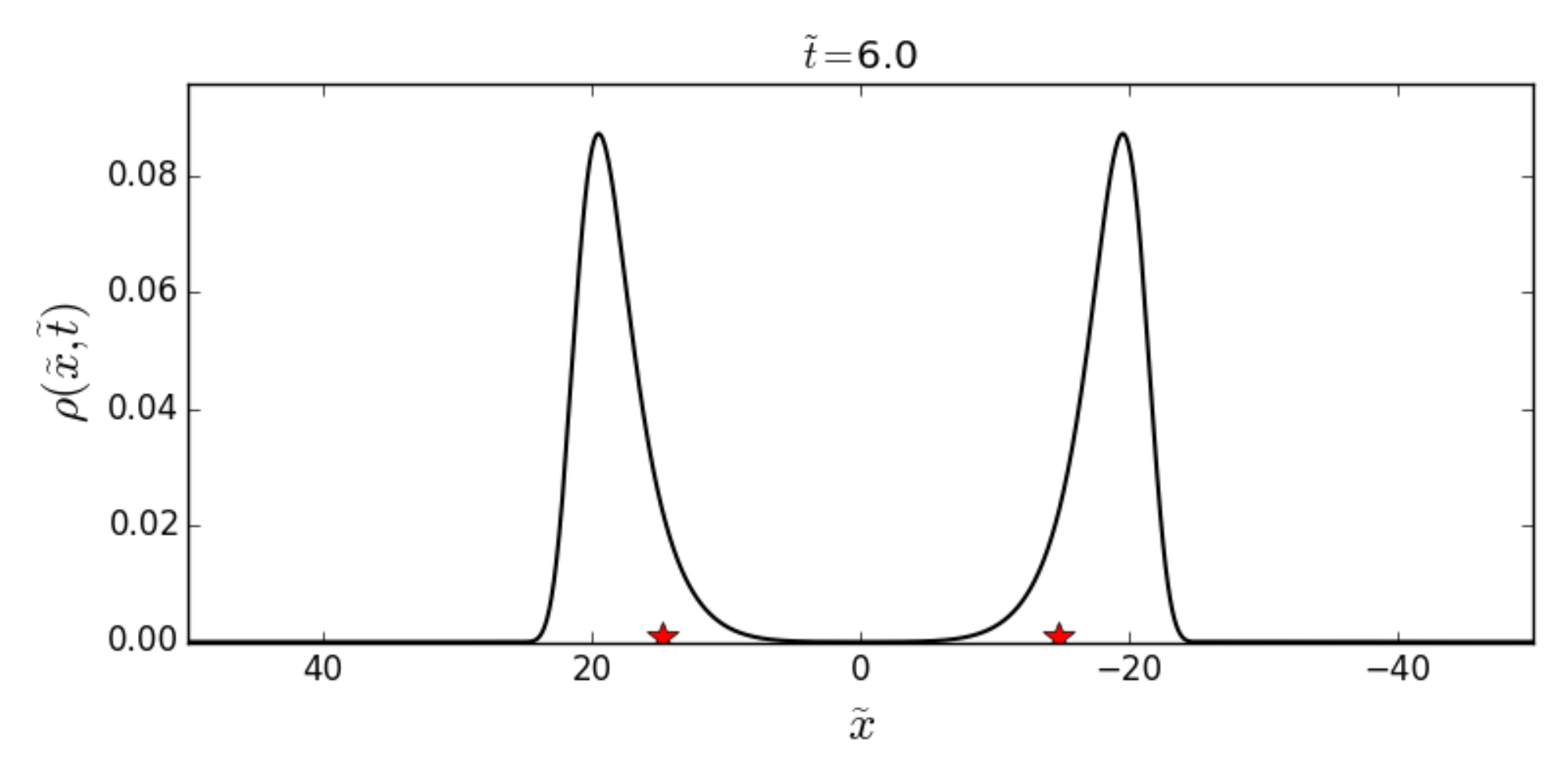} 
        \vspace{-8mm}
   \caption{} 
    \label{fig8:d} 
  \end{subfigure} 
  \caption{ \raggedright  Position-space PDF for numerical Schr\"odinger solution with $\varepsilon=2^{-7},$  at successive 
   times $\tilde{t}=0,$ $2$, $4$, $6$.  Asterisks on the abscissa denote positions of the extremal classical solutions.
   For movie, see \cite{SupplMat}.}
  \label{fig:pos-timeEvol} 
\end{figure*}


\begin{figure*}[!ht] 
  \begin{subfigure}[b]{0.5\linewidth}
    \centering
    \includegraphics[width=1\columnwidth,height=0.5\columnwidth]{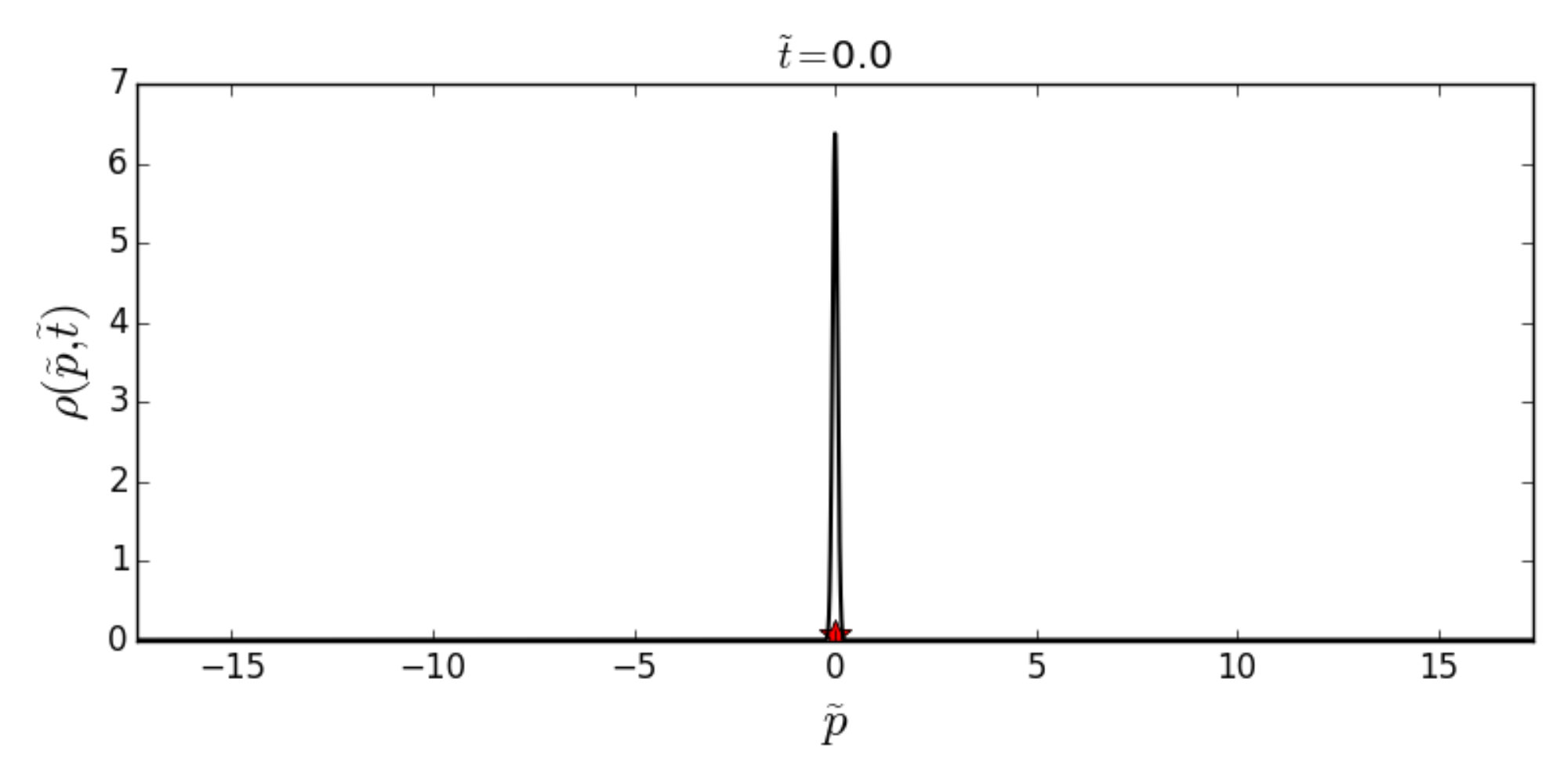} 
        \vspace{-8mm}
   \caption{} 
    \label{fig9:a} 
        \vspace{-1mm}
  \end{subfigure}
  \begin{subfigure}[b]{0.5\linewidth}
    \centering
    \includegraphics[width=1\columnwidth,height=0.5\columnwidth]{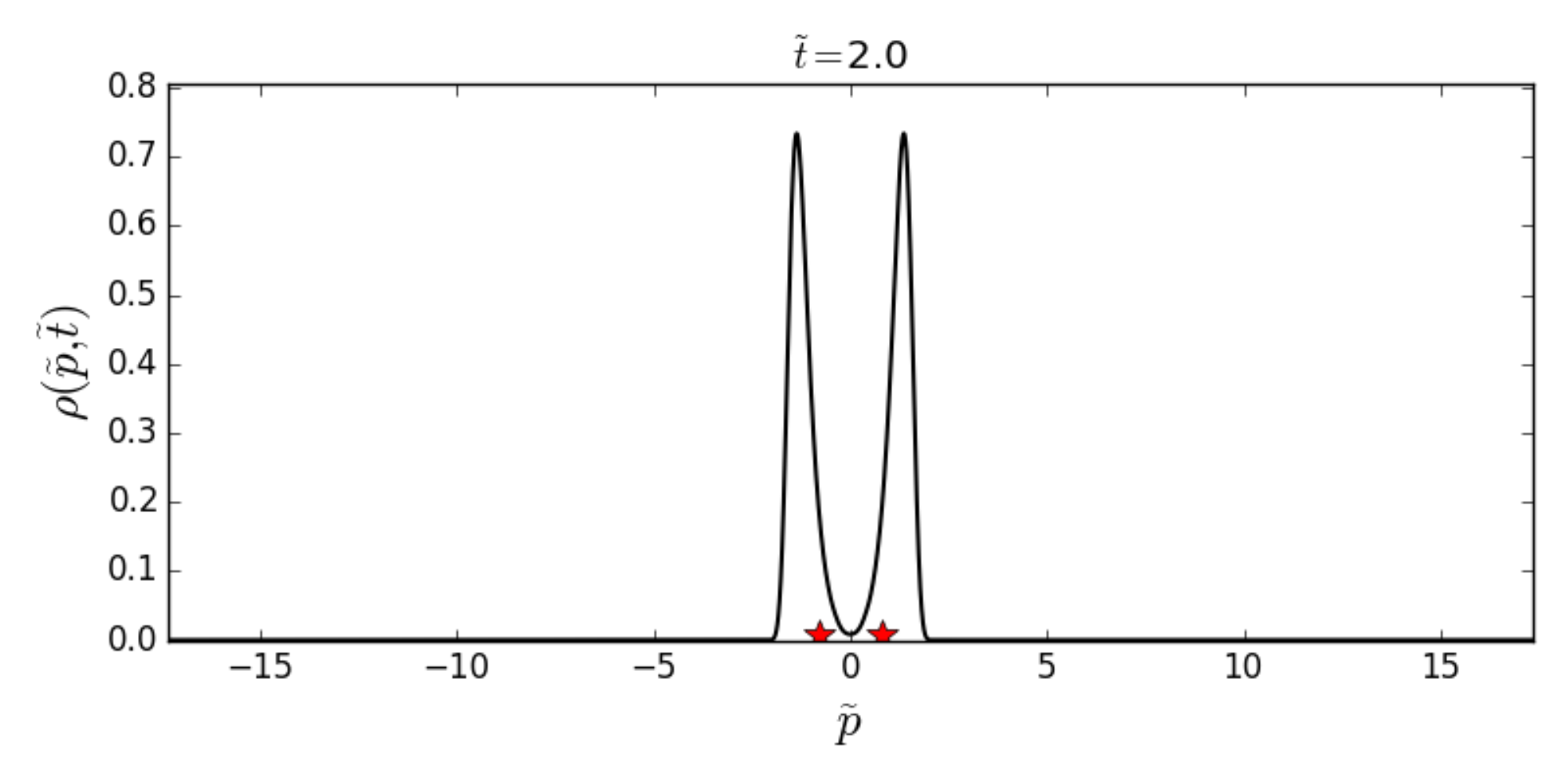} 
   \vspace{-8mm}
   \caption{} 
    \label{fig9:b} 
        \vspace{-1mm}
  \end{subfigure} 
  \begin{subfigure}[b]{0.5\linewidth}
    \centering
    \includegraphics[width=1\columnwidth,height=0.5\columnwidth]{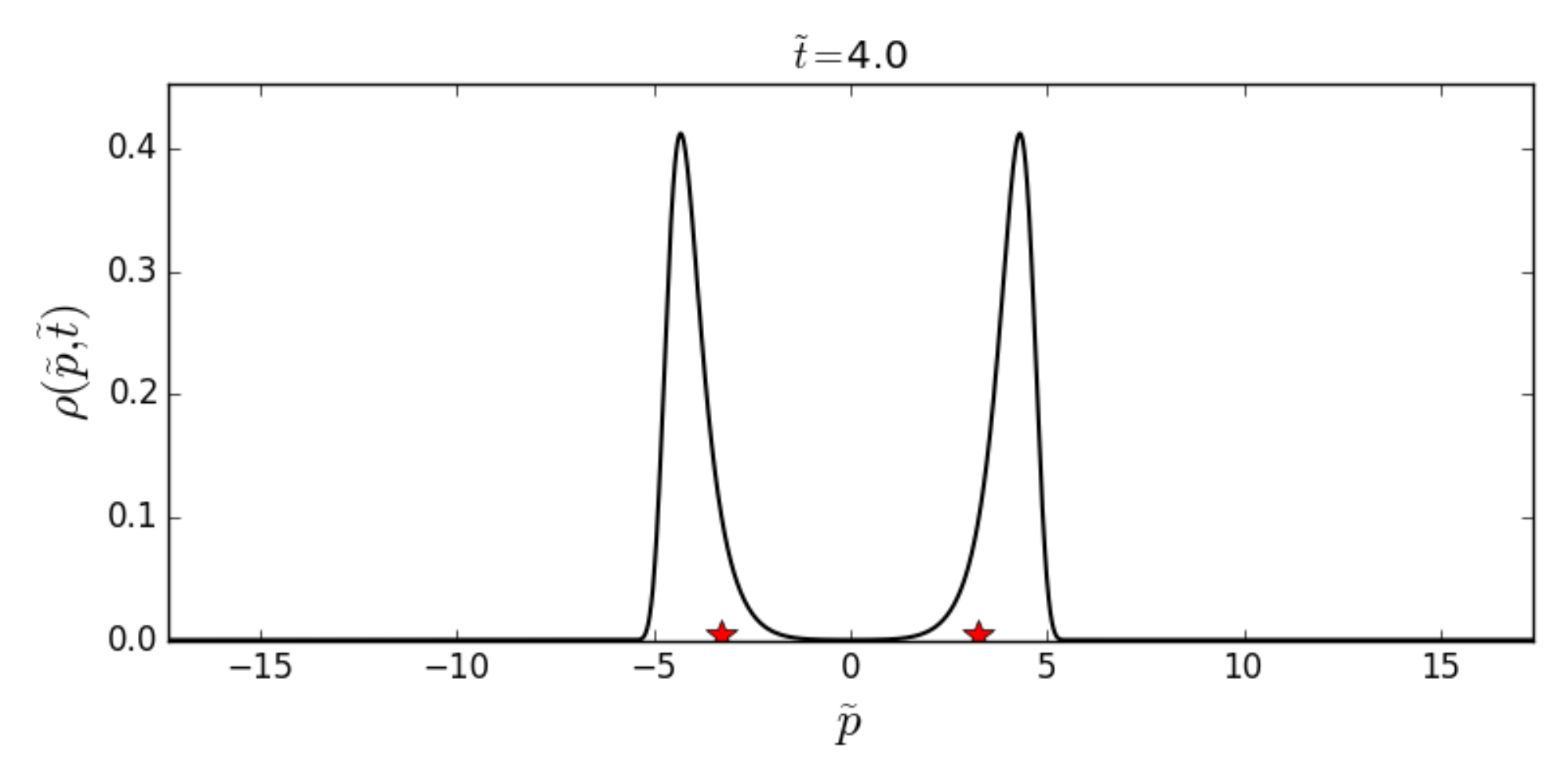} 
        \vspace{-8mm}
   \caption{} 
    \label{fig9:c} 
  \end{subfigure}
  \begin{subfigure}[b]{0.5\linewidth}
    \centering
    \includegraphics[width=1\columnwidth,height=0.5\columnwidth]{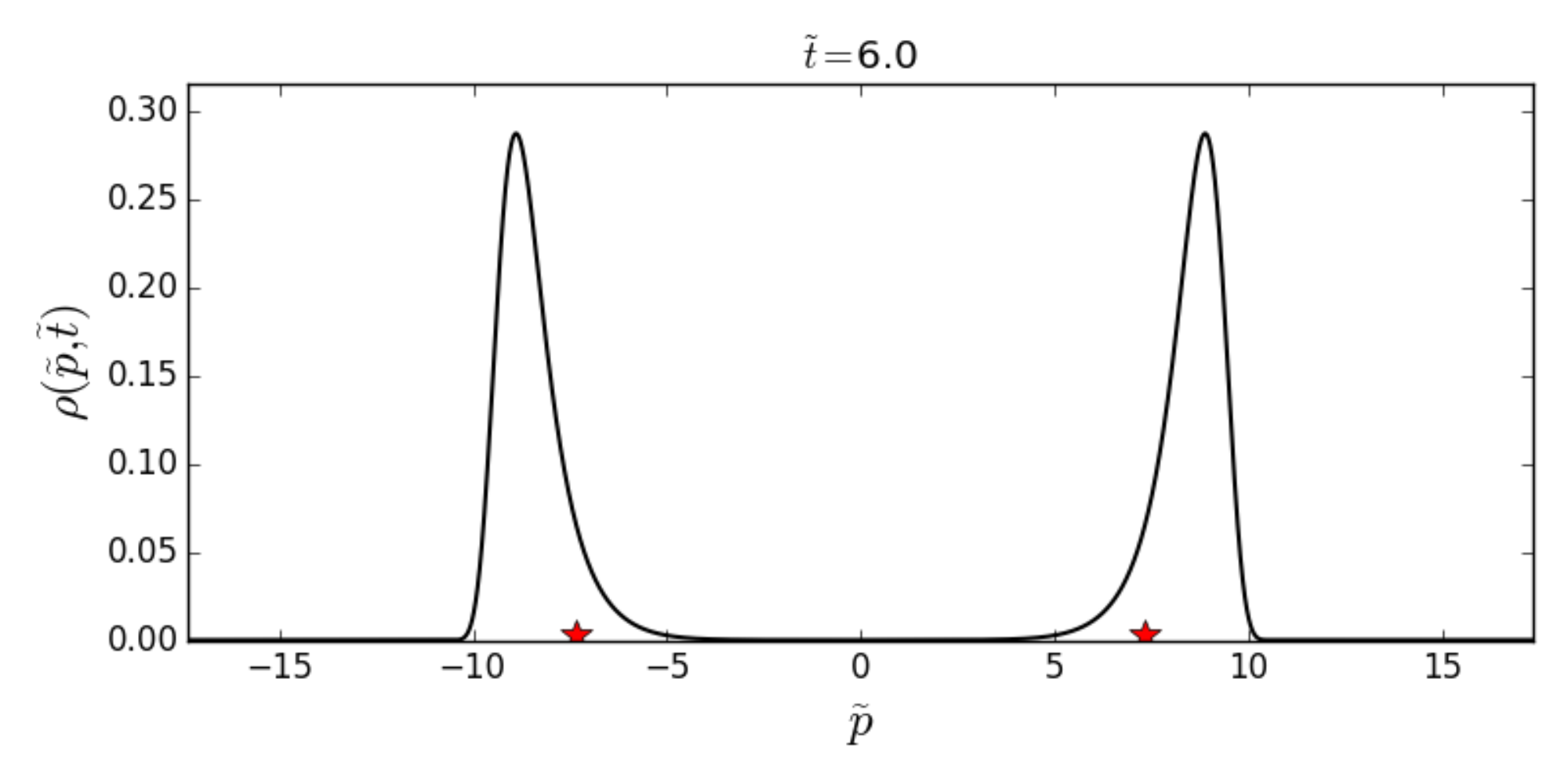} 
        \vspace{-8mm}
   \caption{} 
    \label{fig9:d} 
  \end{subfigure} 
  \caption{ \raggedright  Momentum-space PDF for numerical Schr\"odinger solution with $\varepsilon=2^{-7},$  at successive 
   times $\tilde{t}=0,$ $2$, $4$, $6$.  Asterisks on the abscissa denote momenta of the extremal classical solutions.  For movie, see \cite{SupplMat}.}
  \label{fig:mom-timeEvol} 
\end{figure*}


\begin{figure*}[!ht] 
  \begin{subfigure}[b]{0.5\linewidth}
    \centering
    \includegraphics[width=1\columnwidth,height=0.5\columnwidth]{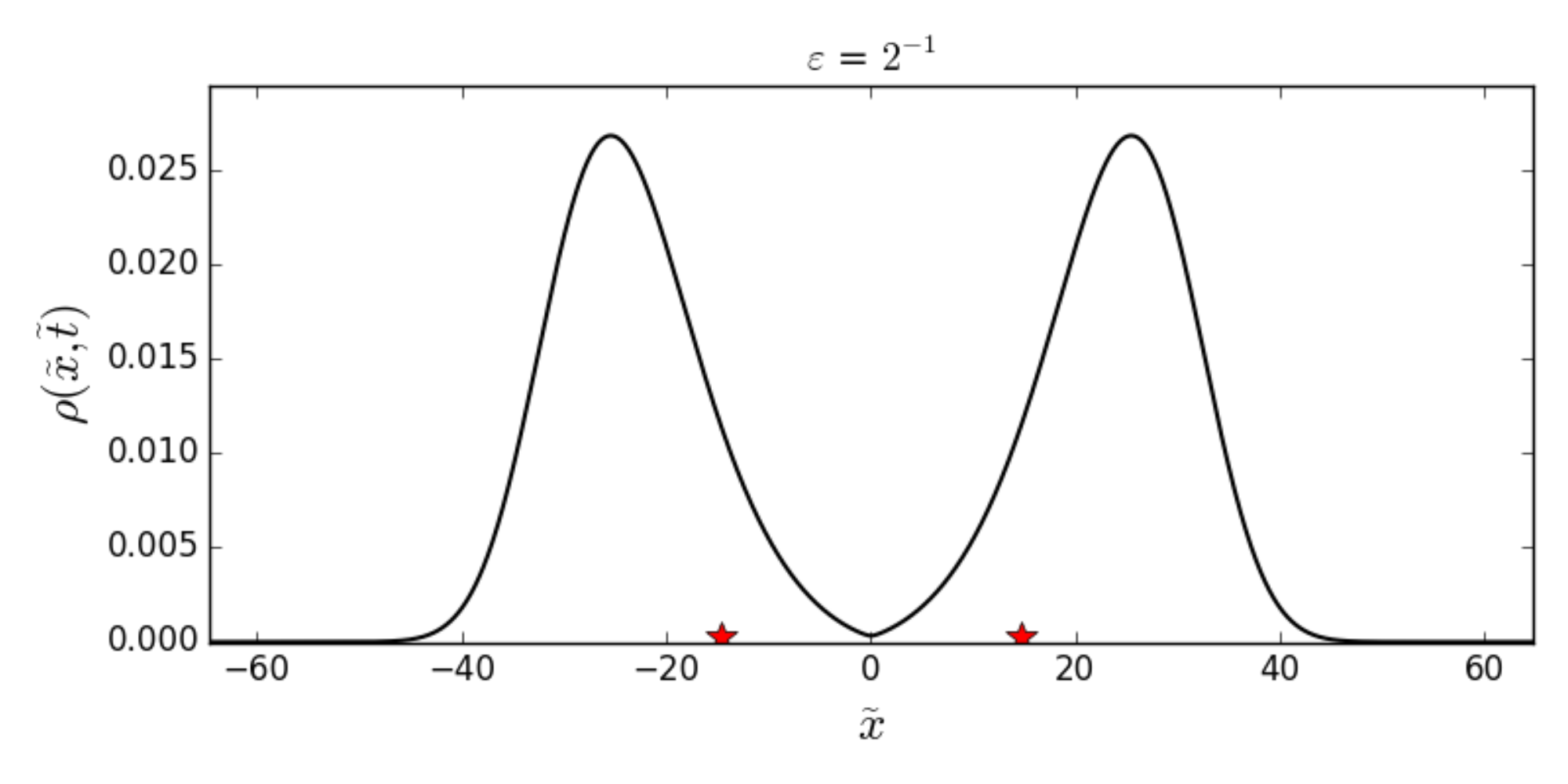} 
        \vspace{-8mm}
   \caption{} 
    \label{fig10:a} 
        \vspace{-1mm}
  \end{subfigure}
  \begin{subfigure}[b]{0.5\linewidth}
    \centering
    \includegraphics[width=1\columnwidth,height=0.5\columnwidth]{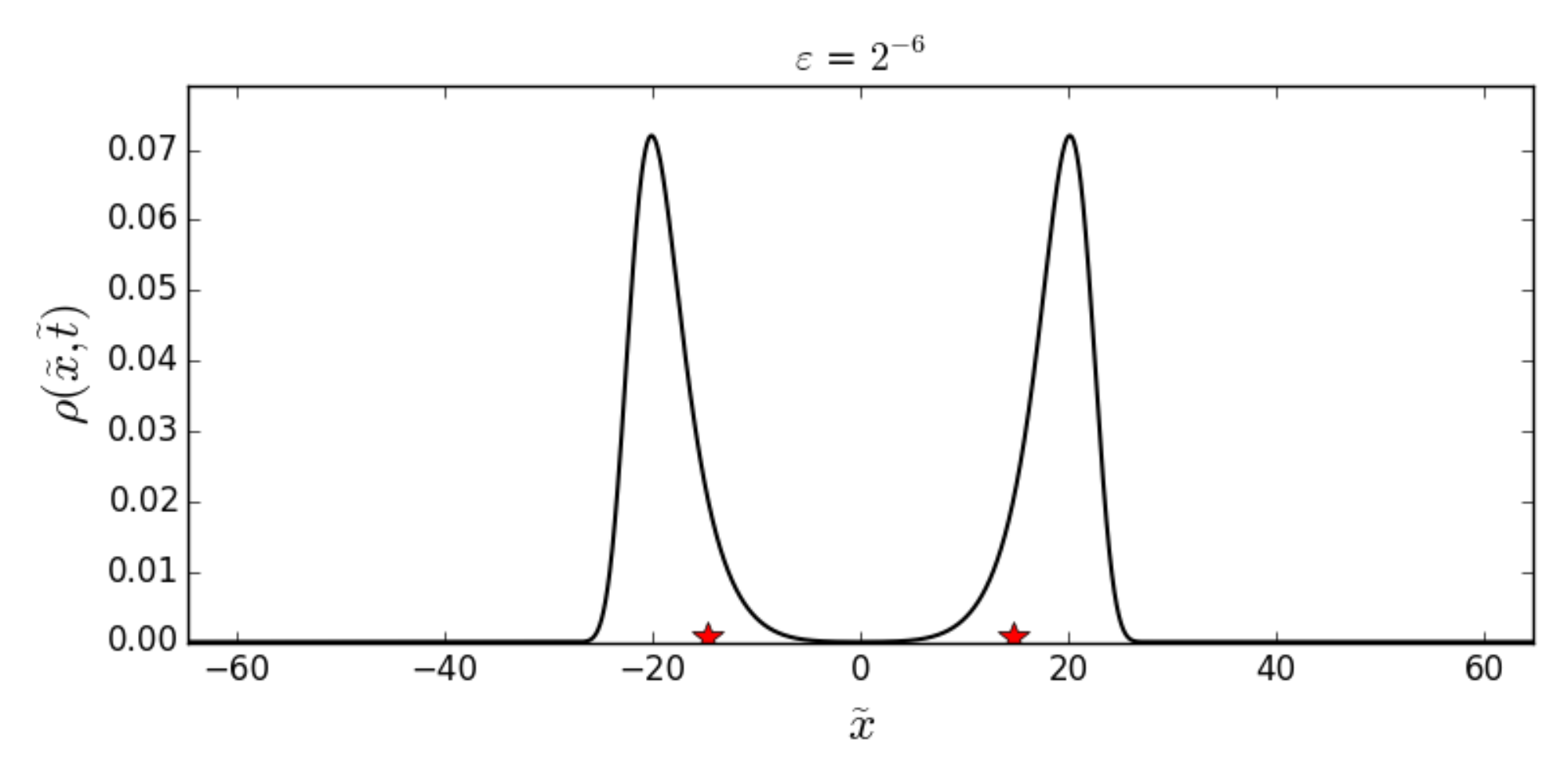} 
   \vspace{-8mm}
   \caption{} 
    \label{fig10:b} 
        \vspace{-1mm}
  \end{subfigure} 
  \begin{subfigure}[b]{0.5\linewidth}
    \centering
    \includegraphics[width=1\columnwidth,height=0.5\columnwidth]{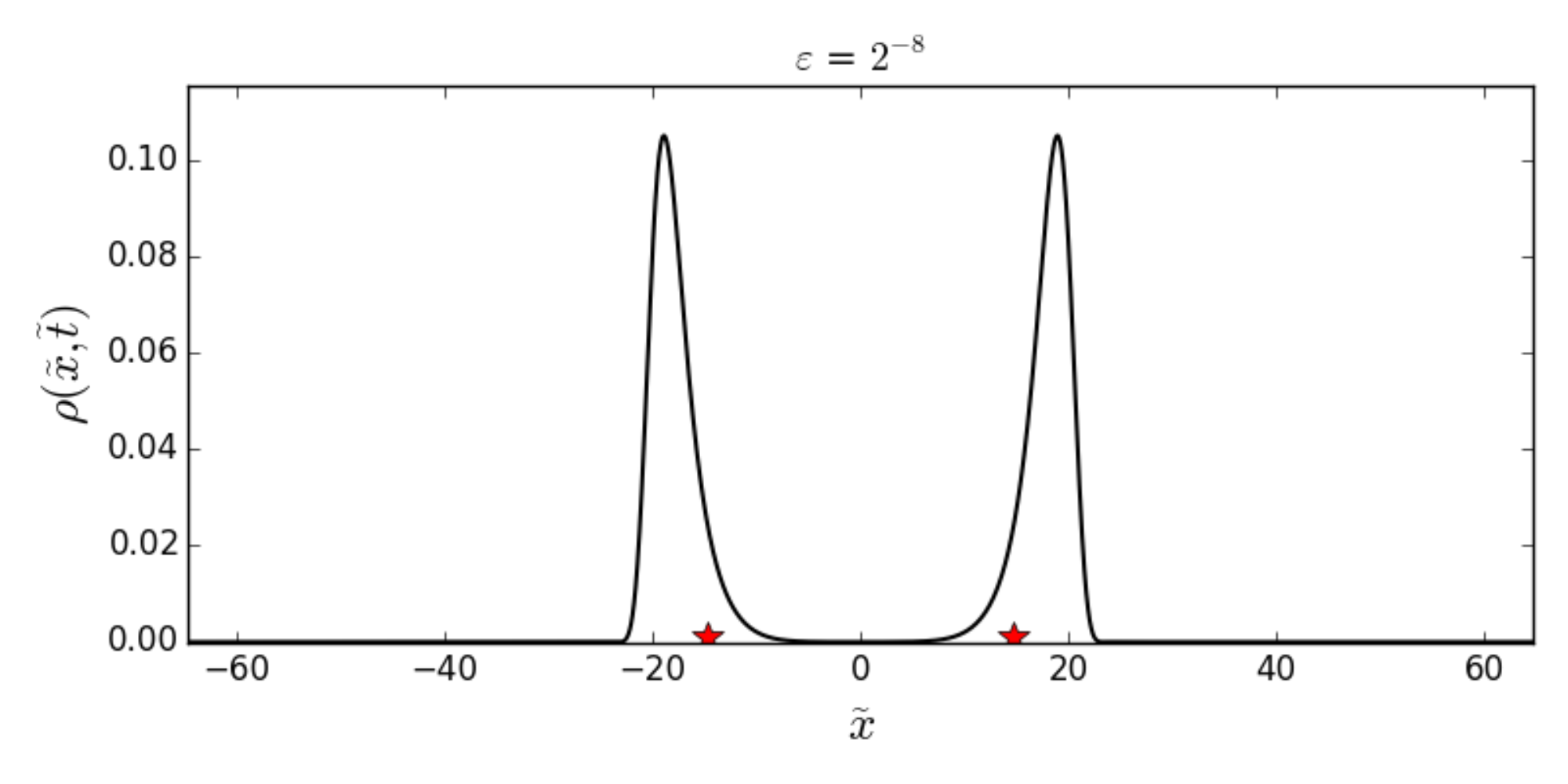} 
        \vspace{-8mm}
   \caption{} 
    \label{fig10:c} 
  \end{subfigure}
  \begin{subfigure}[b]{0.5\linewidth}
    \centering
    \includegraphics[width=1\columnwidth,height=0.5\columnwidth]{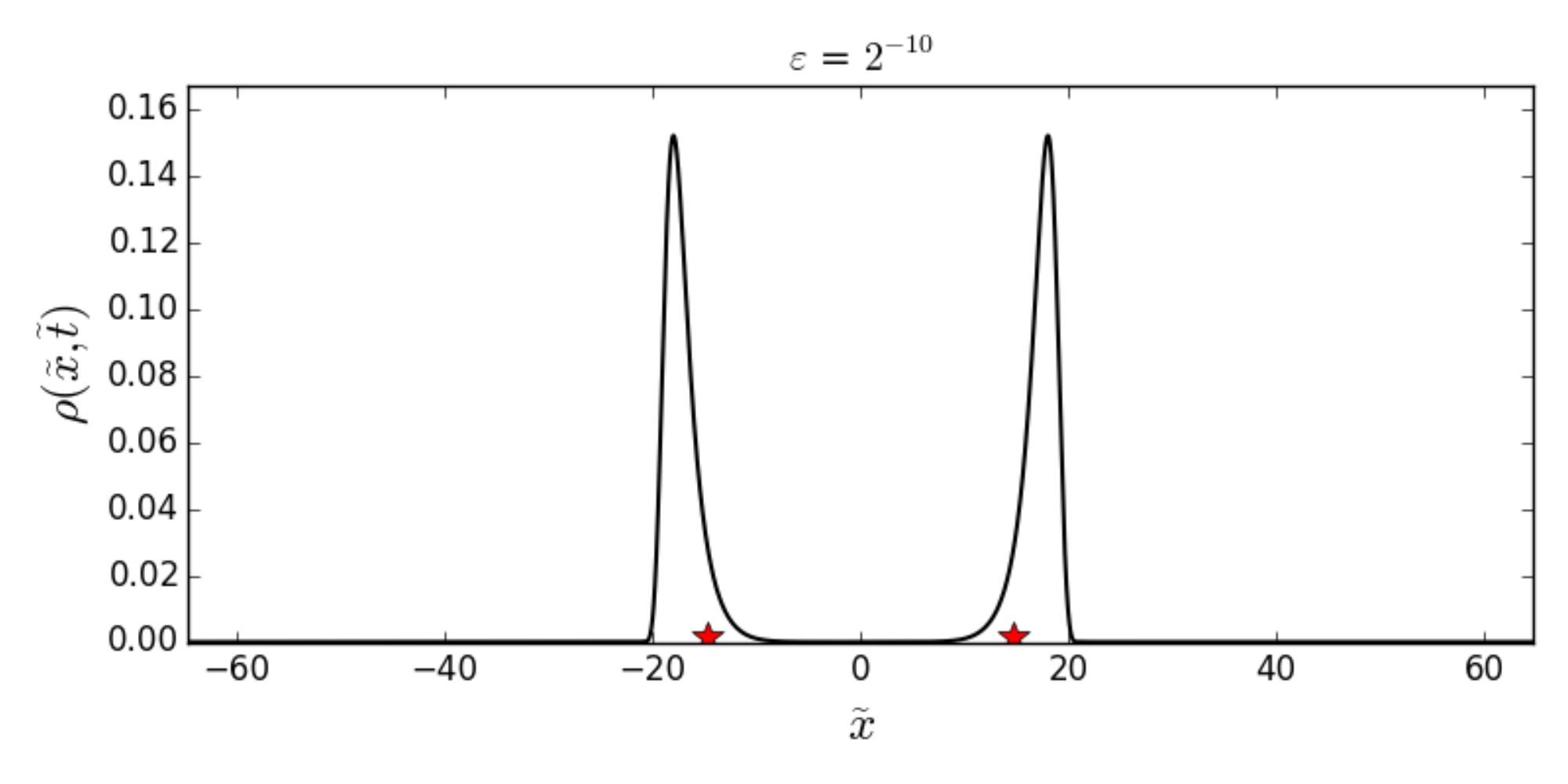} 
        \vspace{-8mm}
   \caption{} 
    \label{fig10:d} 
  \end{subfigure} 
   \caption{ \raggedright  Position-space PDF for Schr\"odinger solution at $\tilde{t}=6,$  for $\varepsilon= 2^{-1}$, $2^{-6}$, $2^{-8}$, $2^{-10},$
   with notations as in Fig.~\ref{fig:pos-timeEvol}. For movie, see \cite{SupplMat}.}
  \label{fig:pos-hbarEvol} 
\end{figure*}

\begin{figure*}[!ht] 
  \begin{subfigure}[b]{0.5\linewidth}
    \centering
    \includegraphics[width=1\columnwidth,height=0.5\columnwidth]{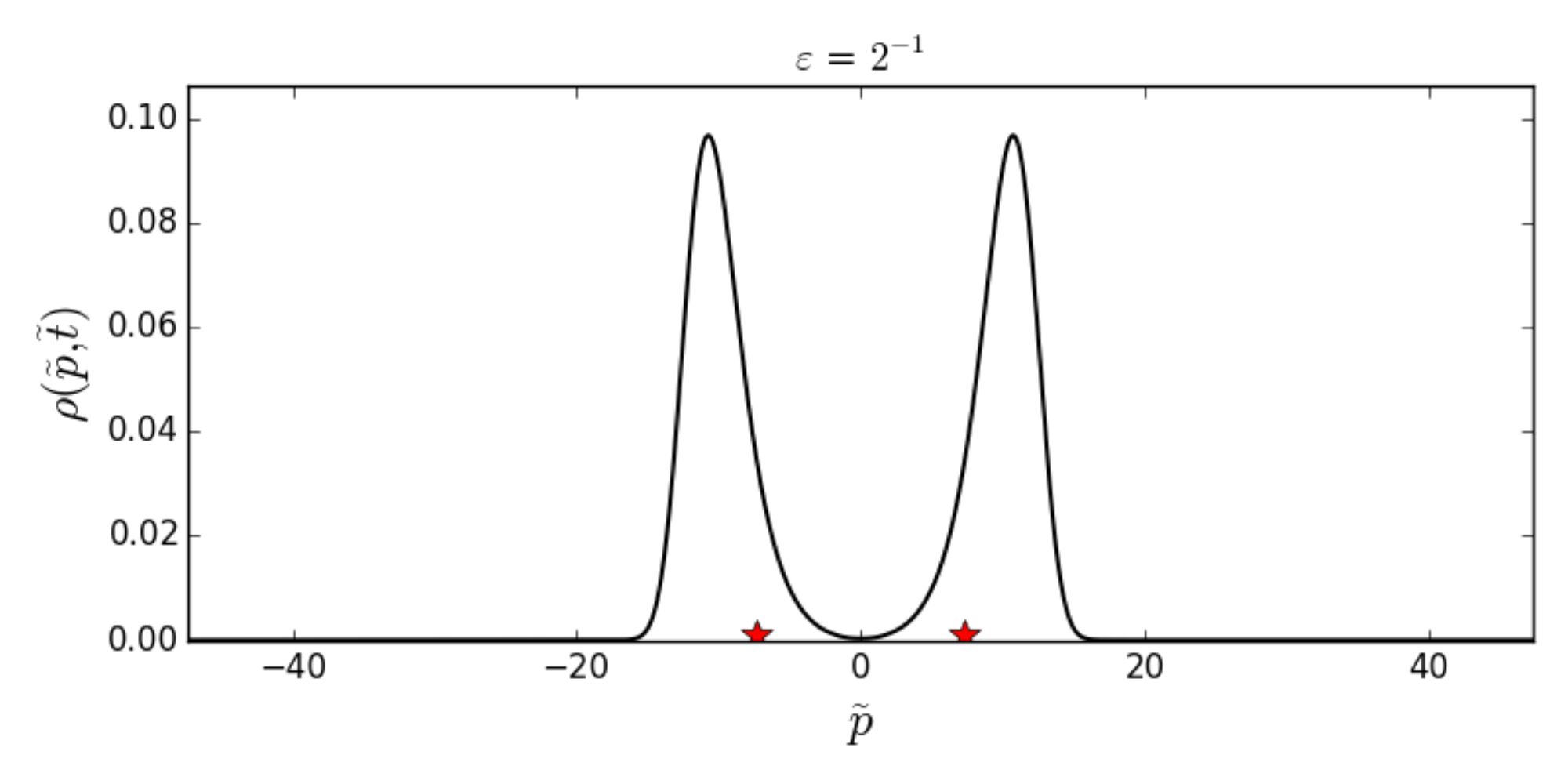} 
        \vspace{-8mm}
   \caption{} 
    \label{fig11:a} 
        \vspace{-1mm}
  \end{subfigure}
  \begin{subfigure}[b]{0.5\linewidth}
    \centering
    \includegraphics[width=1\columnwidth,height=0.5\columnwidth]{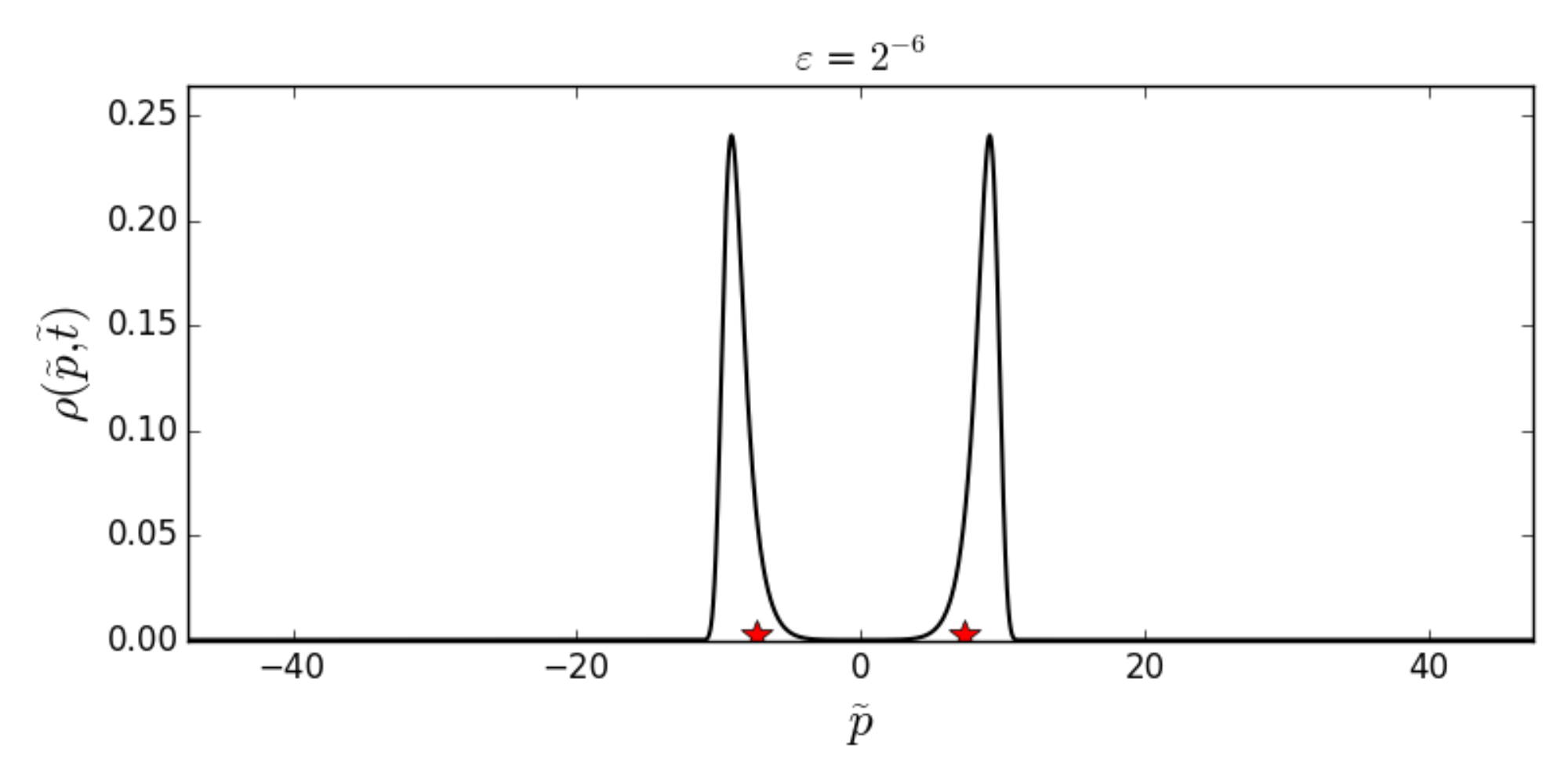} 
   \vspace{-8mm}
   \caption{} 
    \label{fig11:b} 
        \vspace{-1mm}
  \end{subfigure} 
  \begin{subfigure}[b]{0.5\linewidth}
    \centering
    \includegraphics[width=1\columnwidth,height=0.5\columnwidth]{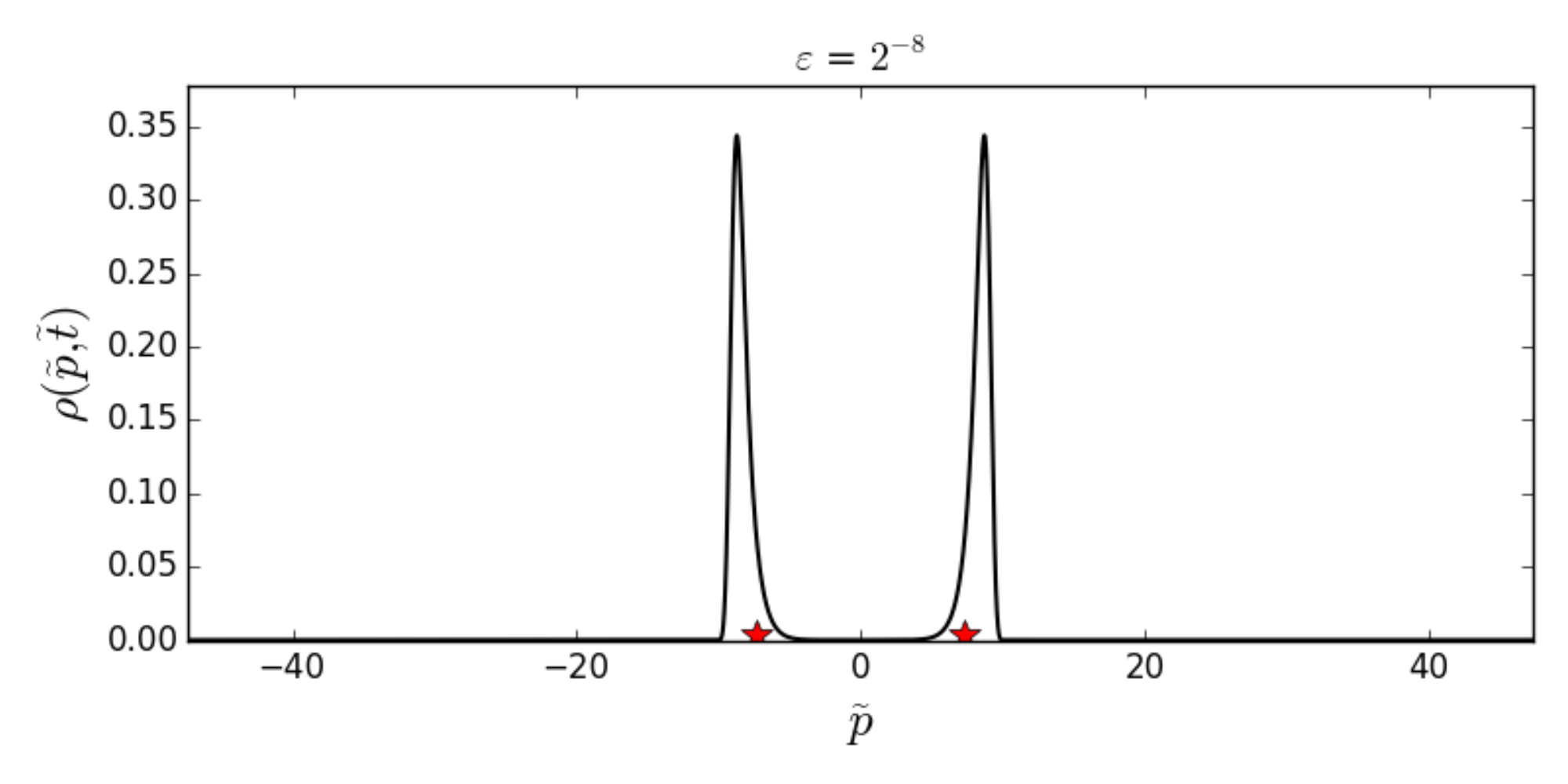} 
        \vspace{-8mm}
   \caption{} 
    \label{fig11:c} 
  \end{subfigure}
  \begin{subfigure}[b]{0.5\linewidth}
    \centering
    \includegraphics[width=1\columnwidth,height=0.5\columnwidth]{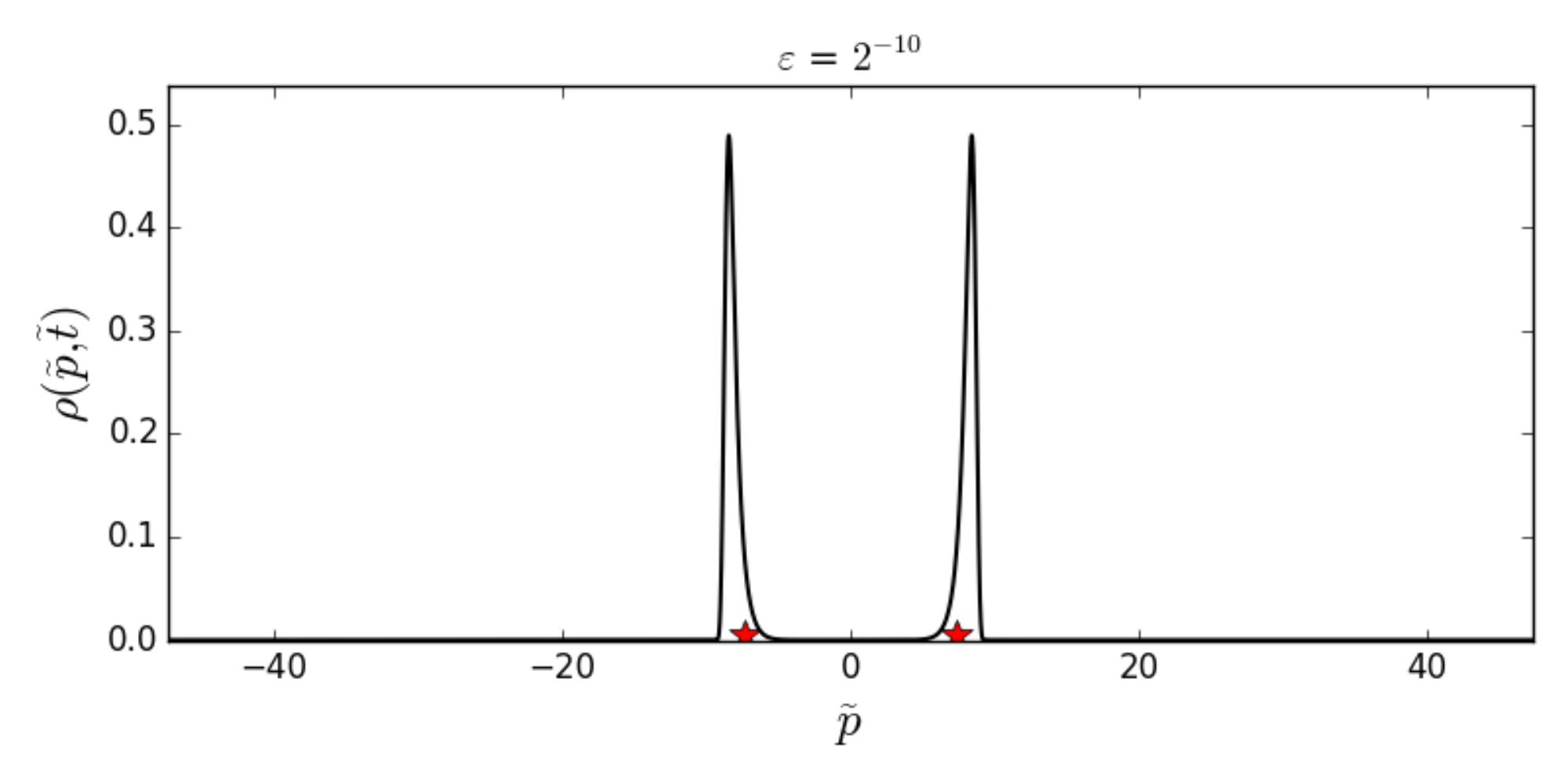} 
        \vspace{-8mm}
   \caption{} 
    \label{fig11:d} 
  \end{subfigure} 
   \caption{ \raggedright  Momentum-space PDF for Schr\"odinger solution at $\tilde{t}=6,$  for $\varepsilon= 2^{-1}$, $2^{-6}$, $2^{-8}$, $2^{-10},$
   with notations as in Fig.~\ref{fig:mom-timeEvol}. For movie, see \cite{SupplMat}.} 
  \label{fig:mom-hbarEvol} 
\end{figure*}

Thus far, we have considered a WKB regime in which, formally, the limit $h/m\rightarrow 0$ is taken first 
with position spread $\Delta x_0=\sigma$ of the wavepacket fixed and then subsequently letting 
$\sigma\rightarrow 0.$ However, this is not the only limit in which deterministic classical dynamics 
is generally expected to emerge. One can also let $h/m$ and $\Delta x_0$ go to zero together, e.g. 
with position spread $\Delta x_0 \propto (h/m)^\beta$ and velocity spread $\Delta v_0 \propto (h/m)^{1-\beta}$ 
for any $0<\beta<1, $ consistent with the Heisenberg uncertainty relations.  For general smooth potentials
and for an initial minimum-uncertainty wavepacket with $\Delta x_0,$ $\Delta v_0$ chosen as above,  
it can be shown that the position and velocity spreads $\Delta x_t,$ $\Delta v_t$ at later times $t$
vanish for the limit $h/m\rightarrow 0$ and the mean values $\langle x\rangle_t$, $\langle v\rangle_t$
evolve according to the classical dynamics \cite{Hagedorn80}. Note that WKB corresponds to the special case $\beta=0.$   
However, if the potential is rough down to a length-scale $\ell$ and if $\ell\rightarrow 0$ as $h/m\rightarrow 0,$ 
then particle motion could remain stochastic in any of these alternate classical limits as well.

To give a careful formulation of such alternative semi-classical limits, we define a ``dimensionless-$\hbar$'' parameter 
\be  \varepsilon = \frac{\hbar}{m C^{1/2}\delta^{(3+\alpha)/2}}, \lb{epsilon} \ee
where $\delta$ is some suitable macro-length scale, which could be the outer length $L$ of the potential, the 
length $D$ of the space domain, or some specified fraction of either of these lengths. The quantity defined in 
(\ref{epsilon}) thus corresponds to the ``inverse P\'eclet number'' $\kappa/UL$ in terms of the turbulent advection -
quantum mechanics analogy discussed in Section \ref{models}. The alternative semi-classical limits are
then defined by taking $\varepsilon\ll 1$ with 
$$ \sigma=\ell/\mu, \quad \ell=\delta \cdot \varepsilon^\beta, \quad 0<\beta<1. $$
This choice of $\ell$ gives the potential additional mass-dependence beyond that in the prefactor $C\propto 1/m.$ 
If the Schr\"odinger equation (\ref{Schrod-by-m}) is non-dimensionalized by introducing 
$\tilde{x}=x/\delta,$ $\tilde{t}=t/(\delta^{1-\alpha}/C)^{1/2}, $ then it becomes
\be i\varepsilon\partial_{\tilde{t}}\psi=-\frac{\varepsilon^2}{2}\triangle_{\tilde{x}}\psi + \tilde{V}(\tilde{x})\psi \lb{Schrod-dim} \ee
with now $\tilde{m}=\tilde{C}=\tilde{\delta}=1$ and $\tilde{\ell}=\varepsilon^\beta$ for $\varepsilon\ll 1.$ 

We numerically study the semi-classical limit in the setting described above with $\beta=1/2$ and $\delta=D/128\pi$. 
We modified a publicly available code \cite{Vanderplas12} to solve the Schr\"{o}dinger equation in one dimension for the Kummer 
potential (\ref{eqn:KummerPot}), in the dimensionless formulation (\ref{Schrod-dim}). The code {\tt pySchrodinger} employs 
the so-called Strang operator-splitting spectral method, which is especially suitable for the semi-classical limit 
problem \cite{Bao02}. This approximation introduces second-order time-discretization 
errors by time-splitting, but has exponential accuracy in spectral computation of spatial derivatives for sufficiently smooth wave-functions.
It has the great advantage of giving evolution which is unitary, time-reversible, and gauge-invariant under spatially constant 
shifts in the potential, to machine precision. In the numerical experiments presented here we study an initial minimum-uncertainty 
wave-packet with $\langle \tilde{x}\rangle=\langle \tilde{v}\rangle=0$ and $\mu=1,$ making a sequence of runs with $\varepsilon= 
2^{-n}$ for all integers  $-7\leq n\leq 10$. We resolve both the spatial and temporal grid to $O(\varepsilon)$ and performed refinement 
studies to verify that our numerical simulations are converged.  

In Fig.~\ref{fig:pos-timeEvol} we show for $\varepsilon=2^{-7}$ a sequence of snapshots at successive times of the position-space PDF, 
along with the positions of the extremal classical solutions. Just as in the WKB limit, the PDF splits into a bimodal distribution at an early time 
and with the two peaks closely tracking the extremal classical solutions. 

We can also study splitting in momentum-space by taking the Fourier transform of the wave-function,
defined in dimensionless form by   
$$ \hat{\psi}(\tilde{p},\tilde{t})=\frac{1}{\sqrt{2\pi\epsilon^2}}\int d\tilde{x}\ \psi(\tilde{x},\tilde{t}) e^{-i\tilde{x}\tilde{p}/\varepsilon} $$
and implemented numerically by a fast Fourier transform (FFT). The momentum-space PDF 
$\rho(\tilde{p},\tilde{t})=|\hat{\psi}(\tilde{p},\tilde{t})|^2$ is plotted in Fig.~\ref{fig:mom-timeEvol} for the same 
cases as in the preceding Fig.~\ref{fig:pos-timeEvol}. For comparison, we also plot the momenta 
$p_\pm(t)=m v_\pm(t)$ of the extremal classical solutions in non-dimensionalized form. We see that there
is splitting of the wave-packet also in momentum-space and the two peaks of the momentum-space PDF closely 
track the momenta of the extremal classical solutions. Although we did not show it earlier, one can obtain similar results 
for the WKB limit. Numerically, the WKB wavefunction can be interpolated to a regular space grid with spacing 
$O(\varepsilon)$ and then transformed by FFT. One observes the same splitting in momentum-space there. 

\begin{figure}[h!]
\includegraphics[width=1\columnwidth,height=0.65\columnwidth]{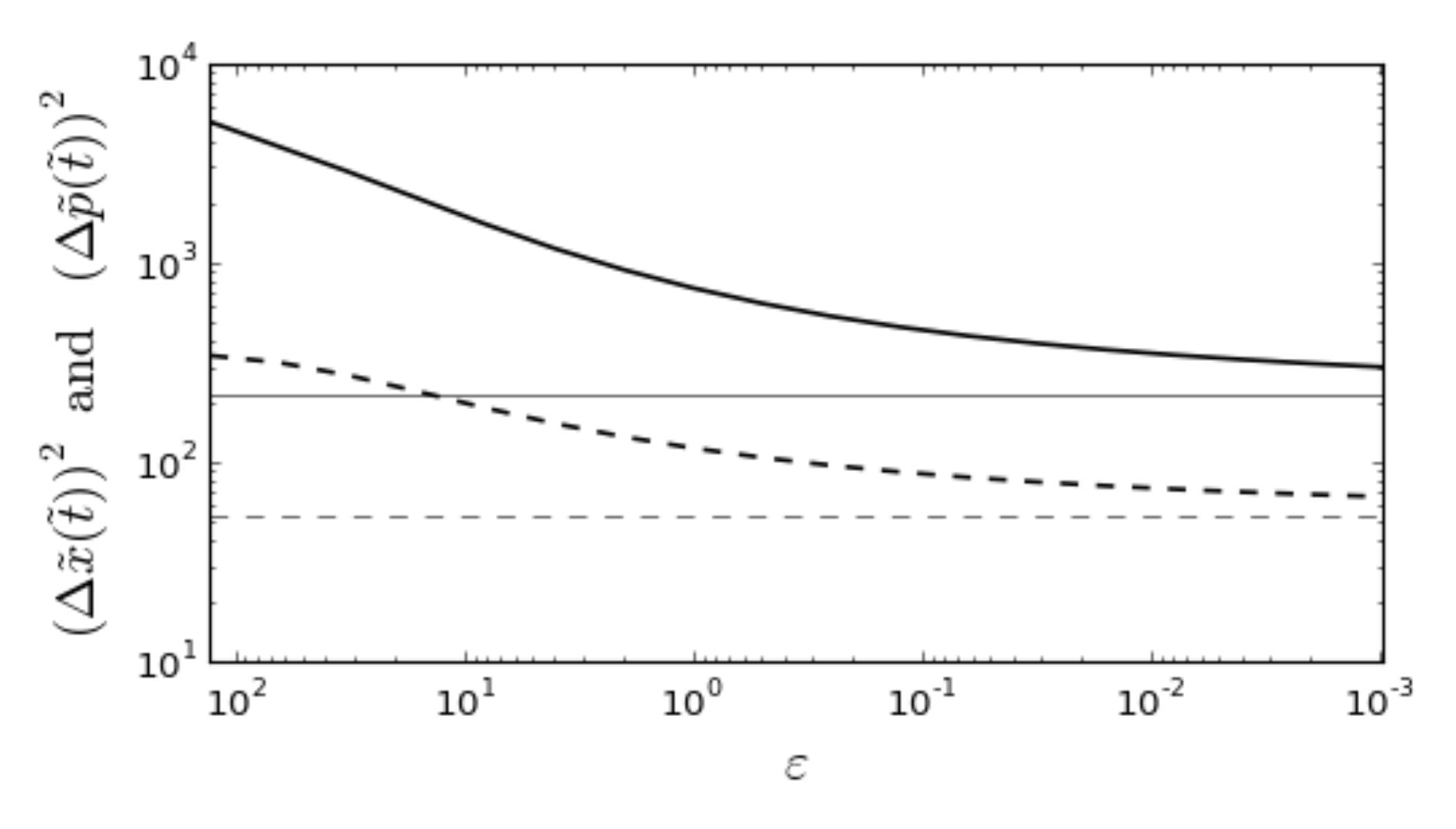}
\caption{ \raggedright  Uncertainties of position and momentum at the fixed time $t$ in Figs.~\ref{fig:pos-hbarEvol}-\ref{fig:mom-hbarEvol},
as functions of $\varepsilon.$ The heavy solid curve is the numerical position-space dispersion as defined by Eq. (\ref{eqn:disp}) and 
the light solid line is the position dispersion of the two classical extremal solutions (\ref{eqn:LimDisp}). The thick dashed curve is the 
numerical momentum-space dispersion and the thin dashed line is the momentum dispersion of the two classical extremal solutions.} 
\label{fig:NumDisp}
\end{figure}

We also study the behavior of the position-space and momentum-space probability distributions at a fixed time $t$ 
and $\ell=\sigma\propto \varepsilon^{1/2}$ for a decreasing sequence of $\varepsilon$ values.  The results are presented 
in Figs.~\ref{fig:pos-hbarEvol}-\ref{fig:mom-hbarEvol}.  In the semiclassical limit $\varepsilon \ll 1$, the probability densities
increasing concentrate around the positions and momenta of the two extremal classical solutions, providing 
good evidence of QSS in both position-space and momentum-space. We can quantify this further by calculating the 
position- and momentum-space uncertainties as a function of $\varepsilon$ at the same fixed time $t.$
These results are shown in Fig. \ref{fig:NumDisp}. The horizontal lines are the corresponding theoretical predictions from 
(\ref{eqn:LimDisp}) for the classical dispersion. We see a definite tendency of the numerical uncertainties to converge to their 
theoretical limit and, most importantly, become asymptotically independent of $\varepsilon,$ corroborating the visual evidence in 
Figs.~\ref{fig:pos-hbarEvol}-\ref{fig:mom-hbarEvol}.

\begin{figure*}[!ht] 
  \begin{subfigure}[b]{0.5\linewidth}
    \centering
    \includegraphics[width=1\columnwidth,height=0.5\columnwidth]{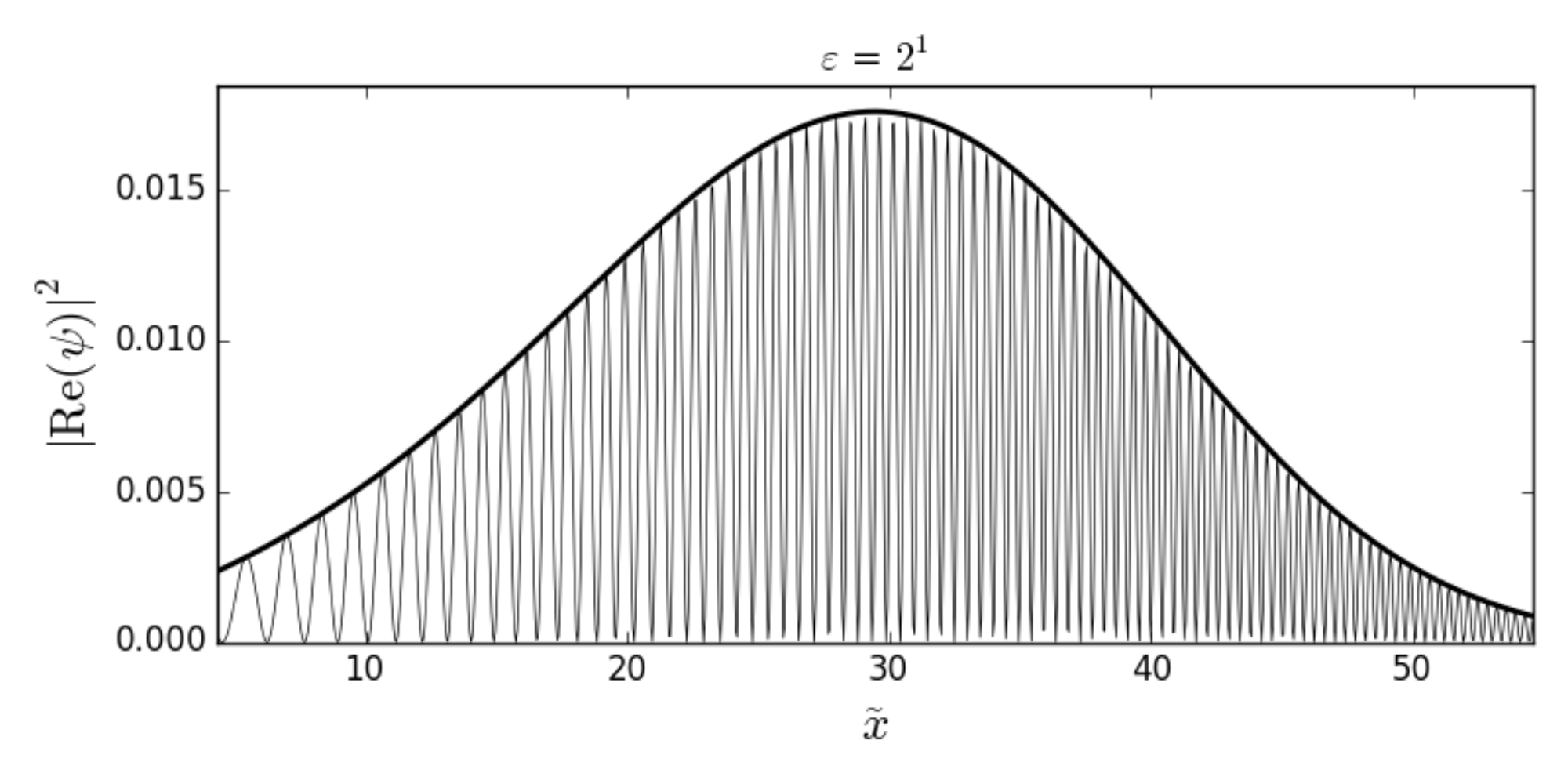} 
        \vspace{-8mm}
   \caption{} 
    \label{fig13:a} 
        \vspace{-1mm}
  \end{subfigure}
  \begin{subfigure}[b]{0.5\linewidth}
    \centering
    \includegraphics[width=1\columnwidth,height=0.5\columnwidth]{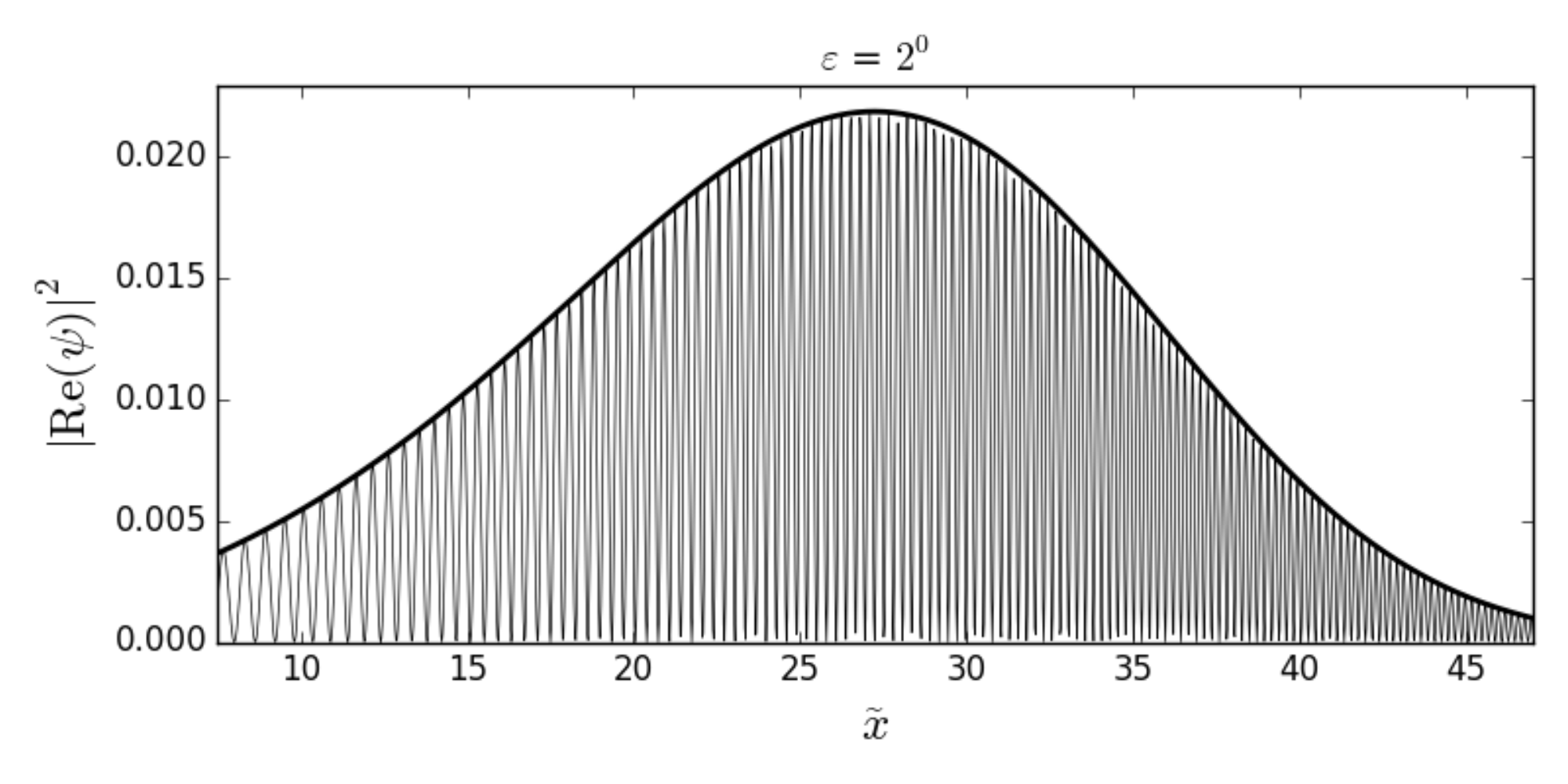} 
   \vspace{-8mm}
   \caption{} 
    \label{fig13:b} 
        \vspace{-1mm}
  \end{subfigure} 
  \begin{subfigure}[b]{0.5\linewidth}
    \centering
    \includegraphics[width=1\columnwidth,height=0.5\columnwidth]{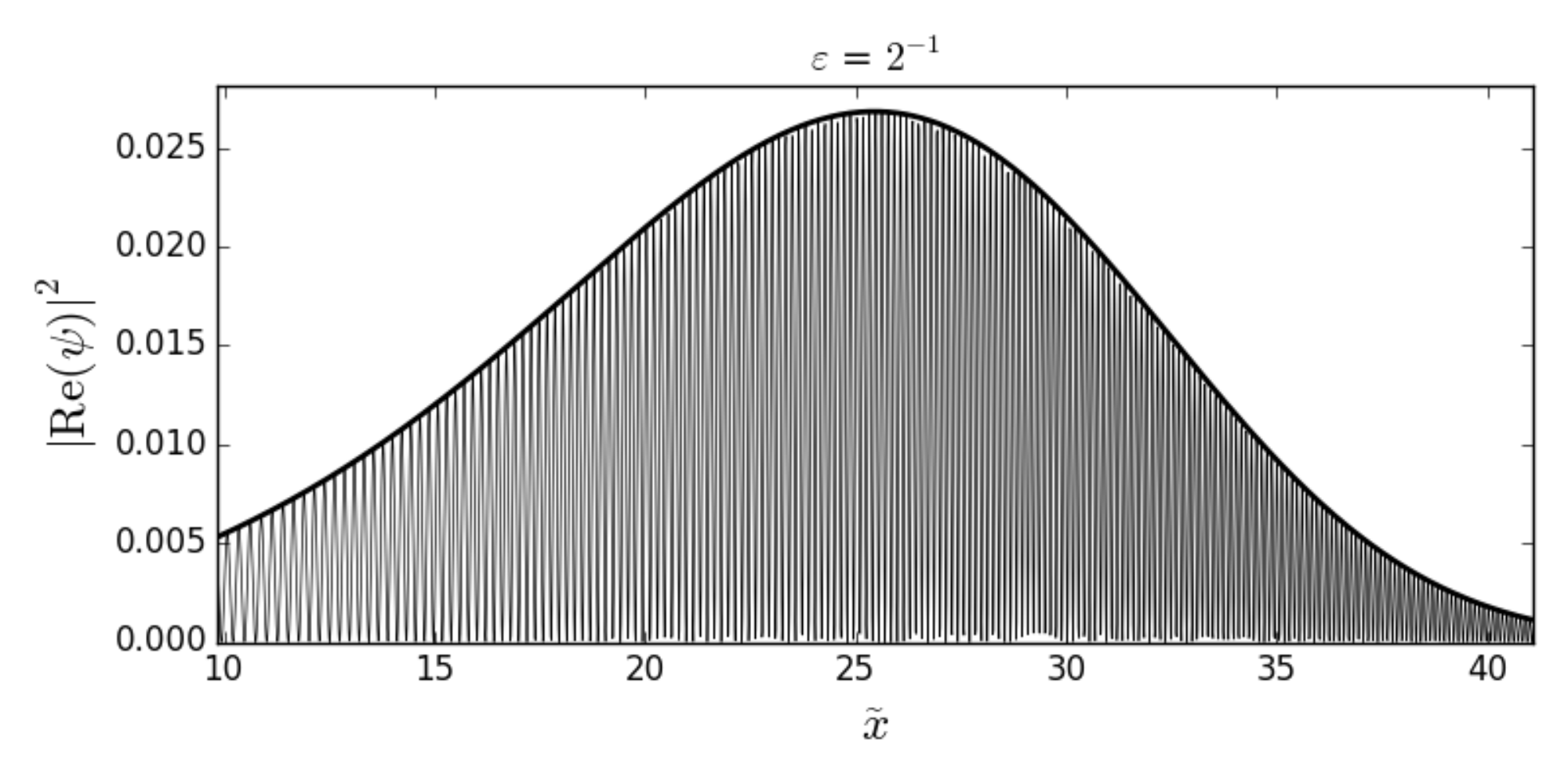} 
        \vspace{-8mm}
   \caption{} 
    \label{fig13:c} 
  \end{subfigure}
  \begin{subfigure}[b]{0.5\linewidth}
    \centering
    \includegraphics[width=1\columnwidth,height=0.5\columnwidth]{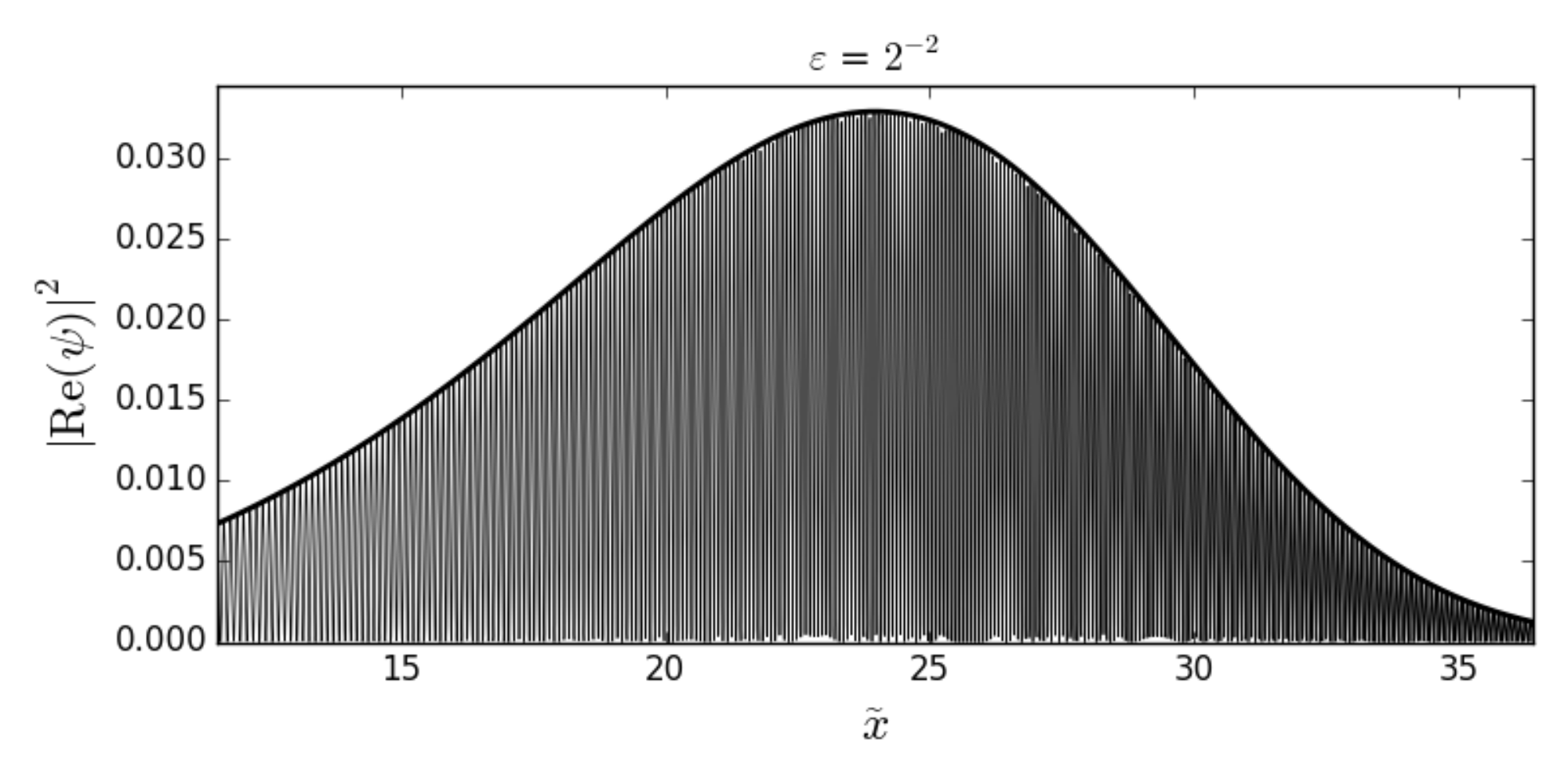} 
        \vspace{-8mm}
   \caption{} 
    \label{fig13:d} 
  \end{subfigure} 
  \caption{ \raggedright  Plot of $|{\rm Re}(\psi)|^2$ near the right peak of the position-space PDF,  for $\varepsilon= 2^{1}$, 
  $2^{0}$, $2^{-1}$, $2^{-2}$, as light black line. The solid black curve is the plot of the position-space PDF $\rho,$ the envelope of the first graph. 
  Each plot window height is a fixed multiple of the maximum value of the peak and each window width is a fixed multiple of the peak half-width. }
  \label{fig:hbarEvol} 
\end{figure*}


The previous results show that quantum systems may remain stochastic (``God still throws dice") in the classical limit, 
where determinism is expected. However, it might appear, even more shockingly, that the particle models discussed here 
become increasingly accurately described as two-state quantum systems in the limit $h/m\rightarrow 0,$ with just two possible 
values for the position (or momentum). If so, then it could be possible by experimental design of such systems to 
construct a  massive ``Schr\"odinger cat" with arbitrarily large $m$! The key question here is whether it is possible to achieve a 
coherent superposition of the two different states of the particle, leading to quantum interference effects. If there is no such
coherence, then any linear superposition of the two possible states really corresponds just to a statistical mixture within classical
probability theory.

There are compelling reasons for lack of coherent superpositions, arising from the quantum uncertainty relations. 
Note that coherence would require separately for each branch of a ``split'' wavefunction that the position-space spread
be comparable to the de Broglie wavelength, $\Delta x\sim \lambda.$ Otherwise, the phase of each branch wavefunction 
would oscillate rapidly within its envelope of width $\Delta x$ and the two branches, even if spatially recombined, would have 
overlap nearly zero. In that case, no interference between them could ever be observed. However, choosing $\Delta x\sim \lambda 
\propto h/m$ in each branch leads by the uncertainty relations to $\Delta v=O(1)$ for each branch individually and thus very 
rapid spatial spreading. In the limits where classical dynamics is expected, $\Delta x\gg\lambda,$ and thus no coherence 
of the branches. Hence, the QSS phenomenon allows indeterminism in the classical limit but not quantum superpositions 
or entanglement.

These arguments can be verified using our numerical Schr\"odinger solutions.  Although the two individual peaks of the 
position-space PDF in Fig.~\ref{fig:pos-hbarEvol} become increasingly narrow $\propto\varepsilon^{1/2}$ as 
$\varepsilon\rightarrow 0,$ they in fact contain increasingly rapid oscillations on the scale of the de Broglie 
wavelength $\sim O(\varepsilon).$ We show this in Fig.~\ref{fig:phase-hbarEvol} by zooming in on one of the two peaks 
(the rightmost) and plotting the function $|{\rm Re}(\psi)|^2 = \rho \cos^2(\varphi),$ whose envelope is the 
position-space PDF but which oscillates with the complex phase $\varphi$ of the wave-function. Clearly, the wave-functions 
become extremely oscillatory within the envelope of a single peak as $\varepsilon$ is decreased. This rapid phase 
oscillation renders the two branches of the wave-function decoherent, as it becomes impossible in practice to 
recombine them spatially in phase. This is also a feature of the WKB wave-function studied previously, as can be seen by 
calculating its phase $S_t(x),$ the classical action (\ref{action}), and observing that it grows with increasing mass $m$
of the particle. 

\section{Possible Experiments}

Such indeterministic, stochastic behavior persisting in the standard classical limits should be observable in laboratory 
experiments. We have examined in this paper only the simplest 1D model which exhibits the QSS phenomenon
and it may not be the easiest to situation to study experimentally. However, the results will readily extend 
to higher-dimensional cases that may be easier to create in the laboratory. The main requirement is that
the repulsive potential of the quantum particle should be (nearly) non-smooth at the maximum energy location.  
For example, the radially symmetric version of the cusp function (\ref{cusp}) as potential energy per unit mass, i.e. 
$$ V(\bx) = -\frac{C}{1+\alpha} r^{1+\alpha}, \quad r=|\bx|, $$
will lead to the same QSS phenomena. By standard arguments, the 3D Schr\"odinger equation for an 
angular momentum $\ell$ initial wave-packet will reduce to a 1D equation with an effective potential 
$V_{\mathrm{eff}}(r) = V(r) + {\hbar^2l(l+1) \over 2m r^2}$. In particular, for an initial $\ell=0$ wavepacket 
concentrated at the origin, the analysis in this paper carries over  and predicts a spherical wave of probability 
amplitude radiating from the origin in the standard classical limits where one naively expects the particle to remain 
sitting at the unstable maximum at the origin. In this case, the randomness involves a continuous infinity of random 
outcomes, corresponding to the possible directions of the outgoing particle, not just two as in the 1D model.

To verify the phenomenon as QSS, there are key signatures for the experimentalist to check. First, there should be super-ballistic 
power-law spreading of the wavepacket as in (\ref{QSS-spread}). Second, this spreading should be independent 
of the initial width of the wavepacket $\sigma$ and inner cutoff $\ell$ of the potential as those are decreased 
(but still much larger than the de Broglie wavelength). Third, the spreading should be independent of $h/m,$ 
which may be testable in some experiments by considering particles with differing masses. (Actually in a fixed potential, 
there will be a particular mass-dependence, through $C$ or classical inertial effects, rather than strict independence.)  
It may be easier for experimentalists to observe the corresponding signatures for momentum-space spreading,
as discussed in section \ref{SchrEq}. Any experiment attempting to observe QSS will need low temperatures and 
low levels of other sources of noise, to ensure that quantum fluctuations are driving the phenomenon. We consider   
here briefly three possible experimental realizations: ultra-cold AMO, ultra-cold neutrons, and optical analogues.

Ultra-cold atomic-molecular-optical (AMO) systems exhibit quantum evolution of neutral particles in an electric dipole potential 
created by the AC Stark effect \cite{MetcalfvanderStraten99,Grimmetal00}. This potential has the form 
$$V(\bx)=\frac{3\pi c^2}{2\omega_0^3}\frac{\Gamma}{\Delta}I(\bx)  $$ 
where $\omega_0$ is the optical transition frequency, $\Gamma$ is the damping rate corresponding to the spontaneous 
decay of the excited level, $I(\bx)$ is the spatial intensity of the laser light with frequency $\omega,$
and $\Delta=\omega-\omega_0$ is the detuning parameter. Achievable potentials have only a small energy scale (mK), 
but there is great flexibility in designing their spatial form (e.g. see \cite{Hendersonetal09,Muldoonetal12}). Attractive electric 
dipole potentials are widely used to create particle traps for atomic bose-condensates  and for quantum computing. 
A stationary repulsive potential for study of QSS could be accomplished with a blue-detuned  ($\Delta>0$) focused 
beam or red-detuned ($\Delta<0$) hollow beam. The QSS  phenomenon requires a large scale ratio, $L/\ell\gg 1,$ say, 
$L/\ell\gtrsim 100,$ which could be achieved perhaps with large-aperture lasers or a 
large number of individual beams. To get the cusp-potential would require a fine control of the spatial intensity distribution 
$I(\bx),$ so that a power-law such as $I(\bx)\propto |\bx|^{1+\alpha}$ could be obtained to a suitable approximation. 
The effects described here should not require an exact power-law but could have discrete steps 
of intensity or some rapid variation that averages out on the dynamical times scales of the particle. 
To achieve such a spatial distribution, one might use methods employed for quantum computing 
applications, such as artificial holograms with acousto-optic deflectors \cite{Hendersonetal09}, spatial light modulators \cite{Muldoonetal12}, 
or other devices. In contrast to quantum computing applications, one does not require rapid manipulation and control of potentials, 
but instead accurate specification of the intensity spatial distribution over a broad enough region. This seems a very promising 
approach to explore QSS in the laboratory.

Another possible system with which to realize QSS is ultra-cold neutrons \cite{RauchWerner15,Byrne13,UtsuroIgnatovich10}. 
These have energies 1-1000 neV, temperatures $\sim$mK, de Broglie wavelengths $\lambda=10$-$1000$ nm, 
velocities $v\sim 5$ m/s. Neutrons interact through through all four fundamental forces ---strong, weak, 
electromagnetic and gravitational---which provide various methods to generate potentials acting upon them.

One possibility is the magnetic dipole potential due to an external magnetic field, $V(\bx)=-\bomu\bdot {\bf B}(\bx)$, which acts on the neutron dipole 
moment with $\mu=60$ neV/T.  (Magnetic dipole forces are also an option for AMO systems). Magnetic field strengths $B$ as large as $30$ T 
are currently achievable with superconducting magnets and generally somewhat smaller strengths $\sim 1$ T for permanent magnets. 
Unless magnetic fields change very rapidly, the alignment of neutron polarization generally follows adiabatically the local magnetic 
direction, with no spin flips.  Thus the dipole interaction is repulsive for spins aligned or magnetic moments anti-aligned with the field (low-field seeking, 
LFS, state),  and attractive for spins anti-aligned/moment aligned (high-field seeking, HFS, state). Magnetic bottles for spin-polarized neutrons, as originally 
proposed in \cite{Vladimirskii61}, have been achieved for both LFS and HFS spin states and for electromagnetic and permanent magnets. 
See e.g. \cite{Brenneretal15} for many references. One could, in principle, use similar methods to generate a (UV-regulated) cusp potential.
An advantage of this approach is that magnetic interactions are ideal conservative and relatively ``clean" compared to methods involving 
material mirrors/walls. However, it is difficult for us to conceive of a practical arrangement of permanent magnets and 
electric currents that would produce a cusp-like potential, with sufficient accuracy.

It is also possible to exploit gravity, since ultra-cold neutrons have kinetic energies comparable to their gravitational potential
energies ($102$ neV per meter height). This can be done in conjunction with material mirrors (essentially, 
the strong interaction with the atomic nuclei, modelled by nuclear optical potentials). Such a gravity + strong force approach has already been 
used to create attractive potential wells and quantum bound states for ultra-cold neutrons \cite{Nesvizhevskyetal02}. 
For classical dynamics the 1D cusp potential can be generated by a particle constrained to move along a suitable space curve 
(``frictionless bead on a wire'') falling under gravity, so long as the required forces are $<mg.$ Quantum mechanically 
one might be able to exploit micron-scale or submicron-scale channels as neutron guides. Polycapillary glass fibers consisting of bundles 
of $\sim 1000$ individual lead-silica glass capillaries, each with diameter $\sim 6\mu $m, have been demonstrated to 
act as guides for thermal neutrons \cite{Chenetal92} and are commercially available \cite{Alvarez-EstradaCalvo14}. It is 
conjectured that neutron guides with inner diameters as small as $5$-$7$ nm may be possible, e.g. using carbon 
nanotubes \cite{CalvoAlvarez-Estrada04,Dabagov11}. When the diameter of the guide is comparable to the wavelength 
of the confined ultra-cold neutron, then simple energetic considerations as well as more detailed 
analysis \cite{Calvo00,Rohwedder02} imply that the wave propagation should be quasi-1D, with little excitation of 
high-order modes transverse to the tube axis. Using gently bent guide tubes to reduce losses \cite{Chenetal92,Calvo00}
and considering times short enough so that neutrons have velocities sufficiently small to remain confined, 
the 1D models of this paper might be obtained.

Another entirely different class of experiments that could realize our 1D models are based on optical 
analogues of Schr\"odinger evolution \cite{Longhi09,DragomanDragoman13}. For example, spatial propagation 
of a monochromatic light beam in guiding dielectric structures which are weakly curved along the propagation direction
are described by the paraxial optical wave equation: 
$$ i\hbar\frac{\partial\psi}{\partial z} =-\frac{\hbar^2}{2n_s}\frac{\partial^2\psi}{\partial x^2} +V(x)\psi-F(z)x\psi. $$
Here $\psi$ is the complex electric field amplitude modified by reference-frame and gauge transformations,  
$z$ is the paraxial propagation distance, $\hbar\equiv \lambda/2\pi$ is the reduced wavelength, $V(x) = [n^2_s -n^2(x)]/(2n_s) 
\simeq n_s - n(x)$, $n(x)$ is the refractive index profile of the guiding structure, $n_s$ is the reference (substrate) refractive 
index, $x_0(z)$ is the axis bending profile, and $F(z)=-n_s^2\ddot{x}_0(z).$ For full details, see \cite{Longhi09}. It 
appears feasible to produce a repulsive, cusp-like potential by manufacturing a dielectric medium with a suitable 
specification of $n(x)<n_s.$ Such experiments are not truly quantum-mechanical, however, but are to be more 
properly considered as ``analogue simulations" of the Schr\"odinger equation. Their value is to be assessed by comparison  
with direct numerical simulations of the Schr\"odinger equation, like those presented in this paper. Optical 
analogue experiments are useful if they can probe a complementary (or wider) range of parameters than those 
achievable by computer simulations.

The specific suggestions made here for experimental realization of QSS may not be the most feasible ones 
and, in any case, are mere sketches of possible approaches.  Other possibilities will surely occur to our 
ingenious experimentalist colleagues. However, we believe that our theoretical 
analysis makes a compelling case that such breakdown of classical determinism will eventually be 
observed for quantum systems in the laboratory. For this not to be true, quantum mechanics itself would 
need to fail in the distinctive situations considered here.

\section{Conclusions}

This paper has developed a novel prediction of quantum mechanics: the breakdown of determinism in the standard 
classical limit (large mass, small initial spreads in position and velocity) when the force-field acting 
on the quantum particle is nearly rough. Like the classical spontaneous stochasticity effect, it is produced by an explosive, 
super-ballistic acceleration of the particle and the resultant ``forgetting" of its initial conditions. Quantum spontaneous 
stochasticity is, in some sense, a diametric opposite of Anderson localization \cite{Anderson58}.
That effect suppresses spreading of the wave-function in a random (disordered) potential,
whereas QSS corresponds to an accelerated spreading. Furthermore, Anderson localization is essentially a 
wave-phenomenon, arising from destructive interference between multiple-scattering paths.  QSS is instead 
an essentially classical particle phenomenon: although the origin of the randomness lies in the quantum 
fluctuations, quantum stochasticity is magnified to macroscopic scales by classical non-smooth dynamics.

In this paper we have confined our attention to the initial-value problem for a quantum particle, described by 
the non-relativistic Schr\"odinger equation, but QSS should appear in other guises and theoretical frameworks. 
The Wigner distribution function method,  previously exploited in \cite{AthanassoulisPaul12}, appears to be a 
powerful method to analyze simultaneously the position-space and momentum-space spreading of the quantum 
particle. There should also be very interesting manifestations of classical non-smooth dynamics in the Feynman 
path-integral representation of quantum scattering amplitudes \cite{Feynman48}, not only in the classical limit but also 
in the fully quantum-mechanical setting. Studying the quantum phenomenon will help to shed further light on classical 
spontaneous stochasticity, which has fundamental importance for astrophysics, geophysics, and even engineering
applications. It is also critical to continue testing quantum mechanics in novel regimes that were previously 
unexplored.

\begin{acknowledgments}

We would like to thank Eberhard Bodenschatz, Kurt Eyink, Igor Kolokolov, Jorg\'e Kurchan, Guy Marcus 
and Thomas O'Conner for very useful conversations and suggestions on possible 
experimental realizations of quantum spontaneous stochasticity.  Abraham Harte informed us 
of the philosophical paper by Norton. TD would like to thank Cristian C. Lalescu and Navid Constantinou
for their help with the Schr\"odinger simulations.
\end{acknowledgments}

\bibliography{QSS.bib}

\end{document}